\documentclass[%
 aip,
 amsmath,amssymb,
 reprint,%
 citeautoscript
]{revtex4-1}
\usepackage[colorlinks=true,linkcolor=blue,citecolor=blue,urlcolor=blue]{hyperref}
\usepackage{graphicx}% Include figure files
\usepackage{dcolumn}% Align table columns on decimal point
\usepackage{bm}% bold math
\usepackage[utf8]{inputenc}
\usepackage[T1]{fontenc}
\usepackage{etoolbox}
\usepackage{xcolor}
\usepackage{booktabs}
\usepackage{subcaption}
\usepackage[version=4,arrows=pgf-filled,
mathfontname=mathsf]{mhchem}
\usepackage{caption}
\graphicspath{{./figures/}}

\makeatletter
\def\@email#1#2{%
 \endgroup
 \patchcmd{\titleblock@produce}
  {\frontmatter@RRAPformat}
  {\frontmatter@RRAPformat{\produce@RRAP{*#1\href{mailto:#2}{#2}}}\frontmatter@RRAPformat}
  {}{}
}%
\makeatother
\begin{document}

\preprint{AIP/123-QED}

% \title[Sample title]{LightPFP: Accelerating the Development of Task-Specific Machine learning interatomic potentials using Universal Potential}
% \title{LightPFP: Accelerating the Development of Task-Specific Machine learning interatomic potentials using Universal Potential}
\title{LightPFP: A Lightweight Route to Ab Initio Accuracy at Scale}

\author{Wenwen Li}
 \email{wenwenli@preferred.jp}
\affiliation{Preferred Networks Inc., Tokyo, Japan.}

\author{Nontawat Charoenphakdee}
 \email{nontawat@preferred.jp}
\affiliation{Preferred Networks Inc., Tokyo, Japan.}

\author{Yong-Bin Zhuang}
\affiliation{Preferred Networks Inc., Tokyo, Japan.}

\author{Ryuhei Okuno}
\affiliation{Preferred Networks Inc., Tokyo, Japan.}

\author{Yuta Tsuboi}
\affiliation{Preferred Networks Inc., Tokyo, Japan.}

\author{So Takamoto}
\affiliation{Preferred Networks Inc., Tokyo, Japan.}

\author{Junichi Ishida}
\affiliation{Matlantis Corporation, Tokyo, Japan.}

\author{Iori Kurata}
\affiliation{Preferred Networks Inc., Tokyo, Japan.}

\author{Ju Li}
\affiliation{Department of Materials Science and Engineering, Massachusetts Institute of Technology, Cambridge, MA 02139, USA}
\affiliation{Department of Nuclear Science and Engineering, Massachusetts Institute of Technology, Cambridge, MA, USA
}

\date{\today}% It is always \today, today,

\begin{abstract}
% ORIGINAL VERSION
% Machine learning interatomic potentials (MLIPs) bridge the accuracy of quantum mechanics with the efficiency of classical simulations. 
% However, a fundamental trade-off persists: universal MLIPs (u-MLIPs) offer broad transferability at high inference costs, while task-specific MLIPs (ts-MLIPs) achieve high efficiency but require expensive density functional theory (DFT)-generated training data.

% VERSION 2
% Atomistic simulation methods have evolved through successive computational levels, each building upon more fundamental approaches: from quantum mechanics to density functional theory (DFT), and subsequently, to machine learning interatomic potentials (MLIPs). 
% Universal MLIPs are broadly transferable but expensive to run at scale, whereas task‑specific MLIPs are faster but require costly DFT labels.

Machine learning interatomic potentials (MLIPs) bridge the gap between quantum-mechanical accuracy and scalable molecular simulations. However, a fundamental trade-off persists: universal MLIPs (u-MLIPs) offer broad transferability but are computationally expensive, whereas task-specific MLIPs (ts-MLIPs) achieve high efficiency at the expense of costly DFT training data.
Here we introduce LightPFP, a data-efficient knowledge distillation framework.
Instead of relying on expensive DFT calculations, LightPFP constructs a distilled ts-MLIP by using a u-MLIP to generate high-quality, material-specific training data, while employing a pre-trained lightweight model to further improve data efficiency.
Across diverse material systems, LightPFP reduces model development time by approximately three orders of magnitude compared with DFT‑based workflows, while maintaining DFT‑level accuracy. Combined with a LAMMPS/Kokkos inference backend, LightPFP achieves more than three orders of magnitude higher inference efficiency than u-MLIP baselines while enabling simulations of systems thousands of times larger.
When the teacher exhibits systematic bias, LightPFP supports few-shot precision correction using as few as 10 high‑accuracy DFT configurations, as demonstrated by reproducing MgO’s experimental melting point.
This u-MLIP-driven distillation establishes a scalable route to high-fidelity, data-efficient MLIPs, accelerating the pace of materials discovery and design.

\end{abstract}

\maketitle

% \begin{quotation}
% The ``lead paragraph'' is encapsulated with the \LaTeX\ 
% \verb+quotation+ environment and is formatted as a single paragraph before the first section heading. 
% (The \verb+quotation+ environment reverts to its usual meaning after the first sectioning command.) 
% Note that numbered references are allowed in the lead paragraph.
% %
% The lead paragraph will only be found in an article being prepared for the journal \textit{Chaos}.
% \end{quotation}

\section{Introduction}
\label{sec:introduction}

\begin{figure}
  \includegraphics[width=\columnwidth]{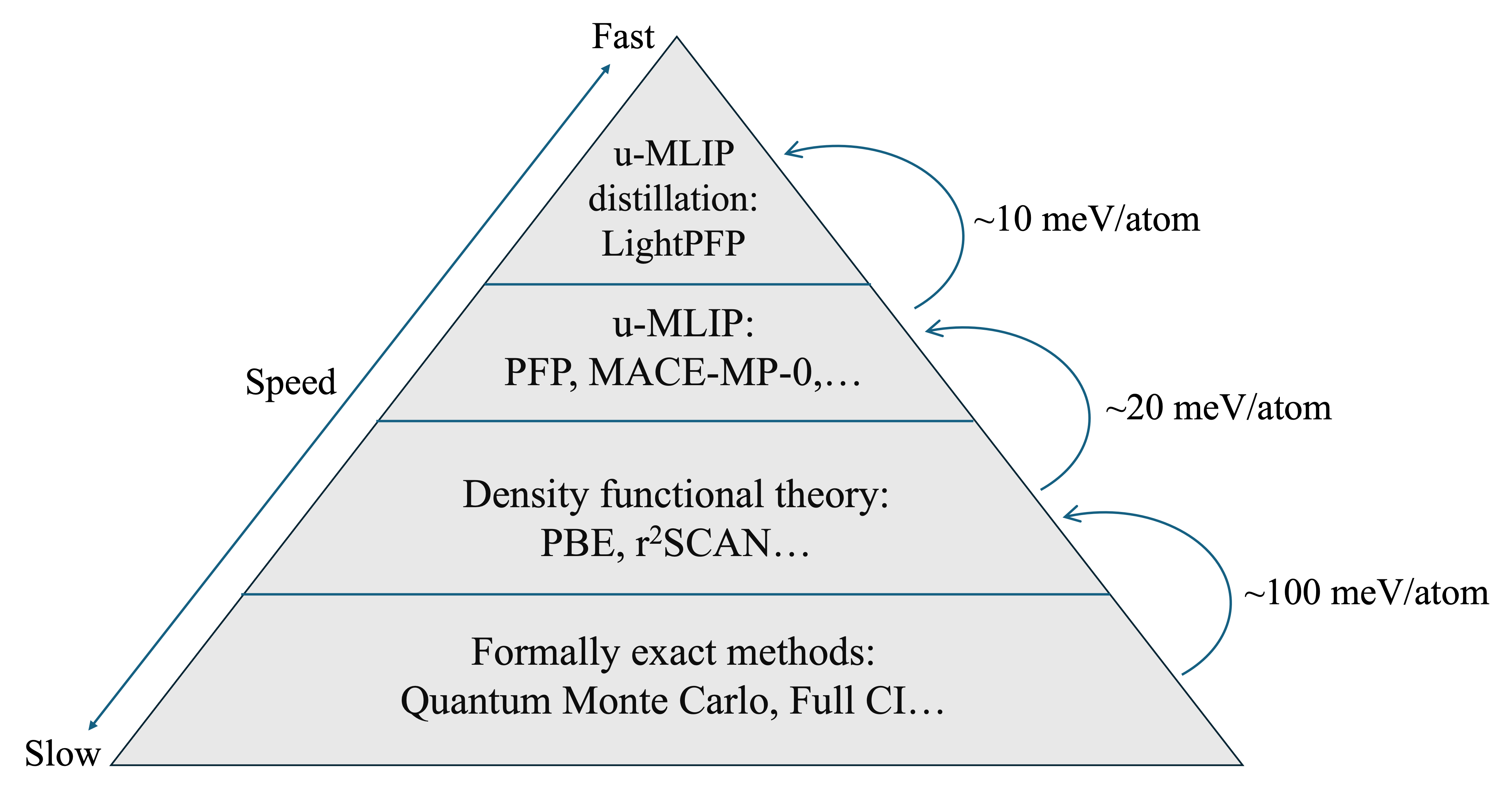} % 
  \caption{Hierarchy of atomistic calculation methods illustrating the trade-off between computational speed (left axis) and error accumulation (right axis).}
  \label{fig:hierarchy}
\end{figure}
The development of accurate and computationally efficient atomistic energy methods is essential for enabling large-scale atomistic simulations in materials science, catalysis, and chemistry.
The evolution of these methods over several decades forms a hierarchical framework (Fig.~\ref{fig:hierarchy}), in which higher-level approaches build upon the computational results of lower-level methods, improving efficiency at the cost of some accuracy.

% \begin{table*}
% %\centering
% \begin{ruledtabular}
% \begin{tabular}{lccccc}
% %\toprule
% \textbf{Perspective} & Morrow \textit{et al.}\cite{morrow2022indirect} & Amin \textit{et al.}\cite{amin2025towards} & Gardner \textit{et al.}\cite{gardner2025distillation} & Zhang \textit{et al.}\cite{zhang2024dpa} & This work \\ 
% \hline
% Teacher is trained from diverse datasets   & × & \checkmark & \checkmark  & \checkmark & \checkmark  \\ 
% Use teacher in data generation & \checkmark & ×  & \checkmark & \checkmark  & \checkmark \\ 
% Use active learning with teacher's labels   & × & ×  & \checkmark & \checkmark & \checkmark  \\ 
% Use student pretraining              & × & × & ×  & × & \checkmark  \\ 
% Does not require teacher's fine-tuning      & × & × & ×  & ×  & \checkmark \\ 

% %\bottomrule
% \end{tabular}
% \end{ruledtabular}
% \caption{Comparisons of the LightPFP framework with existing works related to distillation across different perspectives.}

% % matsumura2025knowledge  % dpa2-workflow-wang2025pre,
% \label{tab:comparison}
% \end{table*}
\begin{table*}
\begin{ruledtabular}
\begin{tabular}{lccccccc}
\textbf{Perspective} 
& Morrow~\cite{morrow2022indirect}
& Amin~\cite{amin2025towards}
& Gardner~\cite{gardner2025distillation}
& Zhang~\cite{zhang2024dpa}
& Wang~\cite{dpa2-workflow-wang2025pre}
& Matsumura~\cite{matsumura2025knowledge}
& This work \\ 
\hline
Teacher is trained from diverse datasets 
& × & \checkmark & \checkmark & \checkmark & \checkmark & \checkmark & \checkmark \\ 
Use teacher in data generation 
& \checkmark & × & \checkmark & \checkmark & \checkmark & \checkmark & \checkmark \\ 
Use active learning with teacher's labels 
& × & × & \checkmark & \checkmark & \checkmark & × & \checkmark \\ 
Use student pretraining 
& × & × & × & × & × & × & \checkmark \\ 
Does not require teacher's fine-tuning 
& × & × & × & × & × & \checkmark & \checkmark \\ 
%\bottomrule
\end{tabular}
\end{ruledtabular}
\caption{
Comparison of knowledge distillation approaches for u-MLIPs.
Columns labeled by surnames correspond to Morrow \textit{et al.}~\cite{morrow2022indirect},
Amin \textit{et al.}~\cite{amin2025towards},
Gardner \textit{et al.}~\cite{gardner2025distillation},
Zhang \textit{et al.}~\cite{zhang2024dpa},
Wang \textit{et al.}~\cite{dpa2-workflow-wang2025pre},
and Matsumura \textit{et al.}~\cite{matsumura2025knowledge}.
}
\label{tab:comparison}
\end{table*}

At the most fundamental level are quantum mechanical methods such as quantum Monte Carlo (QMC) and full configuration interaction (Full CI).
These approaches are formally exact but exhibit unfavorable scaling of computational cost and memory with respect to the number of electrons, which limits their application to systems containing on the order of ten atoms.
The next level consists of density functional theory (DFT), which incorporates information from more fundamental calculations at the lowest rung. 
For example, results from electron gas simulations using QMC\cite{CeperleyA80} are used to parametrize its exchange-correlation functionals (e.g. PBE generalized gradient approximation, or r$^{2}$SCAN meta-GGA approximation). 
DFT enables simulations of systems with a few hundred atoms in molecular dynamics, while single-point calculations can reach on the order of $10^3$ atoms depending on available computational resources. 
This balance between accuracy and computational cost has led to its widespread use in crystal structure discovery and property prediction.  
However, it remains computationally demanding to address certain classes of simulatons, including the direct simulation of extended defects and finite-temperature sampling.

At the next higher level in the hierarchy, machine learning interatomic potentials (MLIPs) are typically trained on large datasets generated from DFT calculations, although in principle higher-level ab initio methods such as coupled-cluster theory or quantum Monte Carlo can also serve as data sources.
These models are trained on chemically diverse structures spanning multiple elements and bonding motifs, and they encode physical symmetries to enable generalization across the periodic table.
Representative examples include PFP~\citep{takamoto2022towards}, M3GNet~\citep{m3gnet}, CHGNet~\citep{chgnet}, MACE-MP-0~\citep{mace,batatia2023foundation}, GRACE~\citep{lysogorskiy2026graph}, PET-MAD~\citep{mazitov2025pet}. 
In particular, PreFerred Potential (PFP) is noted for being trained on a highly complex and diverse DFT database, contributing to its robustness. 
Its applicability without fine-tuning has been demonstrated across a wide range of applications, including battery materials\cite{kong2025exploration,hinuma2025facile,narumi2025tailoring,son2024constructing,kwon2024intelligent,du2023new}, metal-organic frameworks \cite{shimada2024long,koh2024defect}, ceramics\cite{hinuma2024neural,miura2024stress}, catalysts\cite{watanabe2025oxidative}, polymers\cite{honbo2024effects}, nanotubes\cite{hisama2024molecular}, semiconductor processing\cite{kim2024sustained,jin2024atom}, hydrogen storage\cite{kim2024facile}, superconductors\cite{ishikawa2024evolutionary}, memristive  systems\cite{bae2024tunable}.
Despite their broad applicability, computational efficiency remains a limiting factor in large-scale simulations.

To address the limitations of both universal and task-specific MLIPs, we propose \emph{LightPFP}, a framework for constructing ts-MLIPs through knowledge distillation from a universal MLIP. 
LightPFP is designed to enable both efficient training and inference, while maintaining a favorable balance between computational cost and accuracy and avoiding the high training cost associated with DFT-based data generation.
Although implemented here with PFP as the teacher, LightPFP should be viewed as one instance of a broader u-MLIP-driven distillation route, with the appropriate interatomic potential remaining application-dependent according to the desired balance among accuracy, transferability, construction cost, and inference speed.

To assess the position of LightPFP within the hierarchy of atomistic methods, it is useful to examine the trade-off between different sources of error and computational cost in practical atomistic simulations.
LightPFP is positioned between universal MLIPs and classical potentials: it sacrifices the full universality of PFP by focusing on a target domain, but retains PFP-level accuracy with improved computational efficiency, while offering higher accuracy and broader applicability than classical potentials.
For reference, although Fig.~\ref{fig:hierarchy} does not include experimental methods, typical uncertainties in experimental thermochemical measurements of formation or reaction enthalpies are on the order of 1 kcal/mol, often referred to as “chemical accuracy”\cite{pople1999nobel}, corresponding to approximately 43 meV/atom.

The dominant source of error comes from the formally exact$\rightarrow$DFT (Fig.~\ref{fig:hierarchy}), due to the intrinsic limitations of DFT expressivity. 
The subsequent DFT→PFP transfer is already amortized and small\cite{TakamotoOLL23}; as shown later, PFP→LightPFP is even smaller and trains faster. 
One should also consider that practical errors can outweigh level-of-theory limits—for example, small PBC supercells induce electronic and elastic image interactions\cite{LiWCCBHY04}; by enabling much larger cells at comparable cost, LightPFP may help obtain defect energetics and reaction behaviors less affected by such artifacts, even though DFT is intrinsically more accurate.

% In some cases, models are constructed by directly bridging non-adjacent levels of the hierarchy. 
% For example, the recently proposed Multi-task Electronic Hamiltonian Network (MEHnet)~\cite{tang2025approaching} is trained through a direct transfer from CCSD(T) to MEHnet, bypassing intermediate representations such as DFT and universal MLIPs.
% While this approach reduces the number of intermediate steps, it results in a substantially increased cost in the training process.
% Based on the preceding discussion, the DFT→PFP→LightPFP pathway may achieve the preferable universality, practicality, and speed.
More generally, atomistic machine-learning models can also be trained from higher-level references beyond DFT, including coupled-cluster theory and quantum Monte Carlo data. MEHnet is one recent example of this broader direction, using CCSD(T)-level information to build a learned electronic-structure model.~\cite{tang2025approaching} Such high-reference-accuracy routes are complementary to LightPFP, which focuses instead on low-cost task-specific distillation from an already available u-MLIP.

\begin{figure*}
  \centering
  \includegraphics[width=0.9\linewidth]{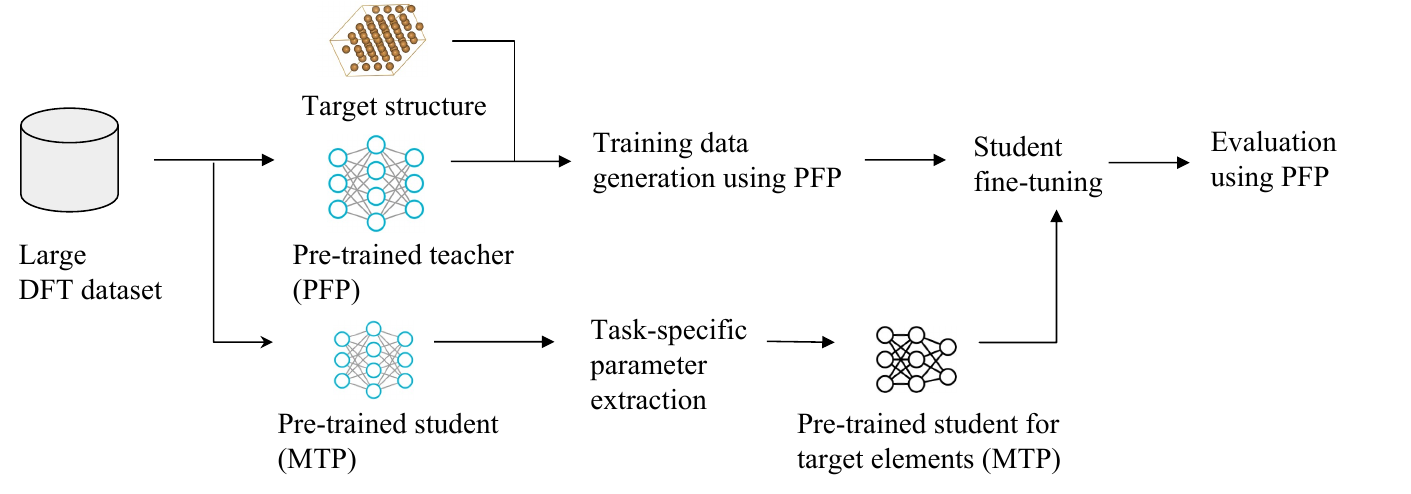} % 
  \caption{Schematic diagram of LightPFP.}
  \label{fig:lpfp-framework}
\end{figure*}

For knowledge distillation of u-MLIPs,~\citet{zhang2024dpa} proposed DPA-2, which is first pretrained on diverse datasets and then fine-tuned on task-specific DFT data. The fine-tuned model then serves as a teacher, whose knowledge is distilled into a smaller student model through active learning. Similarly,~\citet{gardner2025distillation} studied distillation from foundation models, such as MatterSim, MACE-MP-0, and OrbNet, to task-specific MLIPs, such as PaiNN and ACE.

However, these approaches still face two challenges: (1) u-MLIPs must be fine-tuned using DFT datasets, and (2) training student models from scratch requires large amounts of data to ensure robustness. As shown in Table~\ref{tab:comparison}, \emph{LightPFP} addresses these issues by using PFP as an off-the-shelf teacher, without fine-tuning, to generate training data. To further improve data efficiency, we pretrain the student model on existing DFT-labeled u-MLIP datasets before distillation.

A more recent work by~\citet{dpa2-workflow-wang2025pre} proposed an improved DPA-2 workflow, but the core principle summarized in Table~\ref{tab:comparison} remains the same. Another recent work by~\citet{matsumura2025knowledge} also keeps the teacher model off the shelf: the teacher is used to generate candidate training structures, a subset of which is selected for DFT labeling, and the student is then trained from scratch on the resulting DFT-labeled data. However, this work does not conduct active learning; if active learning were incorporated, the required DFT calculations could become a computational bottleneck.

In this paper, we first provide an overview of the LightPFP knowledge‑distillation framework and an assessment of the data efficiency of its distilled, pre‑trained student models. 
We then demonstrate LightPFP across four challenging applications that highlight complementary aspects of the method: (1) \ce{Li+} diffusion in the solid electrolyte \ce{Li6PS5Cl}, illustrating the trade‑off between model-building/inference speed and predictive accuracy; (2) the reaction kinetics of \ce{SiO2} etching by HF vapor, showcasing the integration of model distillation with active learning for complex reactive simulations; and (3) the melting point of \ce{MgO}, demonstrating that when u‑MLIP precision is insufficient, transfer learning with a small, high‑accuracy DFT dataset can substantially improve performance.

\section{Results}
\label{sec:results}

\subsection{LightPFP Framework Overview}

In this section, the overview of LightPFP is presented.
For the teacher model, we employ PFP based on TeaNet architecture\cite{takamoto2022towards,teanet}.
% PFP and LightPFP are both available in Matlantis\cite{matlantis}.
%PFP serves as the source for generating training data, which forms the foundation for training the student model in LightPFP.
As the student model, we adopt the Moment Tensor Potential (MTP), proposed by Novikov \textit{et al.}~\cite{mtp} due to its favorable trade-off between accuracy and efficiency\cite{zuo2020performance}. 
% MTP utilizes polynomial basis functions to represent the local atomic environment, which allows it to adaptively balance accuracy against computational cost through the adjustment of the model complexity parameter. 
The workflow of LightPFP shown in Fig.~\ref{fig:lpfp-framework} begins by defining a target structure and generating training data using PFP, including sampling and labeling. 
Students are pre-trained using Reptile meta-learning algorithm~\cite{reptile} on the diverse datasets described in reference \onlinecite{takamoto2022towards}. 
Finally, the reduced model is then fine-tuned using PFP-generated data, followed by an evaluation to assess its performance.
Importantly, pretrained students only need to prepare  once in advance and can be reused across a wide range of applications.
%LightPFP in Matlantis~\citep{matlantis} offers several pretrained students readily available.
% Due to the large dataset size, student pretraining is expected to take some time.
% Nevertheless, pretrained students can be prepared in advance. 
% LightPFP in Matlantis~\citep{matlantis} offers several pretrained students readily available.
Moreover, the model sizes of pretrained students can be reduced by removing MTP parameters for element pairs that are not present in the structure when they cover more elements than needed for a specific material.
The current LightPFP workflow should be regarded as an accelerated expert-in-the-loop procedure rather than a fully autonomous MLIP construction platform. Although PFP labeling, student fine-tuning, and validation can be executed within a standardized workflow once the target system is defined, several choices remain user-defined, including the student-model hyperparameters, data-sampling methods, active-learning strategies, and model-training hyperparameters. These choices can affect model accuracy, stability, and data efficiency. Automating such expert decisions, for example through adaptive or agent-based workflow control, is an important direction for future work~\cite{li2026lang2mlip}.

\subsection{Data efficiency of pretrained student models}
\label{sec:dataeff}

\begin{figure*}
\includegraphics{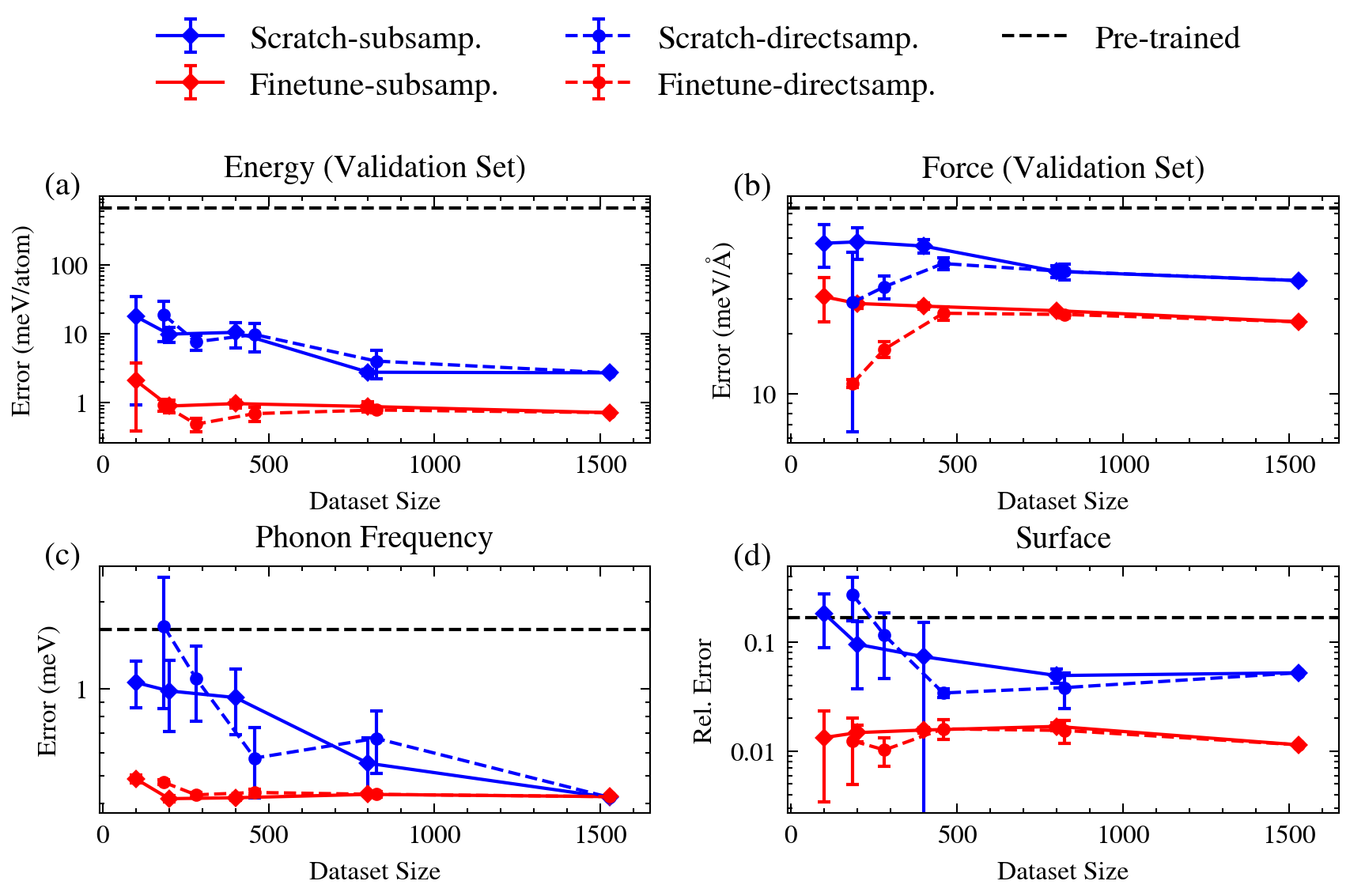}
\caption{Comparison of data efficiency between fine-tuned pretrained and scratch-trained student models. Each size dataset is created five times to obtain the uncertainties of errors.}
\label{fig:ni3al}
\end{figure*}

We first demonstrate the enhanced data efficiency of pretrained student models, using the \ce{Ni3Al} alloy\cite{jozwik2015applications} as an example. 
To this end, we prepared a PFP-labeled dataset of 1,529 structures spanning lattice perturbations and MD snapshots of bulk/defect/surface configurations. 
%a full dataset containing 1529 structures is prepared through the comprehensive sampling involving PFP\cite{takamoto2022towards} in the relevant configuration space. 
%The sampling methods comprise static and dynamic sampling. 
%The static method samples static structures by compressing and deforming their lattice, as well as displacing atomic positions. 
%The dynamic sampling uses MD simulations with initial configurations of both defect-free and defective bulk structures, as well as surface structures. 
%The details of sampling parameters are provided in SI.
% method 1 big one and subsampling 
% method 2 direct sampling 
For testing data efficiency, smaller datasets with sizes ranging from 100 to 850 are created by two methods, subsampling from the full dataset and direct sampling through reducing the length of the MD simulations. 
Structures in the datasets obtained by subsampling tend to be more widely distributed in configuration space, whereas direct sampling is closer to common user practice in real situations (i.e. by decreasing MD steps).
%discuss different evaluation metrics

We compare the performance of fine-tuned pretrained and scratch-trained student models on energy and force errors, as shown in Fig.~\ref{fig:ni3al}a,b. Across all dataset sizes, the fine-tuned pretrained student models outperform the scratch-trained student models. We note that fine-tuning pretrained student models on 100 structures performs almost as well as on 1529 structures. In addition to the standard energy and force testings, we validate the performance of student models on different application tasks, for instance, phonon spectra and surface energies, as shown in Fig.~\ref{fig:ni3al}c,d. Comparable to the energy and force testings, the performance of fine-tuned pretrained models is better than the scratch-trained models. 
Similar performance trend can be observed in other properties (See SI). 
%(see Appendix~\ref{app:ni3al-property-result}).

Moreover, the performance of fine-tuned pretrained student models is more robust in application tasks, whereas scratch-trained models show limited transferability. The force errors from the scratch-trained models on the smaller datasets are lower than on the larger dataset as shown in Fig.~\ref{fig:ni3al}(b). However, the errors on phonon spectra and surface energies are larger as shown in Fig.~\ref{fig:ni3al}(c,d). In contrast, although fine-tuned pretrained student models show a similar trend on force testing, their performance on application tasks are consistently reliable across various dataset sizes.

Fig.~3(a) also shows that the scratch-trained model begins to exhibit early signs of performance saturation, indicating diminishing returns from further increases in data under the current setup. 
This further highlights that pretrained models alleviate the need for extensive dataset coverage, enabling strong and more transferable performance with the available data.

Although the additional data required for a scratch-trained model to match the fine-tuned pretrained model is difficult to quantify, the saturation trend in Fig.~3a--b suggests diminishing returns from naive data scaling, indicating that more sophisticated data generation may be needed to close the gap.

% \section{Applications}
% \label{sec:app}

% In this section, we showcase LightPFP through four applications: (1) \ce{Li+} diffusion in the solid electrolyte \ce{Li6PS5Cl}; (2) the mechanical and grain boundary properties of \ce{AlCoCrFeNi} high-entropy alloy; (3) the reaction kinetics of \ce{SiO2} etching by HF vapor; and (4) the melting point of \ce{MgO}. 

\subsection{\ce{Li6PS5Cl}}
\label{subsec:li6ps5cl}

\begin{table*}
\centering
\caption{Composition of the training dataset for \ce{Li6PS5Cl}}
\begin{ruledtabular}
\begin{tabular}{lccc}
%\hline
\textbf{Sampling} & \textbf{Number of} & \textbf{Number of} & \textbf{Comment} \\ 
\textbf{method} & \textbf{structures} & \textbf{atoms} & \\ 
\hline
\multicolumn{4}{c}{\textbf{LightPFP Dataset (labeled by PFP)}} \\ 
\hline
MD & 1600 & 374400 & NPT MD at 300, 500, 1000, 1500K;\\ 
 &  &  &  1 sample per 100 steps  \\
rattle & 10 & 4160 & Random displacement of atoms \\ 
compress & 22 & 1144 & Compress and stretch lattice \\ 
deform & 48 & 2496 & Deform lattice \\ 
vacancy & 100 & 5100 & Create 1\textasciitilde2 vacancy \\ 
\textbf{Total} & \textbf{1780} & \textbf{387300} & \\ 
\hline
\multicolumn{4}{c}{\textbf{MTP-DFT Dataset (labeled by DFT)}} \\ 
\hline
MD & 800 & 41600 & NPT MD at 300, 500, 1000, 1500K\\ 
 &  &  &  1 sample per 100 steps  \\
rattle & 10 & 520 & Random displacement of atoms \\ 
compress & 22 & 1144 & Compress and stretch lattice \\ 
deform & 48 & 2496 & Deform lattice \\ 
vacancy & 100 & 5100 & Create 1\textasciitilde2 vacancy \\ 
\textbf{Total} & \textbf{980} & \textbf{50860} & \\ 
%\hline
\end{tabular}
\end{ruledtabular}
\label{table:lpscl_dataset}
\end{table*}

\begin{figure*}
\includegraphics[width=\textwidth]{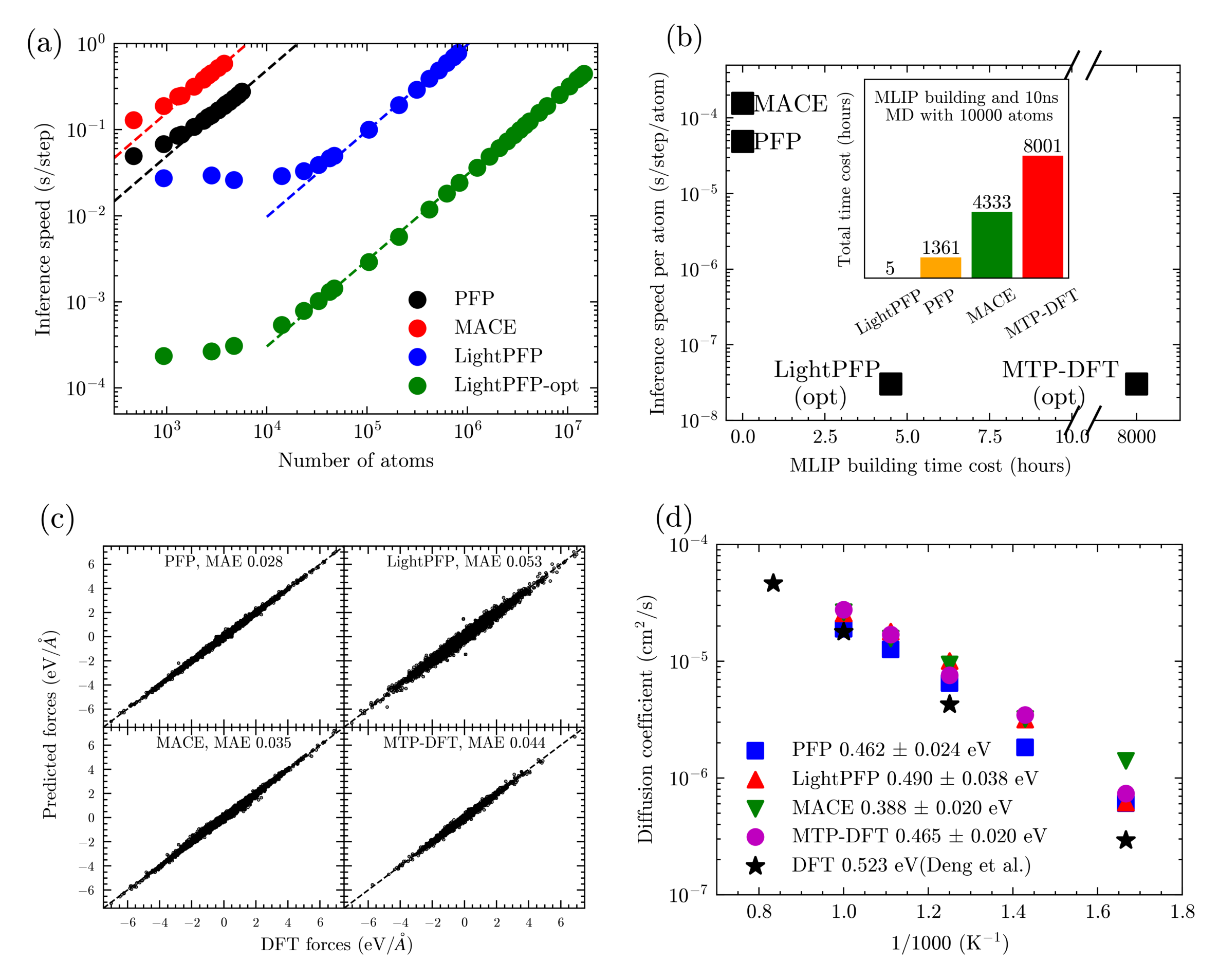}
\caption{Performance comparison of LightPFP and baseline MLIPs for \ce{Li6PS5Cl}.
(a) Molecular dynamics computational speed as a function of the number of atoms for PFP, MACE, LightPFP, and LightPFP-opt, where LightPFP-opt denotes the optimized LAMMPS/Kokkos implementation of LightPFP.
(b) Trade-off between the overall time spent on MLIP building for \ce{Li6PS5Cl}, including data collection and model training, and MD computational speed. Inset: the total time cost to complete both MLIP building and a 10 ns MD simulation of a 10,000-atom system.
(c) Parity plots comparing atomic forces predicted by each MLIP with DFT reference values.
(d) Arrhenius plots of \ce{Li+} diffusivity in \ce{Li6PS5Cl} from \textit{ab initio} MD simulations~\cite{deng2017data} and four MLIPs. The diffusion coefficients are averaged over eight independent trajectories; activation energies $E_a$ are reported in the legend, with error bars denoting the standard error of fitted slope.}
\label{fig:LPSCl}
\end{figure*}

This example focuses on a common solid-state electrolyte, \ce{Li6PS5Cl}, renowned for its high ionic conductivity, with potential applications in solid-state battery development. 
%Extensive experimental and theoretical studies have been conducted on \ce{Li6PS5Cl}. For example,  ~\citet{deng2017data} used \textit{ab initio} MD to calculate the diffusion coefficient and diffusion activation energy (0.52 eV) of Li in \ce{Li6PS5Cl} crystals. 
We demonstrate the advantages of the model distillation method compared to other approaches: (1) directly using universal potentials, and (2) training MLIPs with traditional DFT datasets. We first validate the effectiveness of the model distillation approach.

We compare four strategies for using MLIPs to perform atomistic simulations of \ce{Li6PS5Cl}. These are: directly using the u-MLIP PFP v7.0.0; distilling a compact task-specific MLIP with the MTP architecture from PFP (as described above), yielding LightPFP; using another u-MLIP, MACE-MPA-0\cite{batatia2025crosslearning}; and training an MTP model directly on DFT data. For brevity, we refer to these strategies throughout as PFP, LightPFP, MACE, and MTP-DFT, respectively. LightPFP and MTP-DFT share the same MTP architecture, and software implementation; consequently, their inference-time efficiency is essentially identical.

% data collection for lightPFP
The LightPFP model is obtained by distilling knowledge from the PFP through sampling \ce{Li6PS5Cl} configurations by molecular dynamics and other molecular simulations (such as lattice stretching, compression, deformation, atomic displacement, etc.) and labeling the sampled configurations by PFP to obtain their corresponding energies, forces, and stresses. Dataset acquisition takes 3.5 hours on a single GPU. The dataset composition is listed in Table~\ref{table:lpscl_dataset}. Additionally, our commonly used data collection methods are detailed in supplementary information (SI) I. Subsequently, we perform 1 hour of training, using weights from a pretrained MTP model for initialization.
Ultimately, using only 4.5 hours, we obtain the LightPFP model.

%MTP
MTP-DFT's dataset is labeled with DFT. To reduce cost, we generated trajectories with PFP and then labeled sampled snapshots post hoc, rather than running fully \textit{ab initio} MD. We note that configurations generated by PFP may follows different distribution than those from \textit{ab initio} MD. End-to-end data collection took 100 wall-clock hours on our setup. A simple extrapolation — multiplying MD steps with DFT single-point time — suggests fully DFT-driven MD would require ~8,000 hours under comparable settings. Thus, constructing a ts-MLIP in this traditional DFT-labeled manner is substantially more time- and compute-intensive than the distilled LightPFP route. Dataset composition is listed in Table~\ref{table:lpscl_dataset}. Because Kohn–Sham DFT in plane-wave formalism nominally scales as O(N$^3$), we favored smaller cells; consequently, the MTP-DFT dataset is smaller and skewed toward structures with fewer atoms compared to the LightPFP dataset. After data collection, model training took about one hour.

%Test

After preparing the four models, we first test their computational speed and memory efficiency. 
Figure~\ref{fig:LPSCl}(a) shows the MD inference speed as a function of system size, with all models benchmarked on the same NVIDIA V100 GPU platform with 16 GB memory.
Owing to this lightweight task-specific architecture, LightPFP already achieves an inference speed of $9.70 \times 10^{-7}$ s/step/atom in the original ASE/PyTorch implementation,~\cite{hjorth2017atomic,paszke2019pytorch} where force evaluation relies on PyTorch automatic differentiation. 
This corresponds to approximately 50-fold and 162-fold speedups over PFP ($4.90 \times 10^{-5}$ s/step/atom) and MACE-MPA-0 ($1.56 \times 10^{-4}$ s/step/atom), respectively. 
Because LightPFP has a simple architecture, including a lightweight neural-network readout, we further optimized its inference using a LAMMPS/Kokkos backend.~\cite{thompson2022lammps,trott2021kokkos,meng2026kokkos} 
This architecture makes it practical to implement force computation directly, without relying on PyTorch automatic differentiation, and to design memory-access-optimized kernels in Kokkos/CUDA. 
% Because the distilled model has a simple analytical form, we further optimized LightPFP inference using a LAMMPS/Kokkos backend.~\cite{thompson2022lammps,trott2021kokkos,meng2026kokkos} 
% This implementation uses the same trained LightPFP model but computes forces using explicitly implemented analytical derivatives, while also benefiting from efficient GPU execution and pre/post-processing in LAMMPS/Kokkos.
The optimized LightPFP achieves an inference speed of $3.03 \times 10^{-8}$ s/step/atom and can simulate systems containing up to 14.56 million atoms on a single 16-GB GPU. 
Compared with the u-MLIP baselines, this corresponds to approximately 1,630-fold higher inference efficiency and 2,590-fold larger maximum system size than PFP, and approximately 5,150-fold higher inference efficiency and 3,900-fold larger maximum system size than MACE. 
The efficiency drop of LightPFP below ~10,000 atoms is characteristic of GPU-based MLIPs due to GPU underutilization.
% Note that the inference speed of LightPFP/MTP-DFT is related to the model's hyperparameters (e.g., level max, number of radial basis functions, etc.). 
% In these simulations, we use relatively simple LightPFP/MTP-DFT models because of its limited application scenarios. 
%This is because model distillation makes model training simpler and faster, so its application scenario is for very specific simulations, and the final model does not need to be too complex.

Figure~\ref{fig:LPSCl}(b) summarizes the trade-off between MLIP construction cost and inference efficiency. 
PFP and MACE-MPA-0 require no task-specific construction but are limited by high inference cost, whereas MTP-DFT reaches comparable inference efficiency to LightPFP only after an expensive DFT-based training workflow. 
The LightPFP with optimized LAMMPS/Kokkos backend therefore occupies the most favorable region, combining rapid construction through teacher-model distillation with production-scale inference efficiency. 
The inset further illustrates the total wall-clock time to simulate a 10,000-atom \ce{Li6PS5Cl} system for $10^7$ steps. Under this scenario, LightPFP achieves the shortest total time, completing the task 272–866× faster than u-MLIPs, and its advantage grows with increasing MD length. Conversely, for very short simulations, LightPFP’s initial construction overhead can diminish its advantage relative to u-MLIPs. Despite similar inference speed to LightPFP, MTP-DFT remains slower overall because its total time is dominated by DFT data generation.

While LightPFP offers substantially higher overall efficiency—both in MLIP construction and MD simulation—than existing u-MLIPs and DFT-trained MLIPs, its attainable precision is constrained by two factors: (i) reduced model capacity relative to u-MLIPs and (ii) training on PFP-generated labels rather than DFT, which can propagate the teacher’s deviations from DFT. To quantify these effects, we benchmark force predictions against DFT dataset used for MTP-DFT training (Fig.~\ref{fig:LPSCl} (c)). The uMLIPs, i.e. PFP and MACE, attain the lowest MAE (0.028 eV/\text{\AA} and 0.035 eV/\text{\AA}). As expected, LightPFP exhibits a modestly higher MAE (0.053 eV/\text{\AA}), reflecting both inherited PFP errors and architectural simplification. For comparison, the MTP-DFT trained directly on this dataset achieves 0.044 eV/\text{\AA}; because portions of the DFT set were used for training, only the 10\% held-out test split is shown in the parity plot. Crucially, the gap between LightPFP and PFP/MTP-DFT is small, supporting the feasibility of distilling a reliable u-MLIP into a lightweight model with limited loss in precision.

We next examine how the force-accuracy differences translate into a transport property by computing \ce{Li+} diffusion in \ce{Li6PS5Cl}. 
The NVT MD simulation is performed at 600 - 1000 K with the same settings as the \textit{ab initio} MD in reference~\cite{deng2017data}. 
%The workflow is: (i) relax a 1×1×1 \ce{Li6PS5Cl} cell (52 atoms), optimizing both atomic positions and lattice; (ii) run NVT MD for 100 ps at 600, 700, 800, 900, and 1000 K, using the same settings as the \textit{ab initio} MD in reference~\cite{deng2017data}, with eight independent replicas per temperature for statistics; (iii) extract \ce{Li+} diffusion coefficients from the mean-squared displacement. 
Figure~\ref{fig:LPSCl}(d) compiles diffusion coefficients from the four MLIPs alongside \textit{ab initio} MD results. 
All four MLIPs slightly overestimate diffusion across temperatures; PFP is nearest to DFT, whereas MACE shows the largest overestimation.
%Consistent with the force-MAE trends, all four MLIPs slightly overestimate diffusion relative to the DFT reference at every temperature. Among them, PFP is closest to DFT, while MACE shows the largest overestimation, in line with its higher force MAE.
Despite LightPFP’s slightly higher force MAE, its predictions remain close to PFP and MTP-DFT across temperatures.
% Notably, LightPFP and MTP-DFT exhibit overestimation magnitudes similar to PFP, and the gap between LightPFP and PFP is small across temperatures despite LightPFP’s somewhat larger force MAE. This suggests that the modest force-error increase introduced by distillation has only a limited impact on this property. 
Arrhenius fits yield activation energies of 0.523 eV (DFT), 0.462 eV (PFP), 0.490 eV (LightPFP), 0.388 eV (MACE) and 0.465 eV (MTP-DFT). 
Among the MLIPs tested, LightPFP gives the activation energy closest to the DFT value. Although some of this agreement may be incidental, it suggests that, for this system, distillation error is not the main source of property-level discrepancies; simulation choices, e.g. simulation size and thermostat settings, contributes as much as—or more than—the residual model error. Notably, MACE shows the largest error in activation energy, underscoring the importance of teacher quality: a strong universal teacher can yield a student that, on this benchmark, rivals or even surpasses more complex models trained directly on DFT.
%LightPFP’s activation energy is, in fact, the closest to the DFT value among the MLIPs considered. While some of this agreement may be incidental within statistical and methodological uncertainties, it indicates that, at least for this system, errors introduced by model distillation are not the dominant source of discrepancy in property-level predictions. Instead, differences arising from simulation setup (thermostatting, sampling length) and the DFT reference itself can be comparable to or larger than the residual model error.

We further assessed LightPFP on a chemically disordered system, the AlCoCrFeNi Cantor alloy. 
The LightPFP with LAMMPS/Kokkos backend achieves an inference speed of $3.57 \times 10^{-8}$ s/step/atom, corresponding to 2,070$\times$ and 7,420$\times$ speedups over PFP and MACE, respectively, and supports simulations of up to 14.08 million atoms on a single 16-GB GPU. The model-building speed is also three orders of magnitude faster than that of the MTP-DFT routine.
LightPFP reproduces DFT across the equation of state, elastic constants, surface energies, and grain-boundary energies, with no significant deterioration compared to PFP and MTP-DFT. Details are provided in SI.

\subsection{Dry etching of SiO$_2$: application of active learning}
\label{subsec:dry_etching}

In this example, we consider a more demanding application: dry etching of the \ce{SiO2}(100) surface by HF. 
Dry etching is vital in semiconductor processing yet difficult for atomistic simulation. Modeling spans tens–hundreds of nanometers and couples complex surface chemistry with high-energy bombardment, demanding accurate and robust MLIPs.
%Dry etching is a critical step in semiconductor processing, yet atomistic simulations are particularly challenging. Device-scale simulations require tens to hundreds of nanometers, while the process itself couples complex surface reactions with intense atomic interactions under high-energy bombardment. These demands place stringent requirements on the accuracy and robustness of MLIPs. 
Here, we combine model distillation with active learning to rapidly construct a LightPFP model. 
Because DFT-based active learning is expensive here, we do not compare DFT-labeled MTP models; instead, we focus on the PFP–LightPFP pipeline to demonstrate applicability over relative speed/accuracy.
%Given the prohibitive cost of DFT-based active learning in this setting, we do not construct or compare DFT-labeled MTP models; likewise, we focus on the PFP–LightPFP pipeline rather than benchmarking additional universal models, as our goal is to demonstrate applicability rather than relative speed/accuracy.

An initial dataset was collected via PFP-driven sampling, covering \ce{SiO2} crystal and surface, \ce{HF} molecule, and representative products such as \ce{SiF4} and \ce{H2O}. Dataset collection took 4.5 hours and was used to train an initial LightPFP model. As expected, the initial model was insufficient for dry-etching simulation, having not yet learned the interactions arising from high-velocity \ce{HF} impacts on \ce{SiO2}. We then entered an iterative active-learning loop in which LightPFP runs reactive MD under HF bombardment. In each round, LightPFP ran the MD, and we selected informative configurations by directly comparing LightPFP and PFP predictions and flagging frames with large discrepancies.\cite{vinod2025lfab} The flagged frames were labeled with PFP and immediately used to update LightPFP. Because training was warm-started from a pretrained student MTP, each update converged quickly and finished within 0.5 h. We repeated this MD–selection–update cycle for about 15 iterations, then pooled all collected data for a brief final training stage to obtain a robust production LightPFP model. The total wall-clock time for model construction was approximately 24 h.

% We then entered an iterative active-learning loop in which the current LightPFP model drives reactive MD of the etching process: a \ce{HF} molecule are inserted above the \ce{SiO2} surface with kinetic energies randomly sampled in the range from 20 eV to 80 eV and directed perpendicular to the surface; trajectories are propagated in the NVE ensemble for 200 fs with a 0.2 fs timestep to resolve high-energy collisions, followed by 1,000 fs of NVT dynamics (1 fs/step) to cool to 300 K. This insertion cycle is repeated 100–200 times per iteration.
% To select informative configurations, we directly compare LightPFP and PFP predictions and flag frames with large discrepancies; the selected structures are labeled by PFP and used to re-train LightPFP. To accelerate updates, each training step is capped at 0.5 hours. This is feasible because we warm-start from a pretrained student MTP, so fine-tuning converges rapidly to a satisfactory model for the next MD round. We perform 15 iterations of data collection and model update, completing the end-to-end process within 16 hours. After the active-learning loop, all collected datasets are pooled and used for a longer final training run to obtain a more reliable production LightPFP model. In total, the wall-clock time to build the LightPFP MLIP for this application is approximately 24 hours.

\begin{figure*}
\includegraphics[width=\textwidth]{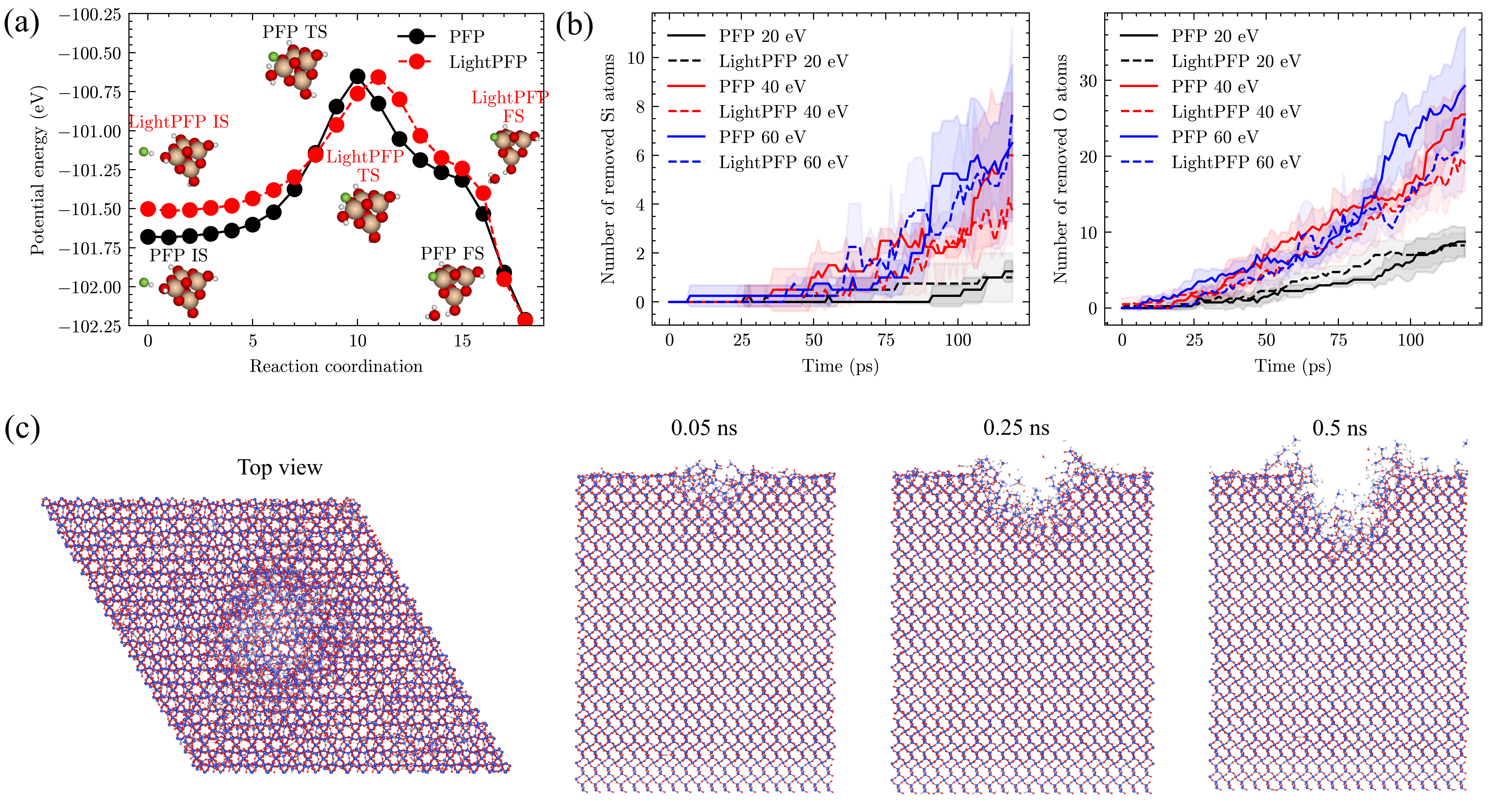}
\caption{Reaction pathway, etch yield, and surface morphology during HF dry etching of SiO$_2$.
(a) Reaction pathway computed with the nudged elastic band (NEB) method for an HF molecule reacting with a SiO$_2$ surface, comparing PFP and LightPFP. Insets show the initial (IS), transition (TS), and final (FS) states.
(b) Cumulative number of removed atoms versus time from molecular dynamics simulations using PFP and LightPFP at three incident kinetic energies (20, 40, and 60~eV). Left: Si atoms; Right: O atoms. Solid lines represent averages over four independent trajectories; shaded regions denote the standard deviation.
(c) Surface morphology evolution of SiO$_2$ during etching with LightPFP at an incident energy of 40~eV. The left panel shows the top view after 0.5~ns (etched region: central $2\times2$~nm square), while the right panels show longitudinal cross-sections at $t=0.05$, $0.25$, and $0.5$~ns.}
\label{fig:SiO2}
\end{figure*}

% \begin{figure}
% \includegraphics[]{SiO2_reaction_path.png}
% \caption{Reaction pathway from NEB calculation for the reaction of an HF molecule with a \ce{SiO2} surface during dry etching, computed using PFP and LightPFP. Atomic structures of the initial state (IS), transition state (TS), and final state (FS) are shown.}
% \label{fig:SiO2_reaction_path}
% \end{figure}

% \begin{figure*}
% \includegraphics[]{SiO2_etching_rate.png}
% \caption{Cumulative number of removed atoms versus time during HF dry etching of a \ce{SiO2} surface, from molecular dynamics simulations using PFP and LightPFP at different incident kinetic energies. (a) Si atoms; (b) O atoms.}
% \label{fig:SiO2_etching_rate}
% \end{figure*}

% \begin{figure*}
% \includegraphics[width=\textwidth]{figures/SiO2_etching_large_scale.png}
% \caption{Surface morphology of \ce{SiO2} during dry etching at a kinetic energy of 40 eV, obtained from a large-scale molecular dynamics simulation with LightPFP. (a) Top view after 0.5 ns of etching; the etched region is the central $2 \times 2$ nm square. 
% Longitudinal cross-sections through the etched region at (b) t~= 0.05 ns, (c) t~= 0.25 ns, and (d) t~= 0.5 ns.}
% \label{fig:SiO2_etching_large_scale}
% \end{figure*}

To validate the reliability of the LightPFP model obtained through active learning, we first examine a representative surface reaction. As shown in Fig.~\ref{fig:SiO2} (a), an HF molecule approaches a dangling OH group on a \ce{SiO2} cluster, displaces an \ce{H2O} molecule, and forms an Si–F bond. We computed the reaction pathway and barriers using the nudged elastic band (NEB) method with both PFP and LightPFP. The initial state (IS), transition state (TS), and final state (FS) structures—shown as insets—agree closely between the two models. The forward/backward barriers are 1.029/1.561 eV for PFP and 0.844/1.560 eV for LightPFP. For reference, literature DFT barriers are 0.929/1.424 eV, while ReaxFF yields 1.848/2.706 eV~\cite{kim2021molecular}. LightPFP’s deviations from DFT are 0.085 eV and 0.136 eV, comparable to PFP’s deviations of 0.100 eV and 0.137 eV, and far smaller than ReaxFF’s errors. Notably, the training data omitted NEB paths or \ce{SiO2} clusters; LightPFP’s agreement arises from related configurations generated during the active-learning MD, demonstrating transferability. We then assess the performance of LightPFP in MD simulations of dry etching at HF incidence energies of 20, 40, and 60 eV. Figure~\ref{fig:SiO2}(b) shows the cumulative number of removed Si and O atoms as a function of simulation time, averaged over four independent trajectories, with shaded regions indicating the standard deviation. LightPFP and PFP yield closely matching etch yields and energy-dependent trends across all three incidence energies, and the overlapping uncertainty bands confirm the statistical consistency between the two models.

To probe scalability and practical applicability, we performed a near feature-scale reactive MD simulation with LightPFP in a 10.06 × 10.06 × 20.00 nm simulation box. HF molecules with 40 eV kinetic energy were injected toward a confined 2 × 2 nm patch; 1,000 molecules were delivered over 0.5 ns. Figure~\ref{fig:SiO2}(c) illustrates the evolution of the surface morphology: by ~0.05 ns, initial removal begins at the impact site; by ~0.25 ns, a recessed pit is clearly formed; by 0.5 ns, the pit develops into a localized crater of ~2 nm depth.
%we performed a near feature-scale reactive MD simulation using LightPFP. The simulation cell measured 10.06 × 10.06 × 20.00 nm along the a, b, and c axes, and contained 72,000 Si and O atoms in a \ce{SiO2}(100) slab. To emulate focused dry etching, HF molecules were accelerated to a kinetic energy of 40 eV and directed toward the surface, with impact points restricted to a 2 × 2 nm patch. Over a total simulation time of 0.5 ns, 1,000 HF molecules were injected. Figure~\ref{fig:SiO2_etching_large_scale} illustrates the evolution of the surface morphology under these conditions. By approximately 0.05 ns, atoms at the bombarded region begin to be removed. A recessed pit is clearly visible by 0.25 ns, and by 0.5 ns the crater reaches a depth of about 2 nm. 
% Because the MD timescale is necessarily short, the HF injection flux used here is higher than in typical experiments; thus absolute etch rates are not directly comparable. Nevertheless, t
The sequence of material removal and the development of a localized crater demonstrates that LightPFP remains stable and predictive in large, high-flux reactive simulations. These large-scale simulations pave the way for feature-scale studies, including aspect-ratio effects, lateral etch selectivity, and the interplay between energy, dose, and local morphology during pattern transfer\cite{oehrlein2024future}.

\subsection{Melting point of MgO: application of few-shot transfer learning}
\label{subsec:mgo}

When the teacher model (i.e., the universal potential) exhibits systematic errors in a given system, a distilled student will generally inherit those deficiencies. To address this limitation, we explore a transfer-learning strategy in which a small amount of high-fidelity DFT data is used to correct the distilled model and enhance its accuracy. 
We test this idea on \ce{MgO}’s melting point: PBE-DFT significantly underestimates it relative to experiment, whereas higher-level functionals (e.g., r$^2$SCAN) are more accurate. Since the PFP model we used was trained on PBE-based datasets, it is less accurate for modeling \ce{MgO} melting point.

\begin{figure*}
\includegraphics[]{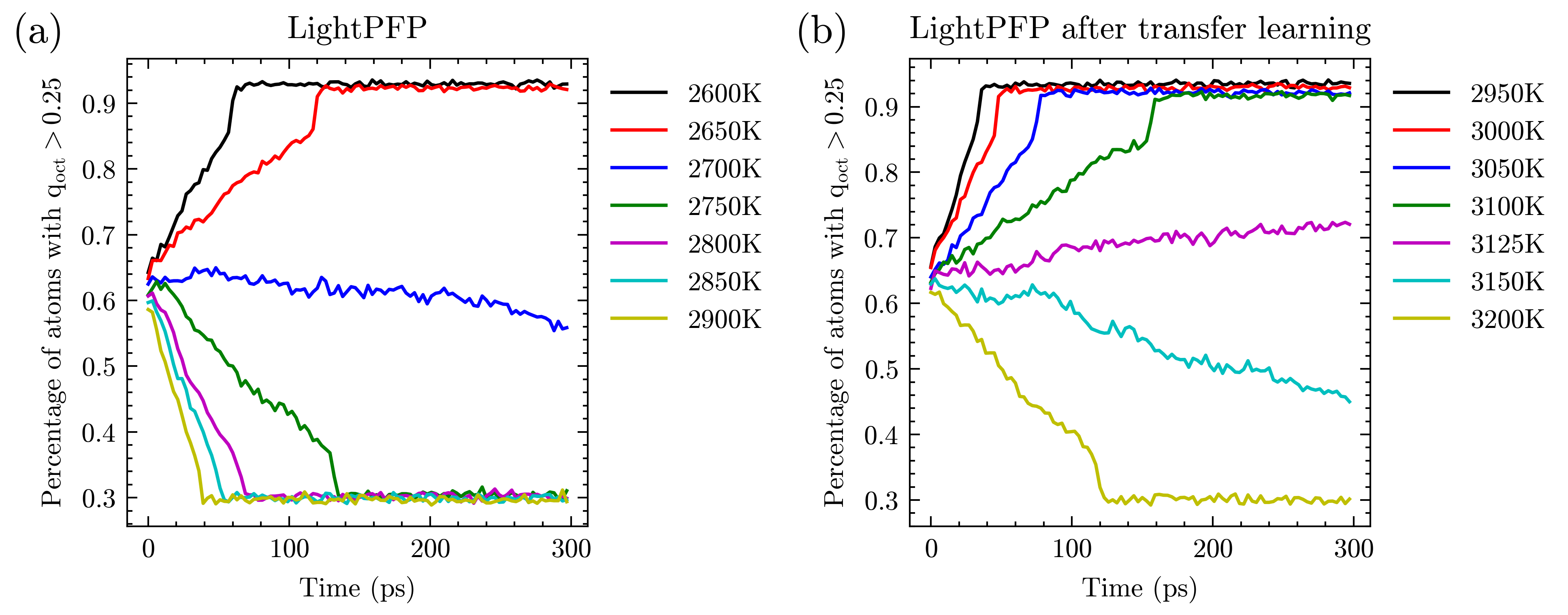}
\caption{Fraction of high-order atoms ($q_{\mathrm{oct}}$ > 0.25) during MD simulations at different temperatures. (a) LightPFP, (b) LightPFP after few-shot transfer learning}
\label{fig:MgO_order}
\end{figure*}

We first follow our standard distillation workflow to construct a LightPFP model using PFP-sampled training data, including crystalline, liquid-phase, and solid–liquid interface configurations of MgO. We then estimated the melting point by monitored the local octahedral order parameter $q_{\mathrm{oct}}$~\cite{zimmermann2017assessing} of atoms a solid–liquid coexistence structure. 
Figure~\ref{fig:MgO_order}(a) shows the fraction of atoms with high order ($q_{\mathrm{oct}}$ > 0.25) versus time. 
% At low temperatures (e.g., 2600 K), this fraction increases toward 1.0, indicating solidification; at higher temperatures it decreases, indicating melting. 
At approximately 2700 K, the fraction remains nearly constant over the trajectory, suggesting solid–liquid equilibrium. As expected for a PBE-level model, this melting point is substantially below the experimental range from 3073 to 3250 K and consistent with prior PBE-trained MLIP studies\cite{liang2018complete,lee2022ab}.

To improve accuracy, the LightPFP model undergoes transfer learning with a small dataset labeled by r$^2$SCAN DFT calculations. The dataset contains 10 structures with 64 atoms per structure. The radial-basis representation of the LightPFP model was frozen, and only the readout neural networks were fine-tuned to mitigate overfitting and catastrophic forgetting.
% To improve accuracy, we applied few-shot transfer learning from PBE to r$^2$SCAN. Using the original LightPFP as a starting point, we sampled small MgO structures (64 atoms each) from MD and selected 10 configurations for r$^2$SCAN single-point calculations. During transfer learning, we froze the LightPFP radial-basis representation and fine-tuned only the readout network to minimize energy, force, and stress errors against the r$^2$SCAN labels. This procedure adapts the model to the r$^2$SCAN potential energy surface. Freezing the representation mitigates overfitting and catastrophic forgetting in the few-shot regime while reducing compute.
% The r$^2$SCAN calculations took 1.25 hours, and fine-tuning required 0.5 hours. 
Re-evaluating with transfer-learned LightPFP under the same MD protocol, we obtained an estimated melting point of 3125 K, in excellent agreement with experiment\cite{liang2018complete}; see Fig.~\ref{fig:MgO_order}(b). Radial distribution function analysis confirms that both pre- and post-transfer models accurately reproduce the structural features of solid and liquid MgO, with the transfer-learned model showing good agreement with SCAN reference data (see SI), indicating that the melting point correction arises from improved energetic balance rather than structural artifacts.
These results suggest that few-shot transfer is most data-efficient when the teacher-level model already provides a useful representation of the target configurational space. However, if the higher-level reference leads to substantially different structures, bonding motifs, or sampled pathways, additional high-fidelity data and more extensive validation may be needed. 
The entire workflow took 10.25 hours, including the initial LightPFP training and transfer learning. Given r$^2$SCAN’s high computational cost relative to PBE, the conventional route to build an r$^2$SCAN-level MLIP is estimated to be 3–4 orders of magnitude slower than our workflow. 

% This case illustrates a general recipe for overcoming teacher limitations. Distill a fast, task-adapted student from a universal potential for broad coverage and efficiency; then, wherever the teacher is biased or undertrained, apply few-shot transfer learning using a higher-fidelity dataset to correct the student. With minimal additional labeling, the student can surpass the teacher for the property of interest. This strategy is agnostic to the source of teacher error and is readily extensible to other universal models and materials systems.

\section{Discussion}

We introduced LightPFP, a knowledge distillation framework that resolves the fundamental trade-off between transferability and efficiency in Machine learning interatomic potentials (MLIPs). By leveraging high-fidelity universal MLIPs (u-MLIPs) as computational engines for data generation—rather than relying on expensive DFT calculations—we enable rapid development of lightweight, task-specific MLIPs (ts-MLIPs) with minimal accuracy loss.
Demonstrations using \ce{Li6PS5Cl} solid-state electrolytes show that our distillation strategy reduces ts-MLIP development time by three orders of magnitude compared with conventional DFT-based approaches. The resulting distilled ts-MLIPs maintain comparable accuracy in critical calculations such as diffusion activation energies while achieving high inference efficiency through their streamlined architecture. Together with the optimized LAMMPS/Kokkos implementation, LightPFP enables inference over 1,000 times faster than its PFP teacher model and increases the maximum accessible system size by over 2,000 times.
% Demonstrations using \ce{Li6PS5Cl} solid-state electrolytes and high-entropy alloys show that our distillation strategy reduces ts-MLIP development time by three orders of magnitude compared to conventional DFT-based approaches. The resulting distilled ts-MLIPs achieve inference speeds one to two orders of magnitude faster than u-MLIPs due to their streamlined architecture, while maintaining comparable accuracy in critical calculations such as diffusion activation energies and surface/grain boundary energies.
For more complex systems, we demonstrate the framework's versatility through reactive ionic etching of \ce{SiO2} surfaces, where combining model distillation with active learning successfully handles intricate chemical processes. The distilled ts-MLIP accurately reproduces the teacher model's predictions for both chemical reactions and etching dynamics in molecular dynamics simulations.
Finally, we show how transfer learning can enhance distilled model accuracy when u-MLIP precision is insufficient. Using \ce{MgO} melting point prediction as a case study, we improved the predicted melting temperature from 2700 K to 3125 K using only 10 additional high-accuracy DFT data points, achieving excellent agreement with experimental values of 3100-3200 K.
We anticipate that this approach will generalize to other universal potentials, providing a scalable, data-efficient foundation for accurate, production-scale materials simulations.

For brevity, we present only four representative examples in the main text. In SI II, we present 11 additional examples of complex simulations achievable by LightPFP, which might be of interest to readers

\section{Methods}

\subsection{Density functional theory}
Density functional theory (DFT) calculations were employed for three main purposes: (1) generating training data for machine learning interatomic potentials (MLIPs) using conventional approaches; (2) evaluating key material properties such as interface energies to benchmark the accuracy of various MLIPs, including u-MLIPs and ts-MLIPs; and (3) producing datasets based on the r$^{2}$SCAN exchange–correlation (xc) functional for transfer learning in LightPFP.

All calculations were performed using spin-polarized DFT as implemented in the Vienna ab initio Simulation Package (VASP, version 6.4.0) with GPU acceleration. The projector augmented-wave (PAW) method and a plane-wave basis set with a kinetic-energy cutoff of 520 eV were employed. For the first two purposes, the Perdew–Burke–Ernzerhof (PBE) generalized gradient approximation was adopted. The pseudopotentials, cutoff energies, and k-point meshes followed the settings of the PFP dataset \cite{takamoto2022towards}, corresponding to a k-point density of approximately 1000 k-points per reciprocal atom.

For the third purpose, calculations were performed using the r$^{2}$SCAN meta-GGA xc functional within the same VASP framework. The functional was activated through the Meta-GGA option, with all other computational parameters—such as the 520 eV kinetic-energy cutoff—kept consistent with the PBE calculations to ensure compatibility. The Brillouin zone was sampled using a k-point grid generated with a KSPACING parameter of $0.5\text{\AA}^{-1}$, ensuring well-converged total energies.

\subsection{Preferred potential (PFP)}
PFP is a commercial universal interatomic potential available via the Matlantis atomic simulation platform.\cite{matlantis} 
It is trained on a large DFT dataset based on PBE\cite{perdew1996generalized}, r$^{2}$SCAN\cite{furness2020accurate} and $\omega$B97X-D\cite{chai2008long} exchange-correlation (xc) functionals. 
The highly disordered training structures, e.g. high temperature MD frames, are included in the dataset to guarantee its reliability in a wide range of applications. 

\subsection{Moment tensor potential (MTP)}

\subsubsection{Basis function}
% \paragraph{Basis function}

Moment tensor potential (MTP) employs a mathematically rigorous descriptor system based on invariant moment tensors that encode atomic environments~\citep{mtp}.
In MTP, energy can be calculated by the sum of the atomic energy functions of each atom $i$ in the structure: $E=\sum_i V_i$, where $V_i = \sum_\alpha \xi_\alpha B_\alpha(\mathbf{n}_i)$. $\xi_\alpha$ denotes a learnable coefficient of MTP, $B_\alpha$ denotes a basis function and $\mathbf{n}_i$ denotes a set of $\mathbf{r}_{ij}$, a relative coordinate position of atom $i$ to its neighbors.
Each basis function $B_\alpha$ comprises of matrix contractions of moment descriptors $M_{\mu, \nu}$, where $\mu$ and $\nu$ are non-negative integers. 
The moment descriptor $M_{\mu, \nu}$ for atom $i$  is defined as:  

\begin{equation*}
M_{\mu,\nu}(\mathbf{n}_i) = \sum_{j} f_{\mu}(\mathbf{r}_{ij}
) \underbrace{\mathbf{r}_{ij} \otimes \mathbf{r}_{ij} \otimes \cdots \otimes \mathbf{r}_{ij}}_{\nu \text{ times}}  
\end{equation*}
where $\mathbf{r}_{ij} = \mathbf{r}_j - \mathbf{r}_i$ is the relative position vector to neighbor j within cutoff radius $R_{\text{cut}}$, ``$\otimes$'' denotes a tensor outer product.
The function $f_\mu$ describing a radial part depending on $\mu$ is expressed as
\begin{equation*}
f_{\mu}(|r_{ij}|, z_i, z_j) = \sum_{\beta=1}^{N_Q}c_{\mu,z_i,z_j}^{(\beta)}Q^{\beta}(|r_{ij}|)
\end{equation*}
where $c_{\mu,z_i,z_j}^{(\beta)}$ is a learnable parameter, $z$ indicates the atomic type, the radial function $Q^{\beta}(|r_{ij}|)$ is the combination of Chebyshev polynomials of the first kind and cutoff function, and $N_Q$ is the number of polynomials.

Moment descriptors are contracted to form rotationally-invariant basis functions $B_\alpha(\mathbf{n}_i)$ that preserve SO(3) symmetry, enabling accurate representation of complex many-body interactions. 
The formulation of MTP achieves high data efficiency—basis functions span a complete polynomial space while avoiding explicit angular dependence, enabling accurate fits with small training sets~\citep{mtp,zuo2020performance}.  
With training data, we fit MTP to learn the parameters $\mathbf{\xi}=\{\xi_i, ... ,\xi_{n_B}\}$, where $n_B$ is the number of basis, and $\mathbf{c}=\{{c_{\mu,z_i,z_j}^{(\beta)}}\}$, where the number of coefficients $n_c$ depends on number of $f_\mu$, element pairs including the pair of itself, and $N_Q$: $n_c = n_{f_\mu} \times n_\text{elem-pair} \times N_Q$. 

\subsubsection{Neural network readout}
% \paragraph{Neural network readout}

We extend the standard MTP architecture by replacing its linear energy predictor with a neural network (NN) employing the multi-layer perceptron architecture parameterized by $\theta$, $\mathcal{M}_{\theta}$. 
The modified energy expression becomes:  

\begin{equation}
E_i^{\text{NN-MTP}} = \mathcal{M}_{\theta} \left( \left\{ B_\alpha(\mathbf{n}_i) \right\}_{\alpha=1}^m \right)  
\end{equation}

where $\theta$ denotes trainable weights, and $\{B_\alpha\}$ are invariant descriptors from the preceding tensor layer. 
This hybrid architecture introduces controlled nonlinearity to enhance the capability to capture subtle correlations in potential energy surfaces (PES) that can be difficult for linear projections to capture. 

\subsection{Pretrained student models}
 
We pretrain the MTP model using the large, diverse DFT dataset used for the training of u-MLIP, PFP. 
The dataset includes both equilibrium and non-equilibrium structures, enabling broad transferability~\cite{takamoto2022towards}. 
Unlike the universal PFP graph neural network, MTPs have limited capacity and are typically material-specific, so we do not fit all data jointly. 
Instead, we adopt Reptile meta-learning~\cite{reptile} to obtain an initialization that adapts rapidly to individual systems: the dataset is split into 12 tasks by structure type; at each meta-iteration we sample one task, train for a single epoch with Adam~\cite{adam} where learning rate is $1e^{-3}$, and batch size is 256), then apply a meta-update with $\beta=0.5$. 
We run 100 meta-iterations until energies, forces, and stresses stabilize across tasks. The pretrained model has a large parameter set due to the large number of supported elements. Owing to MTP’s modularity, parameters can be subset by elements at inference or fine-tuning. This initialization acts as a strong prior from diverse chemistry, improving robustness, reducing overfitting, and speeding convergence when the target dataset is small or undersampled. The details of pretrained models can be found in SI I.

\subsection{Training method}

For all LightPFP models trained in the applications described in Section~\ref{sec:results}, datasets were split 90\%/10\% into training and validation. The validation set was used both for model selection (choosing the checkpoint with the lowest validation loss) and for reporting validation errors. The training objective combined energy, force, and stress terms:

\begin{equation}
L = \alpha \cdot L_{energy} + \beta \cdot L_{force} + \gamma \cdot L_{stress}
\end{equation}
where $L_{energy}$ is the mean squared error (MSE) of the energy per atom, $L_{force}$ is the MSE of the Cartesian force components on each atom (x, y, z), and $L_{stress}$ is the MSE of the stress tensor components. The coefficients $\alpha$, $\beta$ and $\gamma$ weight the energy, force, and stress losses, respectively.

Optimization was performed using Adam~\citep{adam} with a batch size of 128, following a three-stage training procedure.
In the first stage, the loss coefficients for energy, forces, and stress were set to $(10^{-5}, 10, 10^{-5})$. 
In the second stage, they were adjusted to $(1, 0.1, 10)$. 
In the third stage, the loss coefficients were automatically determined to balance the three losses. 
Specifically, we first computed the total validation loss of the second-stage model using the stage-two coefficients as loss weight. 
The coefficient for the energy loss was then calculated as the total weighted validation loss divided by three and further divided by the energy loss from stage two. 
The coefficients for forces and stress were calculated analogously, each using their respective loss from stage two. The three-stage procedure yielded faster convergence than conventional training method without variation of coefficients. A detailed comparison is provided in SI I.

A linear warmup learning rate scheduler was applied, increasing the learning rate from zero to its stage-specific maximum during the first 20\% of epochs in each stage, and then linearly decaying it to approach zero by the final epoch. 
The learning rates for stage 1, stage 2, and stage 3 were set to 0.1, 0.01, and 0.01, respectively.

\section{Data availability}
% The datasets generated and analysed during the current study are available from public repositories.
% The DFT training datasets have been deposited on the Materials Cloud
%  under the accession number DOI: XX.XXXXX/materialscloud:XXXXXX
%  (to be assigned).
Both DFT and PFP-labelled datasets used for model training and evaluation are available in the accompanying GitHub repository https://github.com/pfnet-research/light-pfp.
All other data supporting the findings of this study are available from the corresponding author upon reasonable request.

\section{Code availability}
All codes used for model training and atomistic simulations are available in the LightPFP examples GitHub repository: https://github.com/pfnet-research/light-pfp.
The implementation depends on the PFP and LightPFP packages, which are components of the Matlantis commercial platform developed by Preferred Networks, Inc., and are accessible to users under the platform’s standard licensing terms.

\section{Acknowledgements}
The authors are grateful to Tasuku Onodera, Takashi Kojima, Masanao Goto, Yuta Tanaka, and Yuji Hakozaki (ENEOS Holdings, Inc.) for their valuable suggestions during the development of LightPFP.
We also thank Chikashi Shinagawa (Preferred Networks, Inc.) for insightful discussions regarding the DFT calculation setup.

%\nocite{*}
\bibliography{aipsamp}% Produces the bibliography via BibTeX.

@article{pople1999nobel,
  title={Nobel lecture: Quantum chemical models},
  author={Pople, John A},
  journal={Reviews of Modern Physics},
  volume={71},
  number={5},
  pages={1267},
  year={1999},
  publisher={APS}
}

@ARTICLE{TakamotoOLL23,
	author	    = {Takamoto, S and Okanohara, D and Li, QJ and Li, J},
	title	    = {Towards universal neural network interatomic
	  potential},
	journal     = {J. Materiomics},
	year	    = {2023},
	volume	    = {9},
	pages	    = {447-454}
}

@ARTICLE{CeperleyA80,
        author      = {Ceperley, DM and Alder, BJ},
        title       = {Ground-state of the electron-gas by a stochastic
          method},
        journal     = {Phys. Rev. Lett.},
        year        = {1980},
        volume      = {45},
        pages       = {566-569}
}

@ARTICLE{LiWCCBHY04,
        author      = {Li, J and Wang, CZ and Chang, JP and Cai, W and
          Bulatov, VV and Ho, KM and Yip, S},
        title       = {Core energy and Peierls stress of a screw dislocation
          in bcc molybdenum: A periodic-cell tight-binding study},
        journal     = {Phys. Rev. B},
        year        = {2004},
        volume      = {70},
        pages       = {104113}
}

@article{chai2008long,
  title={Long-range corrected hybrid density functionals with damped atom--atom dispersion corrections},
  author={Chai, Jeng-Da and Head-Gordon, Martin},
  journal={Physical Chemistry Chemical Physics},
  volume={10},
  number={44},
  pages={6615--6620},
  year={2008},
  publisher={Royal Society of Chemistry}
}

@article{furness2020accurate,
  title={Accurate and numerically efficient {r2SCAN} meta-generalized gradient approximation},
  author={Furness, James W and Kaplan, Aaron D and Ning, Jinliang and Perdew, John P and Sun, Jianwei},
  journal={The journal of physical chemistry letters},
  volume={11},
  number={19},
  pages={8208--8215},
  year={2020},
  publisher={ACS Publications}
}

@article{perdew1996generalized,
  title={Generalized gradient approximation made simple},
  author={Perdew, John P and Burke, Kieron and Ernzerhof, Matthias},
  journal={Physical review letters},
  volume={77},
  number={18},
  pages={3865},
  year={1996},
  publisher={APS}
}

@article{morrow2022indirect,
  title={Indirect learning and physically guided validation of interatomic potential models},
  author={Morrow, Joe D and Deringer, Volker L},
  journal={The Journal of Chemical Physics},
  volume={157},
  number={10},
  year={2022},
  publisher={AIP Publishing}
}

@article{zhang2024dpa,
  title={{DPA}-2: a large atomic model as a multi-task learner},
  author={Zhang, Duo and Liu, Xinzijian and Zhang, Xiangyu and Zhang, Chengqian and Cai, Chun and Bi, Hangrui and Du, Yiming and Qin, Xuejian and Peng, Anyang and Huang, Jiameng and others},
  journal={npj Computational Materials},
  volume={10},
  number={1},
  pages={293},
  year={2024},
  publisher={Nature Publishing Group UK London}
}

@article{amin2025towards,
  title={Towards fast, specialized machine learning force fields: Distilling foundation models via energy hessians},
  author={Amin, Ishan and Raja, Sanjeev and Krishnapriyan, Aditi},
  journal={arXiv preprint arXiv:2501.09009},
  year={2025}
}

@article{gardner2025distillation,
  title={Distillation of atomistic foundation models across architectures and chemical domains},
  author={Gardner, John LA and Toit, Daniel F and Mahmoud, Chiheb Ben and Beaulieu, Zo{\'e} Faure and Juraskova, Veronika and Pa{\c{s}}ca, Laura-Bianca and Rosset, Louise AM and Duarte, Fernanda and Martelli, Fausto and Pickard, Chris J and others},
  journal={arXiv preprint arXiv:2506.10956},
  year={2025}
}

@article{mtp,
  title={The MLIP package: moment tensor potentials with MPI and active learning},
  author={Novikov, Ivan S and Gubaev, Konstantin and Podryabinkin, Evgeny V and Shapeev, Alexander V},
  journal={Machine Learning: Science and Technology},
  volume={2},
  number={2},
  pages={025002},
  year={2020},
  publisher={IOP Publishing}
}

@article{takamoto2022towards,
  title={Towards universal neural network potential for material discovery applicable to arbitrary combination of 45 elements},
  author={Takamoto, So and Shinagawa, Chikashi and Motoki, Daisuke and Nakago, Kosuke and Li, Wenwen and Kurata, Iori and Watanabe, Taku and Yayama, Yoshihiro and Iriguchi, Hiroki and Asano, Yusuke and others},
  journal={Nature Communications},
  volume={13},
  number={1},
  pages={2991},
  year={2022},
  publisher={Nature Publishing Group UK London}
}

@article{mace,
  title={{MACE}: Higher order equivariant message passing neural networks for fast and accurate force fields},
  author={Batatia, Ilyes and Kovacs, David P and Simm, Gregor and Ortner, Christoph and Cs{\'a}nyi, G{\'a}bor},
  journal={Advances in neural information processing systems},
  volume={35},
  pages={11423--11436},
  year={2022}
}

@article{m3gnet,
  title={A universal graph deep learning interatomic potential for the periodic table},
  author={Chen, Chi and Ong, Shyue Ping},
  journal={Nature Computational Science},
  volume={2},
  number={11},
  pages={718--728},
  year={2022},
  publisher={Nature Publishing Group US New York}
}

@article{chgnet,
  title={{CHGNet} as a pretrained universal neural network potential for charge-informed atomistic modelling},
  author={Deng, Bowen and Zhong, Peichen and Jun, KyuJung and Riebesell, Janosh and Han, Kevin and Bartel, Christopher J and Ceder, Gerbrand},
  journal={Nature Machine Intelligence},
  volume={5},
  number={9},
  pages={1031--1041},
  year={2023},
  publisher={Nature Publishing Group UK London}
}

@article{zuo2020performance,
  title={Performance and cost assessment of machine learning interatomic potentials},
  author={Zuo, Yunxing and Chen, Chi and Li, Xiangguo and Deng, Zhi and Chen, Yiming and Behler, J{\"o}rg and Cs{\'a}nyi, G{\'a}bor and Shapeev, Alexander V and Thompson, Aidan P and Wood, Mitchell A and others},
  journal={The Journal of Physical Chemistry A},
  volume={124},
  number={4},
  pages={731--745},
  year={2020},
  publisher={ACS Publications}
}

@article{batatia2023foundation,
  title={A foundation model for atomistic materials chemistry},
  author={Batatia, Ilyes and Benner, Philipp and Chiang, Yuan and Elena, Alin M and Kov{\'a}cs, D{\'a}vid P and Riebesell, Janosh and Advincula, Xavier R and Asta, Mark and Avaylon, Matthew and Baldwin, William J and others},
  journal={arXiv preprint arXiv:2401.00096},
  year={2023}
}

@article{zimmermann2017assessing,
  title={Assessing local structure motifs using order parameters for motif recognition, interstitial identification, and diffusion path characterization},
  author={Zimmermann, Nils ER and Horton, Matthew K and Jain, Anubhav and Haranczyk, Maciej},
  journal={Frontiers in Materials},
  volume={4},
  pages={34},
  year={2017},
  publisher={Frontiers Media SA}
}

@article{deng2017data,
  title={Data-driven first-principles methods for the study and design of alkali superionic conductors},
  author={Deng, Zhi and Zhu, Zhuoying and Chu, Iek-Heng and Ong, Shyue Ping},
  journal={Chemistry of Materials},
  volume={29},
  number={1},
  pages={281--288},
  year={2017},
  publisher={ACS Publications}
}

@article{kim2021molecular,
  title={Molecular dynamics simulation of silicon dioxide etching by hydrogen fluoride using the reactive force field},
  author={Kim, Dong Hyun and Kwak, Seung Jae and Jeong, Jae Hun and Yoo, Suyoung and Nam, Sang Ki and Kim, YongJoo and Lee, Won Bo},
  journal={ACS omega},
  volume={6},
  number={24},
  pages={16009--16015},
  year={2021},
  publisher={ACS Publications}
}

@article{kong2025exploration,
  title={Exploration of Lithium-Ion Conductors Based on Local Coordination Environments Using Crystallographic Site Fingerprints},
  author={Kong, Songjia and Matsui, Naoki and Hori, Satoshi and Hirayama, Masaaki and Mori, Kazuhiro and Saito, Takashi and Kanno, Ryoji and Suzuki, Kota},
  journal={Journal of the American Chemical Society},
  year={2025},
  publisher={ACS Publications}
}

@article{hinuma2025facile,
  title={Facile Formation of Two-Phase Domains in a Single Crystalline Li7--x Ti5O12 Particle},
  author={Hinuma, Yoyo and Kitta, Mitsunori},
  journal={ACS Applied Energy Materials},
  year={2025},
  publisher={ACS Publications}
}

@article{narumi2025tailoring,
  title={Tailoring the room-temperature miscibility gap in ordered spinel LiNi 0.5 Mn 1.5 O 4 cathodes by multi-element doping},
  author={Narumi, Shunsuke and Otal, H Eugenio and Nguyen, Tien Quang and Koyama, Michihisa and Zettsu, Nobuyuki},
  journal={Journal of Materials Chemistry A},
  year={2025},
  publisher={Royal Society of Chemistry}
}

@article{shimada2024long,
  title={Long time {CO2} storage under ambient conditions in isolated voids of a porous coordination network facilitated by the “magic door” mechanism},
  author={Shimada, Terumasa and Usov, Pavel M and Wada, Yuki and Ohtsu, Hiroyoshi and Watanabe, Taku and Adachi, Kiyohiro and Hashizume, Daisuke and Matsumoto, Takaya and Kawano, Masaki},
  journal={Advanced Science},
  volume={11},
  number={2},
  pages={2307417},
  year={2024},
  publisher={Wiley Online Library}
}

@article{watanabe2025oxidative,
  title={Oxidative Dehydrogenation of Ethane Combined with CO2 Splitting via Chemical Looping on In2O3 Modified with Ni--Cu Alloy},
  author={Watanabe, Kosuke and Higo, Takuma and Saegusa, Koki and Matsumoto, Sakura and Sampei, Hiroshi and Isono, Yuki and Shimojuku, Akira and Furusawa, Hideki and Sekine, Yasushi},
  journal={ACS Catalysis},
  volume={15},
  number={7},
  pages={5876--5885},
  year={2025},
  publisher={ACS Publications}
}

@article{koh2024defect,
  title={Defect-Driven Evolution of Oxo-Coordinated Cobalt Active Sites with Rapid Structural Transformation for Efficient Water Oxidation},
  author={Koh, Jinseok and Kwon, Choah and Kim, Hyunjeong and Lee, Eunchong and Machida, Akihiko and Nakahira, Yuki and Hwang, Yun Jeong and Sakaki, Kouji and Kim, Sangtae and Cho, Eun Seon},
  journal={ACS nano},
  volume={18},
  number={42},
  pages={28986--28998},
  year={2024},
  publisher={ACS Publications}
}

@article{son2024constructing,
  title={Constructing reversible Li deposition interfaces by tailoring lithiophilic functionalities of a heteroatom-doped graphene interlayer for highly stable Li metal anodes},
  author={Son, Beom Gwon and Kwon, Choah and Cho, YongJun and Jang, Taegyu and Byon, Hye Ryung and Kim, Sangtae and Cho, Eun Seon},
  journal={ACS Applied Materials \& Interfaces},
  volume={16},
  number={25},
  pages={32259--32270},
  year={2024},
  publisher={ACS Publications}
}

@article{kim2024sustained,
  title={Sustained Area-Selectivity in Atomic Layer Deposition of Ir Films: Utilization of Dual Effects of O3 in Deposition and Etching},
  author={Kim, Han and Kim, Taeseok and Chung, Hong Keun and Jeon, Jihoon and Kim, Sung-Chul and Won, Sung Ok and Harada, Ryosuke and Tsugawa, Tomohiro and Kim, Sangtae and Kim, Seong Keun},
  journal={Small},
  volume={20},
  number={46},
  pages={2402543},
  year={2024},
  publisher={Wiley Online Library}
}

@article{jin2024atom,
  title={Atom-by-atom design of Cu/ZrO x clusters on MgO for CO2 hydrogenation using liquid-phase atomic layer deposition},
  author={Jin, Seongmin and Kwon, Choah and Bugaev, Aram and Karakurt, Bartu and Lin, Yu-Cheng and Savereide, Louisa and Zhong, Liping and Boureau, Victor and Safonova, Olga and Kim, Sangtae and others},
  journal={Nature Catalysis},
  volume={7},
  number={11},
  pages={1199--1212},
  year={2024},
  publisher={Nature Publishing Group UK London}
}

@article{kim2024facile,
  title={Facile synthesis of nanoporous Mg crystalline structure by organic solvent-based reduction for solid-state hydrogen storage},
  author={Kim, Hyesun and Kim, HyeonJi and Kim, Wonsik and Kwon, Choah and Jin, Si-Won and Ha, Taejun and Shim, Jae-Hyeok and Park, Soohyung and Jamal, Aqil and Kim, Sangtae and others},
  journal={Nature Communications},
  volume={15},
  number={1},
  pages={10800},
  year={2024},
  publisher={Nature Publishing Group UK London}
}

@article{hinuma2024neural,
  title={Neural Network Potential Molecular Dynamics Simulations of {(La, Ce, Pr, Nd)} 0.95 {(Mg, Zn, Pb, Cd, Ca, Sr, Ba)} 0.05 {F2}. 95},
  author={Hinuma, Yoyo},
  journal={The Journal of Physical Chemistry B},
  volume={128},
  number={49},
  pages={12171--12178},
  year={2024},
  publisher={ACS Publications}
}

@article{hisama2024molecular,
  title={Molecular dynamics of catalyst-free edge elongation of boron nitride nanotubes coaxially grown on single-walled carbon nanotubes},
  author={Hisama, Kaoru and Bets, Ksenia V and Gupta, Nitant and Yoshikawa, Ryo and Zheng, Yongjia and Wang, Shuhui and Liu, Ming and Xiang, Rong and Otsuka, Keigo and Chiashi, Shohei and others},
  journal={ACS nano},
  volume={18},
  number={45},
  pages={31586--31595},
  year={2024},
  publisher={ACS Publications}
}

@article{honbo2024effects,
  title={Effects of Alkyl Side Chain Length on the Structural Organization and Proton Conductivity of Sulfonated Polyimide Thin Films},
  author={Honbo, Tetsuya and Ono, Yutaro and Suetsugu, Kota and Hara, Mitsuo and Taborosi, Attila and Aoki, Kentaro and Nagano, Shusaku and Koyama, Michihisa and Nagao, Yuki},
  journal={ACS Applied Polymer Materials},
  volume={6},
  number={21},
  pages={13217--13227},
  year={2024},
  publisher={ACS Publications}
}

@article{kwon2024intelligent,
  title={Intelligent Stress-Adaptive Binder Enabled by Shear-Thickening Property for Silicon Electrodes of Lithium-Ion Batteries},
  author={Kwon, Ohhyun and Kim, Tae Yong and Kim, Taewon and Kang, Jihyeon and Jang, Seohyeon and Eom, Hojong and Choi, Seyoung and Shin, Junhyeop and Park, Jongkwon and Seol, Myeong-Lok and others},
  journal={Advanced Energy Materials},
  volume={14},
  number={20},
  pages={2304085},
  year={2024},
  publisher={Wiley Online Library}
}

@article{miura2024stress,
  title={Stress-Induced Martensitic Transformation in {Na3YCl6}},
  author={Miura, Akira and Muraoka, Koki and Maki, Kotaro and Kawaguchi, Saori and Hikima, Kazuhiro and Muto, Hiroyuki and Matsuda, Atsunori and Yamane, Ichiro and Shimada, Toshihiro and Ito, Hiroaki and others},
  journal={Journal of the American Chemical Society},
  volume={146},
  number={36},
  pages={25263--25269},
  year={2024},
  publisher={ACS Publications}
}

@article{du2023new,
  title={A new zinc salt chemistry for aqueous zinc-metal batteries},
  author={Du, Haoran and Dong, Yanhao and Li, Qing-Jie and Zhao, Ruirui and Qi, Xiaoqun and Kan, Wang-Hay and Suo, Liumin and Qie, Long and Li, Ju and Huang, Yunhui},
  journal={Advanced Materials},
  volume={35},
  number={25},
  pages={2210055},
  year={2023},
  publisher={Wiley Online Library}
}

@article{ishikawa2024evolutionary,
  title={Evolutionary search for superconducting phases in the lanthanum-nitrogen-hydrogen system with universal neural network potential},
  author={Ishikawa, Takahiro and Tanaka, Yuta and Tsuneyuki, Shinji},
  journal={Physical Review B},
  volume={109},
  number={9},
  pages={094106},
  year={2024},
  publisher={APS}
}

@article{bae2024tunable,
  title={Tunable ion energy barrier modulation through aliovalent halide doping for reliable and dynamic memristive neuromorphic systems},
  author={Bae, Jongmin and Kwon, Choah and Park, See-On and Jeong, Hakcheon and Park, Taehoon and Jang, Taehwan and Cho, Yoonho and Kim, Sangtae and Choi, Shinhyun},
  journal={Science Advances},
  volume={10},
  number={23},
  pages={eadm7221},
  year={2024},
  publisher={American Association for the Advancement of Science}
}

@article{teanet,
  title={{TeaNet}: Universal neural network interatomic potential inspired by iterative electronic relaxations},
  author={Takamoto, So and Izumi, Satoshi and Li, Ju},
  journal={Computational Materials Science},
  volume={207},
  pages={111280},
  year={2022},
  publisher={Elsevier}
}

@misc{matlantis, 
    title = "{Matlantis, software as a service style material discovery tool}",
    howpublished = "\url{https://matlantis.com/}"
}

@article{reptile,
  title={On first-order meta-learning algorithms},
  author={Nichol, Alex and Achiam, Joshua and Schulman, John},
  journal={arXiv preprint arXiv:1803.02999},
  year={2018}
}

@article{jozwik2015applications,
  title={Applications of {Ni3Al} based intermetallic alloys—current stage and potential perceptivities},
  author={Jozwik, Pawel and Polkowski, Wojciech and Bojar, Zbigniew},
  journal={Materials},
  volume={8},
  number={5},
  pages={2537--2568},
  year={2015},
  publisher={MDPI}
}

@article{adam,
  title={Adam: A method for stochastic optimization},
  author={Kingma, Diederik P and Ba, Jimmy},
  journal={arXiv preprint arXiv:1412.6980},
  year={2014}
}

@article{liang2018complete,
  title={Complete thermodynamic description of the {Mg-Ca-O} phase diagram including the {Ca-O, Mg-O and CaO-MgO} subsystems},
  author={Liang, Song-Mao and Schmid-Fetzer, Rainer},
  journal={Journal of the European Ceramic Society},
  volume={38},
  number={14},
  pages={4768--4785},
  year={2018},
  publisher={Elsevier}
}

@article{lee2022ab,
  title={Ab initio construction of full phase diagram of {MgO-CaO} eutectic system using neural network interatomic potentials},
  author={Lee, Kyeongpung and Park, Yutack and Han, Seungwu},
  journal={Physical Review Materials},
  volume={6},
  number={11},
  pages={113802},
  year={2022},
  publisher={APS}
}

@article{morin2004chemisorption,
  title={Chemisorption of benzene on Pt (111), Pd (111), and Rh (111) metal surfaces: a structural and vibrational comparison from first principles},
  author={Morin, C{\'e}line and Simon, D and Sautet, P},
  journal={The Journal of Physical Chemistry B},
  volume={108},
  number={18},
  pages={5653--5665},
  year={2004},
  publisher={ACS Publications}
}

@article{stukowski2014computational,
  title={Computational analysis methods in atomistic modeling of crystals},
  author={Stukowski, Alexander},
  journal={Jom},
  volume={66},
  number={3},
  pages={399--407},
  year={2014},
  publisher={Springer}
}

@article{oehrlein2024future,
  title={Future of plasma etching for microelectronics: Challenges and opportunities},
  author={Oehrlein, Gottlieb S and Brandstadter, Stephan M and Bruce, Robert L and Chang, Jane P and DeMott, Jessica C and Donnelly, Vincent M and Dussart, R{\'e}mi and Fischer, Andreas and Gottscho, Richard A and Hamaguchi, Satoshi and others},
  journal={Journal of Vacuum Science \& Technology B},
  volume={42},
  number={4},
  year={2024},
  publisher={AIP Publishing}
}

@misc{IUPACNISTSolubilityDB,
title = {IUPAC-NIST Solubility Database},
author = {{National Institute of Standards and Technology (NIST)} and {International Union of Pure and Applied Chemistry (IUPAC)}},
url = {https://srdata.nist.gov/solubility/},
doi = {10.18434/T4QC79},
note = {Accessed: 2025-10-13}
}

@misc{pubchem_heptane_solubility,
author = {{National Center for Biotechnology Information}},
title = {PubChem Compound Summary for Heptane},
howpublished = {PubChem},
url = {https://pubchem.ncbi.nlm.nih.gov/compound/Heptane#section=Solubility},
note = {Section: Solubility. Accessed 2025-10-13}
}

@article{carrillo2024probing,
  title={Probing the thermal resistance of solid--liquid interfaces in nanofluids with molecular dynamics},
  author={Carrillo-Berdugo, Iv{\'a}n and Navas, Javier and Grau-Crespo, Ricardo},
  journal={The Journal of Chemical Physics},
  volume={160},
  number={1},
  year={2024},
  publisher={AIP Publishing}
}

@article{hess2002determining,
  title={Determining the shear viscosity of model liquids from molecular dynamics simulations},
  author={Hess, Berk},
  journal={The Journal of chemical physics},
  volume={116},
  number={1},
  pages={209--217},
  year={2002},
  publisher={American Institute of Physics}
}

@misc{NIST_SRD69_nDecane_TFREEZE,
title = {NIST Chemistry WebBook},
publisher = {National Institute of Standards and Technology},
year = {2025},
url = {https://webbook.nist.gov/cgi/cbook.cgi?ID=C124185&Type=TFREEZE},
note = {Entry: {n-Decane} (CAS RN 124-18-5)},
}

@article{jung2023artificial,
  title={Artificial neural network potentials for mechanics and fracture dynamics of two-dimensional crystals},
  author={Jung, Gang Seob and Myung, Hunjoo and Irle, Stephan},
  journal={Machine Learning: Science and Technology},
  volume={4},
  number={3},
  pages={035001},
  year={2023},
  publisher={IOP Publishing}
}

@article{doig2014structure,
  title={Structure and friction of stearic acid and oleic acid films adsorbed on iron oxide surfaces in squalane},
  author={Doig, Michael and Warrens, Chris P and Camp, Philip J},
  journal={Langmuir},
  volume={30},
  number={1},
  pages={186--195},
  year={2014},
  publisher={ACS Publications}
}

@article{luo2020molecular,
  title={Molecular dynamics simulations of polymerized ionic liquids: Mechanism of ion transport with different anions},
  author={Luo, Xubo and Liu, Hongjun and Paddison, Stephen J},
  journal={ACS Applied Polymer Materials},
  volume={3},
  number={1},
  pages={141--152},
  year={2020},
  publisher={ACS Publications}
}

@article{qian2022ab,
  title={Ab initio molecular dynamics simulation of structural and elastic properties of SiO2--P2O5--Al2O3--Na2O glass},
  author={Qian, Yixiao and Song, Bin and Jin, Junteng and Prayogo, Genki I and Utimula, Keishu and Nakano, Kousuke and Maezono, Ryo and Hongo, Kenta and Zhao, Gaoling},
  journal={Journal of the American Ceramic Society},
  volume={105},
  number={11},
  pages={6604--6615},
  year={2022},
  publisher={Wiley Online Library}
}

@article{kim2025pubchem,
  title={PubChem 2025 update},
  author={Kim, Sunghwan and Chen, Jie and Cheng, Tiejun and Gindulyte, Asta and He, Jia and He, Siqian and Li, Qingliang and Shoemaker, Benjamin A and Thiessen, Paul A and Yu, Bo and others},
  journal={Nucleic acids research},
  volume={53},
  number={D1},
  pages={D1516--D1525},
  year={2025},
  publisher={Oxford University Press}
}

@article{si2011abrasive,
  title={Abrasive rolling effects on material removal and surface finish in chemical mechanical polishing analyzed by molecular dynamics simulation},
  author={Si, Lina and Guo, Dan and Luo, Jianbin and Lu, Xinchun and Xie, Guoxin},
  journal={Journal of Applied Physics},
  volume={109},
  number={8},
  year={2011},
  publisher={AIP Publishing}
}

@article{tang2025approaching,
  title={Approaching coupled-cluster accuracy for molecular electronic structures with multi-task learning},
  author={Tang, Hao and Xiao, Brian and He, Wenhao and Subasic, Pero and Harutyunyan, Avetik R and Wang, Yao and Liu, Fang and Xu, Haowei and Li, Ju},
  journal={Nature Computational Science},
  volume={5},
  number={2},
  pages={144--154},
  year={2025},
  publisher={Nature Publishing Group US New York}
}

@article{lysogorskiy2026graph,
  title={Graph atomic cluster expansion for foundational machine learning interatomic potentials},
  author={Lysogorskiy, Yury and Bochkarev, Anton and Drautz, Ralf},
  journal={npj Computational Materials},
  year={2026},
  publisher={Nature Publishing Group UK London}
}

@article{mazitov2025pet,
  title={PET-MAD as a lightweight universal interatomic potential for advanced materials modeling},
  author={Mazitov, Arslan and Bigi, Filippo and Kellner, Matthias and Pegolo, Paolo and Tisi, Davide and Fraux, Guillaume and Pozdnyakov, Sergey and Loche, Philip and Ceriotti, Michele},
  journal={Nature Communications},
  volume={16},
  number={1},
  pages={10653},
  year={2025},
  publisher={Nature Publishing Group UK London}
}

@article{batatia2025crosslearning,
  title={Cross Learning between Electronic Structure Theories for Unifying Molecular, Surface, and Inorganic Crystal Foundation Force Fields},
  author={Batatia, Ilyes and Lin, Chen and Hart, Joseph and Kasoar, Elliott and Elena, Alin M. and Norwood, Sam Walton and Wolf, Thomas and Cs{\'a}nyi, G{\'a}bor},
  journal={arXiv preprint arXiv:2510.25380},
  year={2025}
}

@article{dpa2-workflow-wang2025pre,
  title={Pre-training, fine-tuning, and distillation (PFD): Automatically generating machine learning force fields from universal models},
  author={Wang, Ruoyu and Gao, Yuxiang and Wu, Hongyu and Zhong, Zhicheng},
  journal={arXiv preprint arXiv:2502.20809},
  year={2025}
}

@article{matsumura2025knowledge,
  title={Knowledge Distillation Framework for Accelerating High-Accuracy Neural Network-Based Molecular Dynamics Simulations},
  author={Matsumura, Naoki and Yoshimoto, Yuta and Iwasaki, Yuto and Yamazaki, Meguru and Sakai, Yasufumi},
  journal={arXiv preprint arXiv:2506.15337},
  year={2025}
}

@article{vinod2025lfab,
  title={LFaB: Low fidelity as Bias for Active Learning in the chemical configuration space},
  author={Vinod, Vivin and Zaspel, Peter},
  journal={arXiv preprint arXiv:2508.15577},
  year={2025}
}

@article{hjorth2017atomic,
  title={The atomic simulation environment—a Python library for working with atoms},
  author={Hjorth Larsen, Ask and J{\o}rgen Mortensen, Jens and Blomqvist, Jakob and Castelli, Ivano E and Christensen, Rune and Du{\l}ak, Marcin and Friis, Jesper and Groves, Michael N and Hammer, Bj{\o}rk and Hargus, Cory and others},
  journal={Journal of Physics: Condensed Matter},
  volume={29},
  number={27},
  pages={273002},
  year={2017},
  publisher={IOP Publishing}
}

@article{paszke2019pytorch,
  title={Pytorch: An imperative style, high-performance deep learning library},
  author={Paszke, Adam and Gross, Sam and Massa, Francisco and Lerer, Adam and Bradbury, James and Chanan, Gregory and Killeen, Trevor and Lin, Zeming and Gimelshein, Natalia and Antiga, Luca and others},
  journal={Advances in neural information processing systems},
  volume={32},
  year={2019}
}

@article{thompson2022lammps,
  title={LAMMPS-a flexible simulation tool for particle-based materials modeling at the atomic, meso, and continuum scales},
  author={Thompson, Aidan P and Aktulga, H Metin and Berger, Richard and Bolintineanu, Dan S and Brown, W Michael and Crozier, Paul S and In't Veld, Pieter J and Kohlmeyer, Axel and Moore, Stan G and Nguyen, Trung Dac and others},
  journal={Computer physics communications},
  volume={271},
  pages={108171},
  year={2022},
  publisher={Elsevier}
}

@article{trott2021kokkos,
  title={Kokkos 3: Programming model extensions for the exascale era},
  author={Trott, Christian R and Lebrun-Grandi{\'e}, Damien and Arndt, Daniel and Ciesko, Jan and Dang, Vinh and Ellingwood, Nathan and Gayatri, Rahulkumar and Harvey, Evan and Hollman, Daisy S and Ibanez, Dan and others},
  journal={IEEE Transactions on Parallel and Distributed Systems},
  volume={33},
  number={4},
  pages={805--817},
  year={2021},
  publisher={IEEE}
}

@article{meng2026kokkos,
  title={A Kokkos-accelerated moment tensor potential implementation for LAMMPS},
  author={Meng, Zijian and Zongo, Karim and Torres, Edmanuel and Maxwell, Christopher and Grant, Ryan and B{\'e}land, Laurent Karim},
  journal={SoftwareX},
  volume={33},
  pages={102524},
  year={2026},
  publisher={Elsevier}
}

@article{li2026lang2mlip,
  title={Lang2MLIP: End-to-End Language-to-Machine Learning Interatomic Potential Development with Autonomous Agentic Workflows},
  author={Li, Wenwen and Orimo, Yuki and Charoenphakdee, Nontawat},
  journal={arXiv preprint arXiv:2605.14527},
  year={2026}
}

\end{document}

\title{Supplementary Information \Romannum{1} for: LightPFP: A Lightweight Route to Ab Initio Accuracy at Scale}

\author{Wenwen Li}
 \email{wenwenli@preferred.jp}
\affiliation{Preferred Networks Inc., Tokyo, Japan.}

\author{Nontawat Charoenphakdee}
 \email{nontawat@preferred.jp}
\affiliation{Preferred Networks Inc., Tokyo, Japan.}

\author{Yong-Bin Zhuang}
\affiliation{Preferred Networks Inc., Tokyo, Japan.}

\author{Ryuhei Okuno}
\affiliation{Preferred Networks Inc., Tokyo, Japan.}

\author{Yuta Tsuboi}
\affiliation{Preferred Networks Inc., Tokyo, Japan.}

\author{So Takamoto}
\affiliation{Preferred Networks Inc., Tokyo, Japan.}

\author{Junichi Ishida}
\affiliation{Matlantis Corporation, Tokyo, Japan.}

\author{Iori Kurata}
\affiliation{Preferred Networks Inc., Tokyo, Japan.}

\author{Ju Li}
\affiliation{Department of Materials Science and Engineering, Massachusetts Institute of Technology, Cambridge, MA 02139, USA}
\affiliation{Department of Nuclear Science and Engineering, Massachusetts Institute of Technology, Cambridge, MA, USA
}

\date{\today}

\maketitle

% \section*{Note}
% This supplementary material provides additional methods, figures, tables, and discussions supporting the main manuscript.
\clearpage
\section{Pseudocode}
This supplementary section provides an overview of the five pseudocode algorithms used to describe the LightPFP workflow and benchmark evaluation. 
Algorithm~\ref{alg:lightpfp-detailed} summarizes the overall LightPFP framework, from teacher-driven dataset construction to model initialization, knowledge distillation, and optional loss rebalancing. Algorithm~\ref{alg:generate-dataset} describes how target datasets are generated by applying static and dynamic structural sampling procedures and labeling the resulting structures with the teacher PFP model. Algorithm~\ref{alg:extract-target-parameters} explains how target-element parameters are extracted from a pretrained student model to initialize a task-specific model. Algorithm~\ref{alg:lightpfp-train} details the stage-wise training procedure, including the use of different loss weights for force-focused training, energy--stress refinement, and balanced final optimization. Finally, Algorithm~\ref{alg:data-efficiency-ni3al} outlines the data-efficiency evaluation protocol on the Ni$_3$Al benchmark, where models trained from scratch and from pretrained initialization are compared across different training-set sizes.

\clearpage

\begin{algorithm}[H]
\caption{LightPFP framework for distilling a teacher PFP model into a task-specific MTP model.}
\label{alg:lightpfp-detailed}
 \setstretch{0.85} \setstretch{0.85}
\begin{algorithmic}[1]

\State \textbf{Input:}
\State \quad Target system $\mathcal{S}$; teacher model $M_{\mathrm{PFP}}$
\State \quad Pretrained model $M_{\mathrm{pre}}$
\State \quad MTP parameters $\Theta = (R_{\mathrm{cut}}, \ell_{\max}, C_{\mu}, C_{\nu}, n_q, \Theta_{\mathrm{readout}})$
\State \quad Stage-wise loss coefficients $\Sigma = \{\Sigma^{(s)}\}_{s=1}^{K}$, where $\Sigma^{(s)} = (\alpha^{(s)}, \beta^{(s)}, \gamma^{(s)})$
\State \quad \texttt{use\_balance\_loss} (boolean)
\State \textbf{Output:} Distilled task-specific model $M_{\mathrm{fine}}$

\vspace{2pt}
\State \textbf{Phase 1: Dataset construction}
\State $\mathcal{D}_{\mathrm{target}}, (\mathbf{E}, \mathbf{F}, \boldsymbol{\sigma}) 
\leftarrow \textsc{GenerateDataset}(\mathcal{S}, M_{\mathrm{PFP}})$
\State \quad \textit{Dynamic sampling:} NVT and NPT molecular dynamics
\State \quad \textit{Static sampling:} compression, deformation, displacement, rattle, vacancy, surface, substitution

\vspace{2pt}
\State \textbf{Phase 2: Model initialization}
\State $\mathcal{Z}_{\mathrm{target}} \leftarrow \textsc{GetElements}(\mathcal{S})$
\State $M_{\mathrm{fine}} \leftarrow \textsc{ParamExtract}(M_{\mathrm{pre}},\mathcal{Z}_{\mathrm{target}})$
% \If{$M_{\mathrm{pre}}$ is provided}
%     \State $M_{\mathrm{fine}} \leftarrow \textsc{ParamExtract}(M_{\mathrm{pre}}, \mathcal{Z}_{\mathrm{target}})$
% \Else
%     \State $M_{\mathrm{fine}} \leftarrow \textsc{RandomInit}(\Theta, \mathcal{Z}_{\mathrm{target}})$
% \EndIf

\vspace{2pt}
\State \textbf{Phase 3: Knowledge distillation}
\For{$s = 1$ to $K$} 
    \State $(M_{\mathrm{fine}}, \mathcal{L}_{\mathrm{val}}^{(s)}) 
    \leftarrow \textsc{Train}(M_{\mathrm{fine}}, \mathcal{D}_{\mathrm{target}}, \Sigma^{(s)})$
\EndFor

\vspace{2pt}
\If{\texttt{use\_balance\_loss}}
    \State $\Sigma^{(K+1)} \leftarrow \textsc{BalanceCoeff}(\mathcal{L}_{\mathrm{val}}^{(K)})$ 
    \State $(M_{\mathrm{fine}}, \mathcal{L}_{\mathrm{val}}^{(K+1)}) 
    \leftarrow \textsc{Train}(M_{\mathrm{fine}}, \mathcal{D}_{\mathrm{target}}, \Sigma^{(K+1)})$
\EndIf

\State \textbf{return} $M_{\mathrm{fine}}$

\end{algorithmic}
\end{algorithm}

\clearpage

\begin{algorithm}[H]
\caption{Dataset generation via PFP-labeled structural sampling.}
\label{alg:generate-dataset}
 \setstretch{0.85}
\begin{algorithmic}[1]
\State \textbf{Input:}
\State \quad Initial atomic structure(s) $\mathcal{R}_0$
\State \quad Teacher model $M_{\mathrm{PFP}}$
\State \quad $\mathcal{S}_{\mathrm{cfg}} = [(\mathit{method}_1, \mathit{args}_1), \ldots, (\mathit{method}_K, \mathit{args}_K)]$
\State \qquad \Comment{ordered list of enabled sampling methods}

\State \textbf{Output:}
\State \quad Dataset $\mathcal{D} = \{(\mathcal{R}_k,\, E_k,\, \mathbf{F}_k,\, \boldsymbol{\sigma}_k)\}$

\vspace{4pt}
\State $\mathcal{D} \leftarrow \emptyset$

\For{each initial structure $\mathcal{R} \in \mathcal{R}_0$}

    \State $\mathcal{R} \leftarrow \textsc{Relax}(\mathcal{R},\, M_{\mathrm{PFP}})$
    \Comment{Structure optimization}

    \State $\mathcal{S} \leftarrow \emptyset$
    \vspace{2pt}

    \For{each sampling configuration $(\mathit{method}, \mathit{args}) \in \mathcal{S}_{\mathrm{cfg}}$}

        \If{$\mathit{method} = \textit{Volume scaling}$}
            \State $\mathcal{S} \mathrel{\cup}= \textsc{ScaleCell}(\mathcal{R},\, \mathit{args})$
        \ElsIf{$\mathit{method} = \textit{Strain deformation}$}
            \State $\mathcal{S} \mathrel{\cup}= \textsc{ApplyStrain}(\mathcal{R},\, \mathit{args})$
        \ElsIf{$\mathit{method} = \textit{Single-atom displacement}$}
            \State $\mathcal{S} \mathrel{\cup}= \textsc{Displace}(\mathcal{R},\, \mathit{args})$
        \ElsIf{$\mathit{method} = \textit{Rattle}$}
            \State $\mathcal{S} \mathrel{\cup}= \textsc{Rattle}(\mathcal{R},\, \mathit{args})$
        \ElsIf{$\mathit{method} = \textit{Vacancy}$}
            \State $\mathcal{S} \mathrel{\cup}= \textsc{RemoveAtoms}(\mathcal{R},\, \mathit{args})$
        \ElsIf{$\mathit{method} = \textit{Surface slab}$}
            \State $\mathcal{S} \mathrel{\cup}= \textsc{GenerateSlab}(\mathcal{R},\, \mathit{args})$
        \ElsIf{$\mathit{method} = \textit{Substitution}$}
            \State $\mathcal{S} \mathrel{\cup}= \textsc{SubstituteAtoms}(\mathcal{R},\, \mathit{args})$
        \ElsIf{$\mathit{method} = \textit{MD sampling}$}
            \State $\mathcal{S} \mathrel{\cup}= \textsc{SampleMD}(\mathcal{R},\, M_{\mathrm{PFP}},\, \mathit{args})$
        \EndIf

    \EndFor

    \vspace{4pt}
    \For{each variant $\mathcal{R}' \in \mathcal{S}$}
        \State $(E,\, \mathbf{F},\, \boldsymbol{\sigma}) \leftarrow M_{\mathrm{PFP}}(\mathcal{R}')$
        \State $\mathcal{D} \leftarrow \mathcal{D} \cup \{(\mathcal{R}',\, E,\, \mathbf{F},\, \boldsymbol{\sigma})\}$
    \EndFor

\EndFor

\vspace{2pt}
\State \Return $\mathcal{D}$

\end{algorithmic}
\end{algorithm}

\begin{algorithm}[H]
\caption{ExtractTargetParameters algorithm}
\label{alg:extract-target-parameters}
 \setstretch{0.85}
\begin{algorithmic}[1]

\State \textbf{Input:}
\State \quad Pretrained student model $M_{\mathrm{pre}}$ with parameters 
$\{c^{(\beta)}_{\mu, z_i, z_j}, \xi_\alpha, \Theta_{\mathrm{readout}}\}$
\State \quad Target element set $\mathcal{Z}_{\mathrm{target}}$

\State \textbf{Output:}
\State \quad Initialized student model $M_{\mathrm{init}}$

\vspace{2pt}
\State Extract element set $\mathcal{Z}_{\mathrm{pre}}$ from $M_{\mathrm{pre}}$
\State Construct index mapping $\pi: \mathcal{Z}_{\mathrm{target}} \rightarrow \mathcal{Z}_{\mathrm{pre}}$

\vspace{4pt}
\State \textbf{Extract element-pair-dependent radial parameters}
\For{$z_i, z_j \in \mathcal{Z}_{\mathrm{target}}$}
    \For{all $\mu, \beta$}
        \State $c^{(\beta)}_{\mu, z_i, z_j}(\mathrm{init}) 
        \leftarrow 
        c^{(\beta)}_{\mu, \pi(z_i), \pi(z_j)}(\mathrm{pre})$
    \EndFor
\EndFor

\vspace{4pt}
\State \textbf{Copy element-independent parameters}
\State $\xi_\alpha(\mathrm{init}) \leftarrow \xi_\alpha(\mathrm{pre}) \quad \forall \alpha$
\State $\Theta_{\mathrm{readout}}(\mathrm{init}) \leftarrow \Theta_{\mathrm{readout}}(\mathrm{pre})$

\vspace{4pt}
\State Construct $M_{\mathrm{init}}$ using extracted parameters

\vspace{2pt}
\State \Return $M_{\mathrm{init}}$

\end{algorithmic}
\end{algorithm}

\begin{algorithm}[H]
\caption{StageWiseTraining algorithm}
\label{alg:lightpfp-train}
 \setstretch{0.85}
\begin{algorithmic}[1]

\State \textbf{Input:} 
\State \quad Pretrained student model $M$
\State \quad Labeled dataset $\mathcal{D} = \{(\mathbf{R}, E, \mathbf{F}, \boldsymbol{\sigma})\}$
\State \quad Training hyperparameters $\Theta$ (batch size, $N_{\mathrm{epoch}}$, optimizer settings)

\State \textbf{Output:} 
\State \quad Trained model $M_{\mathrm{fine}}$

\vspace{2pt}
\State \textbf{Preprocessing:}
\State Build neighbor lists for each structure
\State Split $\mathcal{D}$ into training and validation sets (90/10)

\vspace{2pt}
\State Initialize $M$ with pretrained student model

\vspace{2pt}
\For{$s \in \{1,2,3\}$}

    \If{$s=1$}
        \State $(\alpha,\beta,\gamma) \leftarrow (10^{-5}, 10, 10^{-5})$ 
        \Comment{force-focused training}
    \ElsIf{$s=2$}
        \State Load best checkpoint from Stage 1
        \State $(\alpha,\beta,\gamma) \leftarrow (1, 0.1, 10)$ 
        \Comment{energy and stress refinement}
    \Else
        \State Load best checkpoint from Stage 2
        \State Obtain validation losses $(L_E^{(2)}, L_F^{(2)}, L_\sigma^{(2)})$
        \State Compute weighted validation loss:
        \State \quad $L_{\mathrm{val}}^{(2)} \leftarrow \alpha^{(2)} L_E^{(2)} + \beta^{(2)} L_F^{(2)} + \gamma^{(2)} L_\sigma^{(2)}$
        \State $t \leftarrow L_{\mathrm{val}}^{(2)} / 3$
        \State $(\alpha,\beta,\gamma) \leftarrow \left(\dfrac{t}{L_E^{(2)}}, \dfrac{t}{L_F^{(2)}}, \dfrac{t}{L_\sigma^{(2)}}\right)$
        \Comment{balance losses: $\alpha L_E \approx \beta L_F \approx \gamma L_\sigma$}
    \EndIf

    \For{epoch $=1$ to $N_{\mathrm{epoch}}$}
        \For{each mini-batch $B \subset \mathcal{D}$}
            \State Predict $(\hat{E}, \hat{\mathbf{F}}, \hat{\boldsymbol{\sigma}}) \leftarrow M(B)$
            \State $L_E \leftarrow \mathrm{MSE}(\hat{E}, E)$
            \State $L_F \leftarrow \mathrm{MSE}(\hat{\mathbf{F}}, \mathbf{F})$
            \State $L_\sigma \leftarrow \mathrm{MSE}(\hat{\boldsymbol{\sigma}}, \boldsymbol{\sigma})$
            \State $L \leftarrow \alpha L_E + \beta L_F + \gamma L_\sigma$
            \State Update parameters of $M$ using Adam with gradient $\nabla L$
        \EndFor
    \EndFor

    \State Select checkpoint with lowest validation loss for Stage $s$
    \State Update $M$ with selected checkpoint

\EndFor

\vspace{2pt}
\State \Return $M_{\mathrm{fine}} \leftarrow M$

\end{algorithmic}
\end{algorithm}

\clearpage

\begin{algorithm}[H]
\caption{Data-Efficiency Evaluation on the Ni$_3$Al Benchmark}
\label{alg:data-efficiency-ni3al}
 \setstretch{0.85} \setstretch{0.85}
\begin{algorithmic}[1]

\State \textbf{Input:} Ni$_3$Al target system $S$; pretrained student model $M_{\mathrm{pre}}$; universal potential teacher $M_{\mathrm{PFP}}$
\State \textbf{Output:} Data-efficiency curves for comparing $M_{\mathrm{fine}}$ and $M_{\mathrm{scratch}}$

\State \textbf{Define metrics:} $\mathcal{M} = \{\text{energy MAE}, \text{force MAE}, \text{phonon error}, \text{surface energy error}\}$

\State \textbf{Phase 1: Dataset construction}
\State $\mathcal{D}_{\mathrm{full}} \leftarrow \textsc{GenerateDataset}(\text{Ni}_3\text{Al}, M_{\mathrm{PFP}})$
\State \quad \textit{Static sampling:} lattice compression, deformation, atomic displacement
\State \quad \textit{Dynamic sampling:} MD trajectories from defect-free bulk, defective bulk, and surface structures

\State \textbf{Phase 2: Metric calculation}
\For{$N \in \{100, \ldots, 850\}$}
    \For{$t = 1$ to $5$} // \textit{five trials}
        \State $\mathcal{D}_{N}^{(t)} \leftarrow \textsc{SampleSubset}(\mathcal{D}_{\mathrm{full}}, N, t)$

        \State \textbf{Model training:}
        \State $M_{\mathrm{scratch}}^{(t)} \leftarrow \textsc{Train}(\mathcal{D}_{N}^{(t)}, \texttt{random\_init})$
        \State $M_{\mathrm{fine}}^{(t)} \leftarrow \textsc{Train}(\mathcal{D}_{N}^{(t)}, M_{\mathrm{pre}})$

        \State \textbf{Evaluation:}
        \For{each metric $m \in \mathcal{M}$}
            \State $\mu_{\mathrm{scratch}, m}^{(t)} \leftarrow \textsc{Evaluate}(M_{\mathrm{scratch}}^{(t)}, m)$
            \State $\mu_{\mathrm{fine}, m}^{(t)} \leftarrow \textsc{Evaluate}(M_{\mathrm{fine}}^{(t)}, m)$
        \EndFor
    \EndFor
\EndFor

\State \textbf{Phase 3: Result aggregation}
\For{each metric $m \in \mathcal{M}$}
    \State $\mu^m_{\mathrm{scratch}, m}, \sigma_{\mathrm{scratch}, m} \leftarrow \textsc{Aggregate}(\{\mu_{\mathrm{scratch}, m}^{(t)}\}_{t=1}^{5})$
    \State $\mu^m_{\mathrm{fine}, m}, \sigma_{\mathrm{fine}, m} \leftarrow \textsc{Aggregate}(\{\mu_{\mathrm{fine}, m}^{(t)}\}_{t=1}^{5})$
\EndFor

\State Compare performance as a function of dataset size $N$ for all $m \in \mathcal{M}$

\State \textbf{Output:} data-efficiency performance curves

\end{algorithmic}
\end{algorithm}
\clearpage

\section{Error transfer in the DFT→PFP→LightPFP pipeline}

As mentioned in the main text, the dominant error arises from formally exact → DFT due to DFT’s intrinsic limitations; by contrast, DFT → PFP transfer error is already small (with the costly training completed), and PFP → LightPFP is even smaller with fast, overnight training.
Considering independent sources of error $e_1$, $e_2$, ..., $e_m$ often do not add up linearly but quadratically (if statistically uncorrelated due to different ``physics"):

\begin{equation}
    e = \sqrt{e_1^2 + e_2^2 + ... + e_m^2},
\end{equation}
if $|e_1|\sim 100$ meV/atom dominates over $|e_2|, ..., |e_m|$, then the leading-order contributions of $|e_2|, ..., |e_m|$ to the total error would be even smaller than it seems, based on Taylor expansion:
\begin{equation}
    e \approx e_1 + \frac{e_2^2 + ... + e_m^2}{2e_1},
\end{equation}

and likely become practically negligible. In other words, if the DFT$\rightarrow$PFP and PFP$\rightarrow$LightPFP
neural network trainings are done well, 
LightPFP may represent DFT much better than how DFT reflects reality.

\clearpage

\section{Pretrained student model}

\paragraph{Meta learning pretrained student model}
We employed a comprehensive dataset to train PFP, a universal potential-based graph neural networks. 
This dataset comprises 86 different elements, covering nearly the entire periodic table and encompassing both equilibrium structures and numerous disordered structures that deviate from equilibrium states. 
The dataset includes not only bulk phases but also complex structures such as surfaces, adsorption configurations, and clusters. 
This comprehensive coverage is the fundamental reason why PFP exhibits broad applicability across diverse materials simulations.
For dataset details, please refer to~\citet{takamoto2022towards}.

However, compared to PFP~\citep{takamoto2022towards}, moment tensor potentials (MTPs) are compact models with limited parameters and constrained expressive power, typically applicable only to single materials systems. 
Consequently, using MTPs to fit all datasets simultaneously presents significant challenges. 
Therefore, our MTP pretraining strategy aims to optimize the model to facilitate subsequent fine-tuning for individual tasks, instead of maximizing accuracy across all datasets. 
To achieve this objective, we employed the Reptile meta-learning algorithm~\citep{reptile}.

The Reptile algorithm operates by iteratively sampling tasks from a task distribution and updating model parameters to enhance the model's ability to rapidly adapt to new tasks. 
In our implementation, we partitioned the complete dataset into 12 specific tasks based on structural types, as detailed in Table~\ref{table:tasks}. 
% In our implementation, we partitioned the complete dataset into 12 specific tasks based on structural types. 
During each inner loop iteration, we select a task (i.e., a dataset containing specific structural types such as single molecules) to train the MTP model. 
Given the substantial size of each task's dataset, we limit training to one epoch per inner loop before proceeding to the parameter update. 
The model parameters are then updated according to the following formula:
\begin{equation*}
\delta \theta = \theta_{i} - \theta,
\end{equation*}
\begin{equation*}
\theta \leftarrow \theta + {\beta} \delta \theta,
\end{equation*}
where $\theta$ represents the MTP parameters, $\theta_{i}$ denotes the parameters after the $i$-th inner loop, and ${\beta}$ is a hyperparameter in the Reptile algorithm that controls the magnitude of the meta-update step during training. In our implementation, ${\beta}$ is set to $0.5$.
We iteratively repeat the task sampling and inner-loop/meta-update procedures for $100$ iterations until convergence of energy, forces, and stress is observed across all datasets.

We employed the Adam optimization method with a learning rate of $1 \times 10^{-3}$.
The model was trained for 1 epochs with a batch size of 256.
Total pretraining time was approximately 100 hours.

For example, pretrained student model with hyperparameter (levmax=8, $C_{\mu}$=1, $C_{\nu}$=1, $n_q$=16) contains 86×86×4×16 training parameters for the radial function $c$ and additional 27 coefficients for the basis functions $\xi$. 
The modular structure of MTP enables selective parameter extraction during inference or fine-tuning, significantly enhancing computational efficiency. 
The extraction procedure is straightforward, depending on elements used for the task.
For example, when handling a material containing only H and O elements, we can extract the relevant subset of the radial function parameter tensor—specifically a 2×2×4×16 matrix corresponding to these elements, while maintaining the coefficients of the basis function unchanged. 
Consequently, although the pretrained model may contain numerous parameters, it automatically reduces to a compact, element-specific model equivalent in size to those trained from scratch for the particular material system.

\begin{table}
    \centering
\begin{tabular}{cl}
\hline
\textbf{Task index} & \textbf{Dataset description} \\ \hline

1 & Single substance structures \\ 
2 & Bulk structures with two elements \\ 
3 & Crystal structures from materials project database \\ 
4 & Structures with random atomic position \\ 
5 & Bulk structures with defects \\ 
6 & Single molecules \\ 
7 & Structures composed by multiple molecules \\ 
8 & Molecules with element substitution \\ 
9 & Adsorption structures of surfaces and molecules \\ 
10 & Random combinations of surfaces and molecules \\ 
11 & Slabs \\ 
12 & Clusters \\ \hline
\end{tabular}
\caption{Task definitions for Reptile meta learning algorithm based on datasets trained for 
PFP~\citep{takamoto2022towards}.}
\label{table:tasks}
\end{table}

\paragraph{Dataset description}
The dataset used in for LightPFP pretraining was extending the crystal portion of the PreFerred Potential (PFP) dataset in reference ~\cite{takamoto2022towards}. The original PFP crystal dataset contains approximately $3 \times 10^{6}$
density functional theory (DFT)-calculated structures under periodic boundary conditions and covers 45 elements; in the present work, this dataset was further expanded to cover 86 elements while retaining broad structural and chemical diversity. The dataset contains a wide variety of material configurations, including three-dimensional bulk crystals, elemental solids, binary and multicomponent compounds, alloy structures, ionic and covalent crystals, metallic systems, finite clusters, two-dimensional slabs, surfaces with different terminations, and adsorbate–surface configurations. The adsorbed systems include small molecules or fragments placed on crystalline surfaces, covering both optimized adsorption geometries and randomly initialized adsorption configurations. In addition, the dataset includes molecular structures treated under periodic boundary conditions, disordered atomic configurations, high-temperature molecular dynamics snapshots, and intentionally non-equilibrium or metastable structures. Examples of such non-equilibrium configurations include distorted bulk crystals, expanded or compressed cells, strained cells, randomly displaced atomic positions, and hypothetical crystal prototypes that are not necessarily stable in their ideal forms. This design allows the dataset to sample not only local environments near equilibrium structures, but also configurations relevant to surface reactions, adsorption, diffusion, phase transitions, alloy disordering, structural deformation, and high-temperature dynamics. For each structure, DFT total energies and atomic forces are provided as training targets. The diversity of bulk, surface, cluster, adsorption, molecular, and disordered systems follows the construction strategy of the original PFP crystal dataset, which was designed to improve the transferability of neural network potentials across different material classes and atomic environments.

\paragraph{Benchmark of meta-learning pretrained students}

To provide a more comprehensive evaluation of pretrained students beyond the results presented in Section II.B and Figure 3 of the main text, we conduct an extended benchmark comparing different transfer learning approaches. Specifically, we investigate the effectiveness of meta-learning by contrasting it with a multi-task learning baseline for creating pretrained students. In the multi-task approach, all training datasets from different material systems, as listed in Table~\ref{table:tasks}, are combined and fed simultaneously into a single training process, using identical datasets as those employed in meta-learning. The subsequent fine-tuning protocol remains consistent across both methods.

\begin{figure}[h]
\includegraphics[width=0.7\columnwidth]{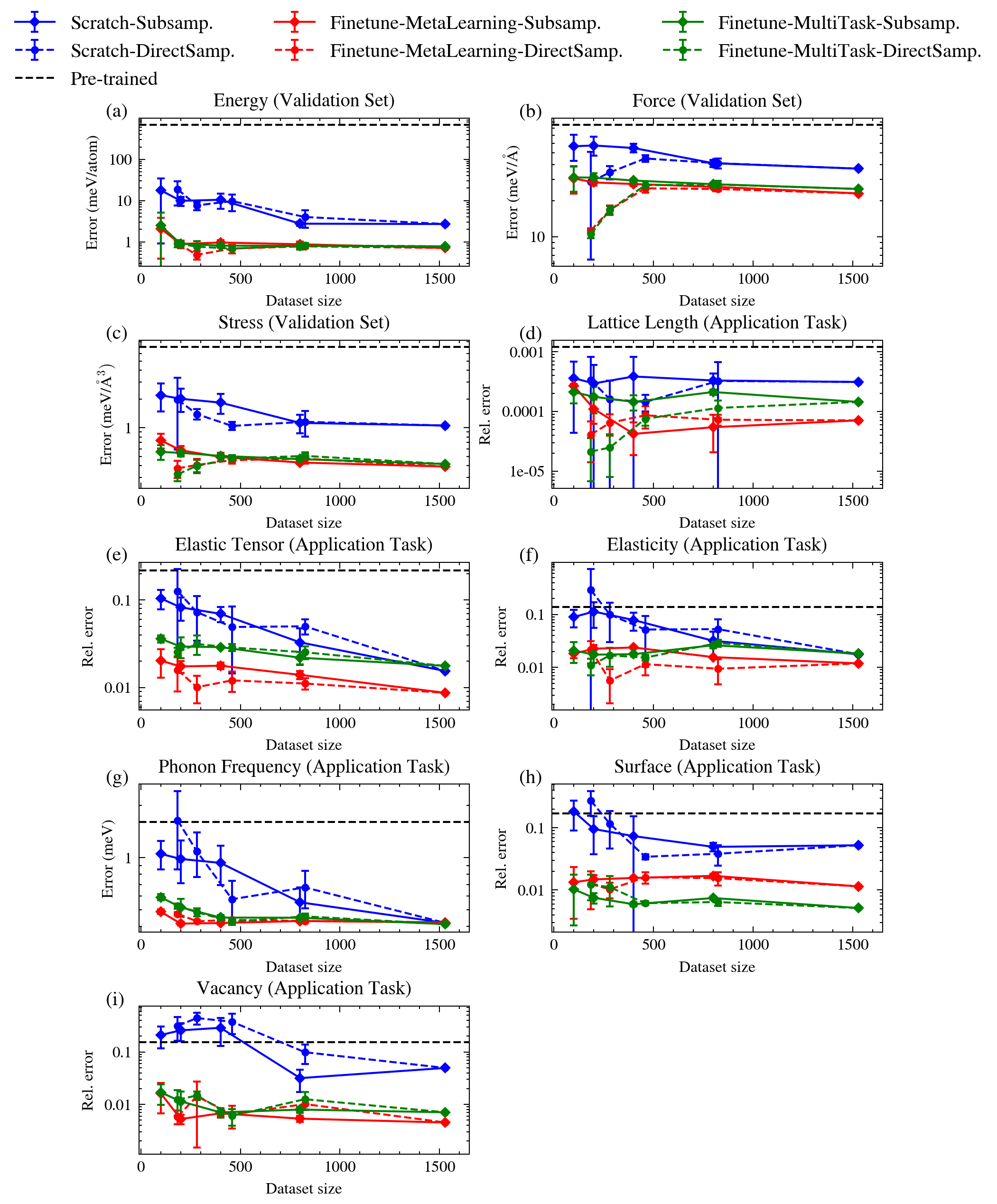}
\caption{Detailed comparison of data efficiency between fine-tuned pretrained and scratch-trained student models.}
\label{fig:data_efficient_full}
\end{figure}

Figure~\ref{fig:data_efficient_full} presents the comprehensive benchmark results across nine different metrics as a function of training dataset size. In addition to the energy, force, phonon frequency, and surface energy metrics shown in the main text, we include stress (panel c), lattice length (panel d), elastic tensor (panel e), elasticity moduli (panel f), and vacancy formation energy (panel i) to provide a more complete picture of model performance across diverse physical properties. The results consistently demonstrate that fine-tuning from pretrained students—whether obtained via meta-learning or multi-task learning—substantially outperforms training from scratch across all evaluated properties, providing strong and comprehensive evidence for our conclusions.

% Importantly, beyond overall performance, the results highlight differences in adaptability under limited data regimes. Meta-learned pretrained students consistently achieve lower errors at smaller dataset sizes, indicating faster adaptation to new target systems compared to the multi-task baseline. The advantage of meta-learning is particularly pronounced in the low-data regime, while the gap gradually narrows as the dataset size increases.

Interestingly, we observe that the meta-learning approach yields pretrained students that achieve superior performance compared to the multi-task approach in the majority of benchmarks. For directly fitted quantities such as energy (panel a), force (panel b), and stress (panel c), the meta-learning models exhibit only marginal improvements over their multi-task counterparts. However, when evaluated on derived physical properties that are more relevant to practical applications, meta-learning demonstrates clear advantages. Specifically, meta-learning shows significantly better performance in predicting lattice constants (panel d), elastic tensors (panel e), elasticity moduli (panel f), and vacancy formation energies (panel i). With the exception of surface energy (panel h), where both approaches achieve comparable performance and the multi-task model shows slightly lower error, the overall trend consistently favors meta-learning. These results suggest that meta-learning's ability to capture transferable knowledge across different material systems leads to improved generalization for complex physical properties, even when the improvements in fundamental energy and force predictions are modest. This is consistent with the objective of meta-learning, which explicitly optimizes for rapid adaptation across tasks, whereas multi-task learning focuses on joint performance without explicitly encouraging fast task-specific adaptation.

\paragraph{Specific type pretrained student model}
In addition to the pretrained LightPFP model that covers almost all materials, we also tried pretrained LightPFP models for special types of materials. As an illustrative example, we consider our organic pretrained student model, which is specifically designed for organic molecular systems. The training process begins with dataset construction. We randomly sample molecular information, such as SMILES representations, from PubChem\cite{kim2025pubchem} and generate corresponding three-dimensional conformers. Several molecules are then randomly placed into a simulation box, ensuring that the overall density falls within an appropriate range. An optimization algorithm is employed to adjust atomic positions without breaking chemical bonds, thereby minimizing atomic overlap between molecules. From these initial configurations, we perform molecular dynamics (MD) simulations using PFP at temperatures randomly selected between 300 and 3000 K. Each simulation runs for 1000 steps, and configurations are sampled every 100 steps. This procedure is repeated many times to obtain a diverse collection of molecular configurations. The resulting dataset is subsequently used for training a Moment Tensor Potential (MTP) model, yielding a pretrained MTP tailored for organic systems.

\begin{figure}[h]
\includegraphics[width=0.5\columnwidth]{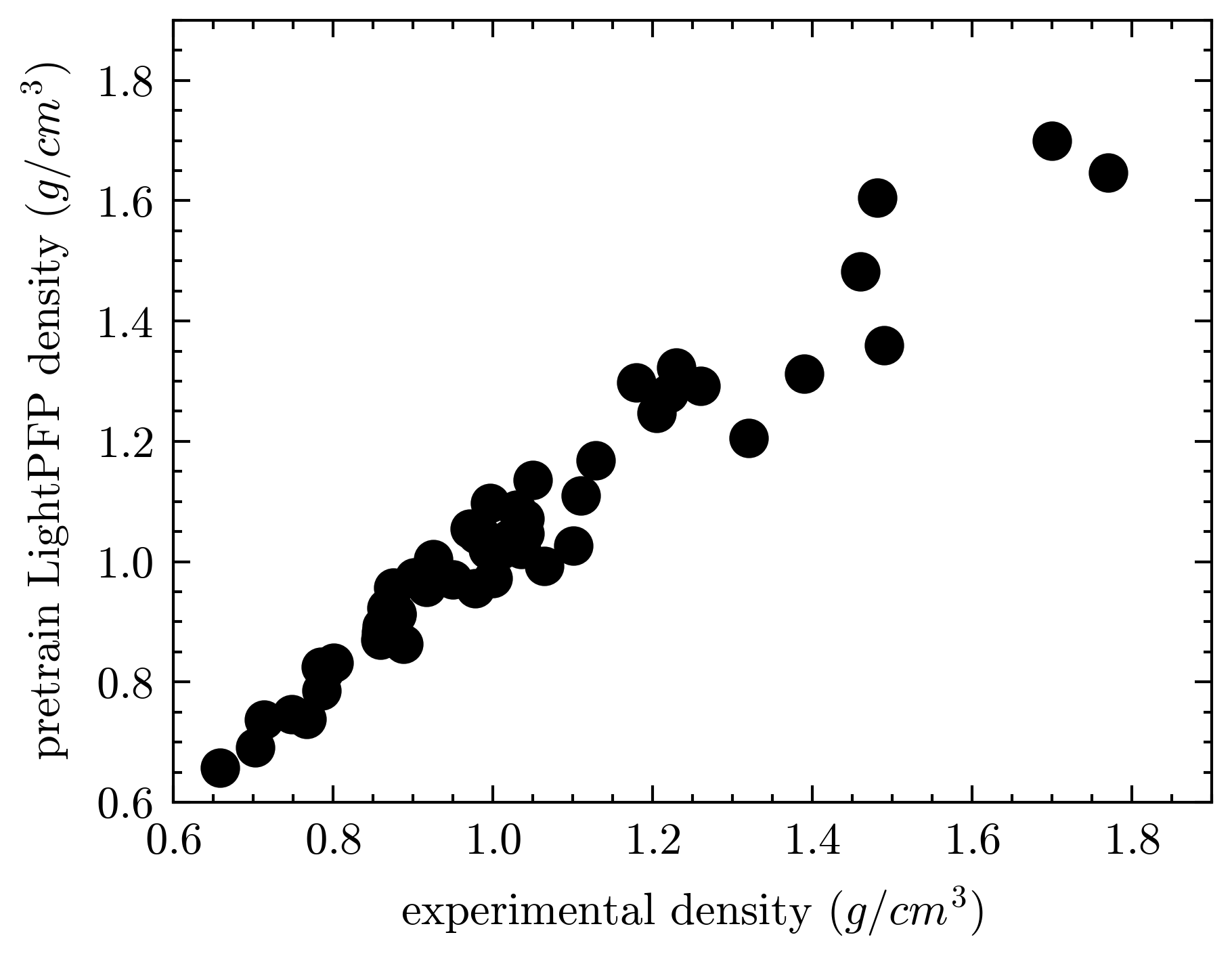}
\caption{Learning curves for different training strategies.}
\label{fig:molecule_density}
\end{figure}

We observe that when the model is restricted to a specific class of materials, the pretrained MTP demonstrates a notable capability for direct application without fine-tuning. As shown in Fig.~\ref{fig:molecule_density}, the pretrained model accurately predicts the densities of various organic molecules, exhibiting strong agreement with experimental results despite the absence of these molecules from the training dataset. This finding suggests a promising new direction: by constraining the material domain, one can develop lightweight machine-learned interatomic potentials (MLIPs) with reduced generalization compared to universal MLIPs (uMLIPs), yet capable of fast inference and requiring no additional training. For instance, pretrained student models can be constructed for specific material classes such as alloys, oxides, perovskites, and metal–organic frameworks (MOFs).

\clearpage

\section{Training structure sampling methods}

\begin{figure}[h]
\includegraphics[width=\textwidth]{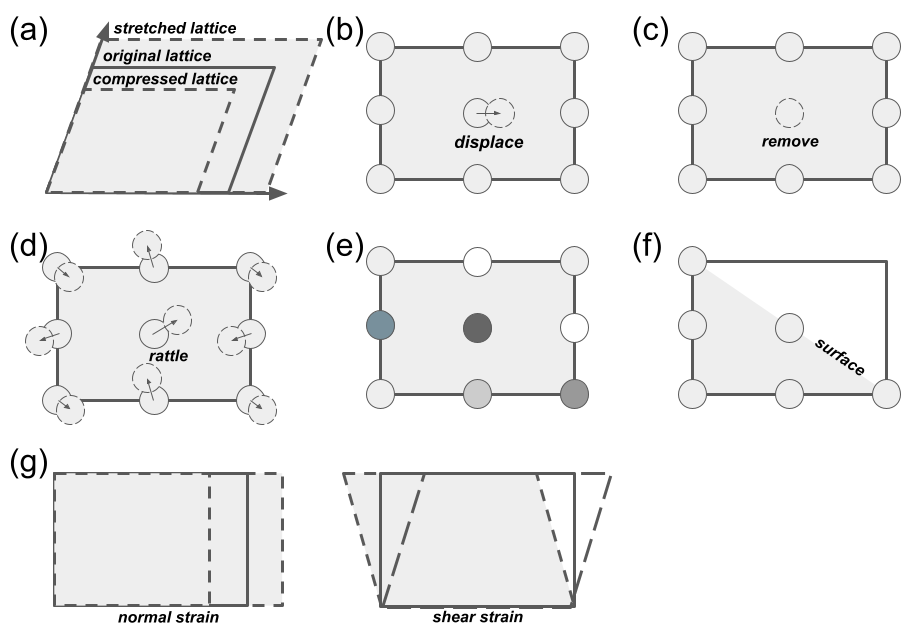}
\caption{Illustration of sampling methods used in the LightPFP dataset generation. (a) Uniform compression/stretch sampling, (b) Displacement sampling, (c) Vacancy sampling, (d) Rattle sampling, (e) Substitution sampling, (f) Surface sampling and (g) Deformation sampling. Reproduced from the Matlantis manual (accessed 12 Nov 2025).}
\label{fig:sampling_method}
\end{figure}

A robust training dataset for machine learning interatomic potentials is built by systematically sampling diverse yet physically meaningful configurations around one or more initial structures. Sampling methods can be combined and run independently to cover thermal, mechanical, defect, surface, and chemical degrees of freedom. These strategies balance relevance to the target material with diversity across configuration space, improving both accuracy and robustness of the potential. The illustration of these sampling methods are shown at Figure~\ref{fig:sampling_method}.

\subsection{Molecular dynamics sampling}

Molecular dynamics (MD) sampling generates structures by propagating the atomic system under finite-temperature dynamics, optionally at controlled pressure. By choosing ensembles such as NVT (constant volume) or NPT (constant pressure), and by varying temperature, one can explore configurational space from near-equilibrium states to highly disordered regimes. To prevent collecting redundant configurations, snapshots are taken at a fixed stride along the trajectory. See Algorithm \ref{alg:sample-md}.

\begin{algorithm}[H]
\caption{\textsc{SampleMD}}
\label{alg:sample-md}
 \setstretch{0.85}
\begin{algorithmic}[1]

\State \textbf{Input:}
\State \quad Structure $\mathcal{R}$, teacher model $M_{\mathrm{PFP}}$
\State \quad Ensemble $\in \{\mathrm{NVT}, \mathrm{NPT}\}$, temperature $T$, (optional) pressure $P$
\State \quad Thermostat/barostat scheme (e.g., Langevin, Nos\'e--Hoover)
\State \quad Equilibration steps $N_{\mathrm{eq}}$, production steps $N_{\mathrm{MD}}$, snapshot stride $\tau$

\State \textbf{Output:}
\State \quad Set of MD snapshots $\mathcal{S}$

\vspace{4pt}
\State Attach $M_{\mathrm{PFP}}$ to $\mathcal{R}$
\State Initialize atomic velocities from Maxwell--Boltzmann distribution at temperature $T$
\State Equilibrate $\mathcal{R}$ for $N_{\mathrm{eq}}$ MD steps under the specified ensemble
\State $\mathcal{S} \leftarrow \emptyset$

\For{$t = 1, \ldots, N_{\mathrm{MD}}$}
    \State Propagate $\mathcal{R}$ by one MD step (Ensemble, $T$, $P$)
    \If{$t \bmod \tau = 0$}
        \State $\mathcal{S} \leftarrow \mathcal{S} \cup \{\mathcal{R}^{(t)}\}$
    \EndIf
\EndFor

\State \Return $\mathcal{S}$

\end{algorithmic}
\end{algorithm}

\clearpage
\subsection{Uniform compression/stretch sampling}

Uniform compression/stretch sampling produces structures by isotropically scaling the lattice vectors of the input periodic structure, keeping the lattice angles unchanged and preserving fractional atomic coordinates. This method targets the volume–energy relationship and can be augmented by fixed-cell relaxation or MD runs starting from the scaled configurations to enrich the dataset at specific densities. See Algorithm \ref{alg:compress}.

\begin{algorithm}[H]
\caption{\textsc{ScaleCell}}
\label{alg:compress}
 \setstretch{0.85}
\begin{algorithmic}[1]
\State \textbf{Input:}
\State \quad Structure $\mathcal{R}$, teacher model $M_{\mathrm{PFP}}$
\State \quad Supercell multipliers $(n_x, n_y, n_z)$
\State \quad Scale range $[s_{\min}, s_{\max}]$, scale interval $\Delta s$
\State \quad (If \textit{md}) MD parameters $\mathit{args}_{\mathrm{MD}}$
\State \quad Flags: \textit{opt} (fixed-volume relaxation), \textit{md} (MD sampling)
\State \textbf{Output:}
\State \quad Set of sampled structures $\mathcal{S}$
\State $\mathcal{S} \gets \emptyset$
\State $\mathcal{R}_0 \gets \mathcal{R} \times (n_x, n_y, n_z)$ \Comment{expand to supercell}
\State $C_0 \gets \mathrm{cell}(\mathcal{R}_0)$ \Comment{reference lattice vectors}
\For{$s \in \{s_{\min},\, s_{\min}+\Delta s,\, \ldots,\, s_{\max}\}$}
    \State $\mathcal{R}' \gets \mathcal{R}_0$ with cell set to $s \cdot C_0$, fractional coordinates fixed
    \State $\mathcal{S} \gets \mathcal{S} \cup \{\mathcal{R}'\}$ \Comment{always keep the scaled structure}
    \If{\textit{opt}}
        \State $\mathcal{R}'' \gets \textsc{RelaxFixedVolume}(\mathcal{R}', M_{\mathrm{PFP}})$ \Comment{relax atoms and cell shape at fixed volume}
        \State $\mathcal{S} \gets \mathcal{S} \cup \{\mathcal{R}''\}$
    \EndIf
    \If{\textit{md}}
        \State $\mathcal{S} \gets \mathcal{S} \cup \textsc{SampleMD}(\mathcal{R}', M_{\mathrm{PFP}},\, \mathit{args}_{\mathrm{MD}})$ \Comment{NVT snapshots at this density}
    \EndIf
\EndFor
\State \Return $\mathcal{S}$
\end{algorithmic}
\end{algorithm}

\clearpage
\subsection{Deformation sampling}

Deformation (strain) sampling applies prescribed normal and shear strain components to the unit cell, changing both lattice lengths and angles while maintaining periodicity. By scanning the six independent components of the strain tensor, one obtains structures spanning elastic distortions relevant to mechanical properties. Atomic positions may be further optimized under fixed cell shape to produce relaxed strained configurations, helping the model learn stress–strain behavior and elastic responses. See Algorithm \ref{alg:sample-deformed}.

\begin{algorithm}[H]
\caption{\textsc{ApplyStrain}}
\label{alg:sample-deformed}
 \setstretch{0.85}
\begin{algorithmic}[1]
\State \textbf{Input:}
\State \quad Structure $\mathcal{R}$, teacher model $M_{\mathrm{PFP}}$
\State \quad Supercell multipliers $(n_x, n_y, n_z)$
\State \quad Strain range $[\delta_{\min}, \delta_{\max}]$, strain interval $\Delta\delta$
\State \quad Flag: \textit{opt} (fixed-cell relaxation)
\State \textbf{Output:}
\State \quad Set of deformed structures $\mathcal{S}$
\vspace{4pt}
\State $\mathcal{S} \leftarrow \emptyset$
\State $\mathcal{R}_0 \leftarrow \mathcal{R} \times (n_x, n_y, n_z)$ \Comment{expand to supercell}
\State $C_0 \leftarrow \mathrm{cell}(\mathcal{R}_0)$
\For{each strain component $(i, j)$ with $1 \le i \le j \le 3$}
    \For{$\delta \in \{\delta_{\min},\, \delta_{\min}+\Delta\delta,\, \ldots,\, \delta_{\max}\}$}
        \State Set strain tensor $\epsilon$ with $\epsilon_{ij} = \delta$, all other components zero
        \State $F \leftarrow \sqrt{2\epsilon + I}$ \Comment{deformation gradient}
        \State Apply $C_0 \cdot F$ as new cell of $\mathcal{R}_0$, atoms scaled accordingly $\to \mathcal{R}'$
        \State $\mathcal{S} \leftarrow \mathcal{S} \cup \{\mathcal{R}'\}$
        \If{\textit{opt}}
            \State $\mathcal{R}'' \leftarrow \textsc{RelaxFixedCell}(\mathcal{R}', M_{\mathrm{PFP}})$
            \State $\mathcal{S} \leftarrow \mathcal{S} \cup \{\mathcal{R}''\}$
        \EndIf
    \EndFor
\EndFor
\State \Return $\mathcal{S}$
\end{algorithmic}
\end{algorithm}

\clearpage
\subsection{Displacement sampling}

Single-atom displacement sampling perturbs one atom from its equilibrium position along a Cartesian direction by a controlled amplitude. Such localized perturbations probe the curvature of the potential energy surface and the force constants around equilibrium, which are essential for learning vibrational responses. Each displaced configuration is generated independently from the same starting structure to map local force landscapes efficiently. See Algorithm \ref{alg:sample-displaced}.

\begin{algorithm}[H]
\caption{\textsc{Displace}}
\label{alg:sample-displaced}
 \setstretch{0.85}
\begin{algorithmic}[1]
\State \textbf{Input:}
\State \quad Structure $\mathcal{R}$, teacher model $M_{\mathrm{PFP}}$
\State \quad Supercell multipliers $(n_x, n_y, n_z)$
\State \quad Displacement amplitude $\delta$
\State \quad Number of samples $N$
\State \textbf{Output:}
\State \quad Set of displaced structures $\mathcal{S}$
\vspace{4pt}
\State $\mathcal{S} \leftarrow \emptyset$
\State $\mathcal{R}_0 \leftarrow \mathcal{R} \times (n_x, n_y, n_z)$ \Comment{expand to supercell}
\State $\mathcal{D} \leftarrow \{(i, \alpha, d) : i \in \mathcal{R}_0,\ \alpha \in \{x, y, z\},\ d \in \{-1, +1\}\}$
\State $N' \leftarrow \min(N,\, |\mathcal{D}|)$ \Comment{clip to maximum unique displacements}
\State $\mathcal{D} \leftarrow$ random subset of $\mathcal{D}$ of size $N'$ (without replacement)
\For{each $(i, \alpha, d) \in \mathcal{D}$}
    \State Displace atom $i$ by $d \cdot \delta$ along axis $\alpha$, all other atoms fixed $\to \mathcal{R}'$
    \State $\mathcal{S} \leftarrow \mathcal{S} \cup \{\mathcal{R}'\}$
\EndFor
\State \Return $\mathcal{S}$
\end{algorithmic}
\end{algorithm}

\clearpage
\subsection{Rattle sampling}

Rattle sampling introduces random displacements to all atoms simultaneously, drawing each component of the displacement from a specified distribution (e.g., Gaussian). This global perturbation broadens coverage of non-equilibrium configurations and can reveal failure modes of the model under larger distortions. Because it may produce unphysical configurations with extreme forces, filtering based on maximum force thresholds and optional relaxation steps are recommended to maintain data quality. This method is recommended for the molecular systems since it provided useful information of bond breaking. See Algorithm \ref{alg:sample-rattle}.

\begin{algorithm}[H]
\caption{\textsc{Rattle}}
\label{alg:sample-rattle}
 \setstretch{0.85}
\begin{algorithmic}[1]
\State \textbf{Input:}
\State \quad Structure $\mathcal{R}$
\State \quad Supercell multipliers $(n_x, n_y, n_z)$
\State \quad Displacement standard deviation $\sigma$
\State \quad Number of samples $N$
\State \textbf{Output:}
\State \quad Set of rattled structures $\mathcal{S}$
\vspace{4pt}
\State $\mathcal{S} \leftarrow \emptyset$
\State $\mathcal{R}_0 \leftarrow \mathcal{R} \times (n_x, n_y, n_z)$ \Comment{expand to supercell}
\For{$k = 1, \ldots, N$}
    \State Draw $\Delta \mathbf{r}_i \sim \mathcal{N}(0,\, \sigma^2 I)$ independently for each atom $i \in \mathcal{R}_0$
    \State Displace each atom $i$ by $\Delta \mathbf{r}_i$ $\to \mathcal{R}'$
    \State $\mathcal{S} \leftarrow \mathcal{S} \cup \{\mathcal{R}'\}$
\EndFor
\State \Return $\mathcal{S}$
\end{algorithmic}
\end{algorithm}

\clearpage
\subsection{Vacancy sampling}

Vacancy sampling creates point-defect structures by randomly removing one or more atoms from the initial configuration. These defective structures can be complemented with fixed-cell relaxations or MD to sample local reconstructions and thermally activated defect configurations. By including vacancy-containing data, the model gains sensitivity to defect energetics and local structural changes associated with missing atoms. See Algorithm \ref{alg:sample-vacancy}.

\begin{algorithm}[H]
\caption{\textsc{RemoveAtoms}}
\label{alg:sample-vacancy}
 \setstretch{0.85}
\begin{algorithmic}[1]
\State \textbf{Input:}
\State \quad Structure $\mathcal{R}$, teacher model $M_{\mathrm{PFP}}$
\State \quad Supercell multipliers $(n_x, n_y, n_z)$
\State \quad Number of vacancies per structure $n_v$
\State \quad Number of samples $N$
\State \quad (If \textit{md}) MD parameters $\mathit{args}_{\mathrm{MD}}$
\State \quad Flags: \textit{opt} (cell-relaxed), \textit{md} (MD sampling)
\State \textbf{Output:}
\State \quad Set of vacancy structures $\mathcal{S}$
\vspace{4pt}
\State $\mathcal{S} \leftarrow \emptyset$
\State $\mathcal{R}_0 \leftarrow \mathcal{R} \times (n_x, n_y, n_z)$ \Comment{expand to supercell}
\For{$k = 1, \ldots, N$}
    \State Randomly select $n_v$ distinct atoms from $\mathcal{R}_0$ (without replacement)
    \State Remove the selected atoms $\to \mathcal{R}'$
    \State $\mathcal{S} \leftarrow \mathcal{S} \cup \{\mathcal{R}'\}$
    \If{\textit{opt}}
        \State $\mathcal{R}'' \leftarrow \textsc{RelaxWithCell}(\mathcal{R}', M_{\mathrm{PFP}})$
        \State $\mathcal{S} \leftarrow \mathcal{S} \cup \{\mathcal{R}''\}$
    \EndIf
    \If{\textit{md}}
        \State $\mathcal{S} \leftarrow \mathcal{S} \cup \textsc{SampleMD}(\mathcal{R}', M_{\mathrm{PFP}},\, \mathit{args}_{\mathrm{MD}})$
    \EndIf
\EndFor
\State \Return $\mathcal{S}$
\end{algorithmic}
\end{algorithm}

\clearpage
\subsection{Surface sampling}

Surface sampling constructs slab models by cleaving the periodic bulk along specified Miller indices and introducing a vacuum layer to isolate the surfaces. Symmetry analysis can be used to avoid duplicate surfaces generated by equivalent indices in high-symmetry crystals. Subsequent fixed-cell relaxations and MD on slab geometries enrich the dataset with surface reconstructions and thermal fluctuations. This approach is intended for periodic crystalline inputs and targets accurate description of surface energetics and structure. See Algorithm \ref{alg:sample-surface}.

\begin{algorithm}[H]
\caption{\textsc{GenerateSlab}}
\label{alg:sample-surface}
 \setstretch{0.85}
\begin{algorithmic}[1]
\State \textbf{Input:}
\State \quad Periodic structure $\mathcal{R}$, teacher model $M_{\mathrm{PFP}}$
\State \quad In-plane supercell multipliers $(n_x, n_y)$
\State \quad Maximum Miller index $h_{\max}$
\State \quad Minimum slab thickness $d_{\mathrm{slab}}$, minimum vacuum thickness $d_{\mathrm{vac}}$
\State \quad (If \textit{md}) MD parameters $\mathit{args}_{\mathrm{MD}}$
\State \quad Flags: \textit{opt} (fixed-cell relaxation), \textit{md} (MD sampling)
\State \textbf{Output:}
\State \quad Set of surface slab structures $\mathcal{S}$
\vspace{4pt}
\State $\mathcal{S} \leftarrow \emptyset$
\State $\mathcal{H} \leftarrow$ symmetrically distinct Miller indices $(h, k, l)$ with $|h|, |k|, |l| \le h_{\max}$
\For{each $(h, k, l) \in \mathcal{H}$}
    \State Cleave slab $\mathcal{R}_{\mathrm{slab}}$ from $\mathcal{R}$ along $(h, k, l)$ with slab thickness $\ge d_{\mathrm{slab}}$ and vacuum thickness $\ge d_{\mathrm{vac}}$
    \State $\mathcal{R}' \leftarrow \mathcal{R}_{\mathrm{slab}} \times (n_x, n_y, 1)$ \Comment{expand only in in-plane directions}
    \State $\mathcal{S} \leftarrow \mathcal{S} \cup \{\mathcal{R}'\}$
    \If{\textit{opt}}
        \State $\mathcal{R}'' \leftarrow \textsc{RelaxFixedCell}(\mathcal{R}', M_{\mathrm{PFP}})$
        \State $\mathcal{S} \leftarrow \mathcal{S} \cup \{\mathcal{R}''\}$
    \EndIf
    \If{\textit{md}}
        \State $\mathcal{S} \leftarrow \mathcal{S} \cup \textsc{SampleMD}(\mathcal{R}', M_{\mathrm{PFP}},\, \mathit{args}_{\mathrm{MD}})$
    \EndIf
\EndFor
\State \Return $\mathcal{S}$
\end{algorithmic}
\end{algorithm}

\clearpage
\subsection{Substitution sampling}

Element substitution sampling generates chemically disordered structures by stochastically replacing atoms in the initial structure with user-specified species at defined probabilities. This method captures configurational variability in multicomponent systems, such as alloys, by sampling diverse local chemistries. Fixed-cell relaxation and MD can be applied after substitution to explore thermally accessible configurations, improving robustness and transferability across compositional variations. This method is useful for the solid-solution and high-entropy alloy. See Algorithm \ref{alg:sample-substitution}.

\begin{algorithm}[H]
\caption{\textsc{SubstituteAtoms}}
\label{alg:sample-substitution}
 \setstretch{0.85}
\begin{algorithmic}[1]
\State \textbf{Input:}
\State \quad Structure $\mathcal{R}$, teacher model $M_{\mathrm{PFP}}$
\State \quad Supercell multipliers $(n_x, n_y, n_z)$
\State \quad Candidate elements $\{e_1, \ldots, e_M\}$ with substitution probabilities $\{p_1, \ldots, p_M\}$, $\sum_{m=1}^{M} p_m \le 1$
\State \quad Number of samples $N$
\State \quad (If \textit{md}) MD parameters $\mathit{args}_{\mathrm{MD}}$
\State \quad Flags: \textit{opt} (cell-relaxed), \textit{md} (MD sampling)
\State \textbf{Output:}
\State \quad Set of substituted structures $\mathcal{S}$
\vspace{4pt}
\State $\mathcal{S} \leftarrow \emptyset$
\State $\mathcal{R}_0 \leftarrow \mathcal{R} \times (n_x, n_y, n_z)$ \Comment{expand to supercell}
\For{$k = 1, \ldots, N$}
    \State $\mathcal{R}' \leftarrow \mathcal{R}_0$
    \For{each atom $i \in \mathcal{R}'$}
        \State Draw $c_i$ from the categorical distribution over $\{e_1, \ldots, e_M, \varnothing\}$ with probabilities $\{p_1, \ldots, p_M, 1 - \sum_m p_m\}$
        \If{$c_i \neq \varnothing$}
            \State Replace the element of atom $i$ with $c_i$
        \EndIf
    \EndFor
    \State $\mathcal{S} \leftarrow \mathcal{S} \cup \{\mathcal{R}'\}$
    \If{\textit{opt}}
        \State $\mathcal{R}'' \leftarrow \textsc{RelaxWithCell}(\mathcal{R}', M_{\mathrm{PFP}})$
        \State $\mathcal{S} \leftarrow \mathcal{S} \cup \{\mathcal{R}''\}$
    \EndIf
    \If{\textit{md}}
        \State $\mathcal{S} \leftarrow \mathcal{S} \cup \textsc{SampleMD}(\mathcal{R}', M_{\mathrm{PFP}},\, \mathit{args}_{\mathrm{MD}})$
    \EndIf
\EndFor
\State \Return $\mathcal{S}$
\end{algorithmic}
\end{algorithm}

\clearpage

\section{LightPFP model and LAMMPS/Kokkos implementation}
\subsection{LightPFP model}
  \label{sec:supp_lightpfp_kokkos}

  The hyperparameters of the LightPFP models used in the results section are
  listed in Table~\ref{tab:lightpfp_hyperparameters}.

  \begin{table}[ht]
  \centering
  \begin{tabular}{l c c c c c c}
  \hline
  Material & Cutoff & levmax & $C_{\mu}$ & $C_{\nu}$ & $n_q$ & Neural-network readout \\
  \hline
  Ni$_3$Al & 6.0 & 8 & 1 & 1 & 16 & None \\
  Li$_6$PS$_5$Cl & 6.0 & 8 & 1 & 1 & 16 & [16, 16, 1] \\
  MgO & 6.0 & 8 & 1 & 1 & 16 & [16, 16, 1] \\
  SiO$_2$-HF & 6.0 & 8 & 1 & 1 & 16 & [16, 16, 1] \\
  \hline
  \end{tabular}
  \caption{Hyperparameters of the LightPFP models used in this work.}
  \label{tab:lightpfp_hyperparameters}
  \end{table}

  The complexity of a moment tensor potential is specified by the cutoff radius,
  the maximum level \texttt{levmax}, the level coefficients $C_{\mu}$ and
  $C_{\nu}$, and the number of radial basis functions $n_q$.  Once these
  hyperparameters are defined, each admissible basis function must have a level
  smaller than or equal to \texttt{levmax}.  As described in the main text, the
  basis function $B_{\alpha}$ is constructed from contractions of moment
  descriptors $M_{\mu,\nu}$:
  \begin{equation}
  M_{\mu,\nu}(\mathbf{n}_i)
  =
  \sum_{j \in \mathcal{N}(i)}
  f_{\mu}(\mathbf{r}_{ij})
  \underbrace{
  \mathbf{r}_{ij} \otimes \mathbf{r}_{ij} \otimes \cdots \otimes \mathbf{r}_{ij}
  }_{\nu \ \mathrm{times}},
  \end{equation}
  where $\mathcal{N}(i)$ is the neighbor list of atom $i$ and
  $\mathbf{r}_{ij} = \mathbf{r}_{i} - \mathbf{r}_{j}$.

  The level of each moment descriptor is defined as
  \begin{equation}
     \mathrm{lev}(M_{\mu,\nu})
     =
     2 + C_{\mu}\mu + C_{\nu}\nu .
  \end{equation}
  For the setting used here, \texttt{levmax}$=8$, $C_{\mu}=1$,
  $C_{\nu}=1$, and $n_q=16$, which yields 27 admissible scalar basis
  functions.  The basis definitions are listed in
  Table~\ref{tab:lightpfp_basis}.

  \begin{table}[ht]
  \centering
  \begin{tabular}{llc}
  \hline
  Basis index & Moment tensor component & Level \\
  \hline
  $B_{0}$  & $M_{0,0}$ & 2 \\
  $B_{1}$  & $M_{0,0} M_{0,0}$ & 4 \\
  $B_{2}$  & $M_{0,0} M_{0,0} M_{0,0}$ & 6 \\
  $B_{3}$  & $M_{0,0} M_{0,0} M_{0,0} M_{0,0}$ & 8 \\
  $B_{4}$  & $M_{1,0}$ & 3 \\
  $B_{5}$  & $M_{0,0} M_{1,0}$ & 5 \\
  $B_{6}$  & $M_{0,0} M_{0,0} M_{1,0}$ & 7 \\
  $B_{7}$  & $M_{1,0} M_{1,0}$ & 6 \\
  $B_{8}$  & $M_{0,0} M_{1,0} M_{1,0}$ & 8 \\
  $B_{9}$  & $M_{2,0}$ & 4 \\
  $B_{10}$ & $M_{0,0} M_{2,0}$ & 6 \\
  $B_{11}$ & $M_{0,0} M_{0,0} M_{2,0}$ & 8 \\
  $B_{12}$ & $M_{1,0} M_{2,0}$ & 7 \\
  $B_{13}$ & $M_{2,0} M_{2,0}$ & 8 \\
  $B_{14}$ & $M_{3,0}$ & 5 \\
  $B_{15}$ & $M_{0,0} M_{3,0}$ & 7 \\
  $B_{16}$ & $M_{1,0} M_{3,0}$ & 8 \\
  $B_{17}$ & $M_{4,0}$ & 6 \\
  $B_{18}$ & $M_{0,0} M_{4,0}$ & 8 \\
  $B_{19}$ & $M_{5,0}$ & 7 \\
  $B_{20}$ & $M_{6,0}$ & 8 \\
  $B_{21}$ & $M_{0,1} \cdot M_{0,1}$ & 6 \\
  $B_{22}$ & $M_{0,0} (M_{0,1} \cdot M_{0,1})$ & 8 \\
  $B_{23}$ & $M_{0,1} \cdot M_{1,1}$ & 7 \\
  $B_{24}$ & $M_{0,1} \cdot M_{2,1}$ & 8 \\
  $B_{25}$ & $M_{1,1} \cdot M_{1,1}$ & 8 \\
  $B_{26}$ & $M_{0,2} : M_{0,2}$ & 8 \\
  \hline
  \end{tabular}
  \caption{Definitions of the 27 scalar basis functions used when
  \texttt{levmax}$=8$, $C_{\mu}=1$, and $C_{\nu}=1$.  The symbols
  ``$\cdot$'' and ``$: $'' denote vector dot product and Frobenius inner product,
  respectively.}
  \label{tab:lightpfp_basis}
  \end{table}

  For models without a neural-network readout, the atomic energy is expressed as a
  linear combination of the basis functions,
  \begin{equation}
      E_i
      =
      \sum_{\alpha=0}^{26}
      \zeta_{\alpha} B_{\alpha}^{(i)}
      +
      \Delta E_{s_i},
  \end{equation}
  where $s_i$ is the species of atom $i$.  For models with a neural-network
  readout, the same 27-dimensional basis vector is passed to a small multilayer
  perceptron:
  \begin{equation}
      E_i
      =
      \mathrm{NN}\!\left(
      B_{0}^{(i)}, B_{1}^{(i)}, \ldots, B_{26}^{(i)}
      \right)
      +
      \Delta E_{s_i}.
  \end{equation}
  For the models denoted by [16, 16, 1] in
  Table~\ref{tab:lightpfp_hyperparameters}, this corresponds to a
  $27 \rightarrow 16 \rightarrow 16 \rightarrow 1$ readout network.

\subsection{Original ASE/PyTorch implementation}
  In the original LightPFP implementation, the model is evaluated through an
  ASE/PyTorch calculator.  At each force evaluation, atomic positions, cell
  vectors, and periodic boundary conditions are passed from ASE to the LightPFP
  estimator.  
  The estimator constructs or reuses a neighbor list, materializes the
  edge vectors $\mathbf{r}_{ij}$ as PyTorch tensors, evaluates the MTP basis and
  the readout network, and obtains forces by automatic differentiation:
  \begin{equation}
      \mathbf{F}_i
      =
      - \frac{\partial E}{\partial \mathbf{r}_i}.
  \end{equation}
  The edge-level force contributions are then accumulated onto atoms using tensor
  scatter operations, and the energy, forces, and virial are copied back to the
  ASE calculator.
  In addition, the LightPFP implementation used here evaluates the radial
  functions through a spline representation rather than the original polynomial
  radial basis form of the moment tensor potential.  For each species pair
  $(s_i,s_j)$ and radial channel $\mu$, the radial function is tabulated on a
  uniform grid with spacing $\Delta r$.  If $r_{ij}$ lies in the grid interval
  $[r_k,r_{k+1})$, the radial function is evaluated as a cubic polynomial,
  \begin{equation}
      f_{\mu}^{s_i s_j}(r_{ij})
      =
      c_{0,k}^{s_i s_j \mu}
      + c_{1,k}^{s_i s_j \mu} \delta r
      + c_{2,k}^{s_i s_j \mu} \delta r^2
      + c_{3,k}^{s_i s_j \mu} \delta r^3,
      \qquad
      \delta r = r_{ij} - r_k .
  \end{equation}
  Thus, instead of evaluating the radial basis expansion on the fly, the
  ASE/PyTorch implementation first determines the spline interval for each edge,
  gathers the corresponding coefficients
  $(c_0,c_1,c_2,c_3)$, and evaluates the cubic spline value as part of the
  PyTorch computation graph.  The spline evaluation remains differentiable with
  respect to the edge vector through $r_{ij}$, so the forces can still be obtained
  by automatic differentiation.  This spline form is the same mathematical
  parameterization used by the LAMMPS/Kokkos implementation, although the latter
  evaluates both the spline values and their derivatives explicitly in CUDA
  kernels.

\subsection{LAMMPS/Kokkos implementation}
  To accelerate this workflow, we implemented LightPFP as a custom
  LAMMPS/Kokkos pair style.  The PyTorch checkpoint is first exported into a
  potential file containing the species mapping, cutoff radius, spline
  coefficients, elemental energy shifts, and readout weights.  The LAMMPS pair
  style is specialized to the fixed model dimensions in
  Table~\ref{tab:lightpfp_hyperparameters}.  Thus, for the
  \texttt{levmax}$=8$ models, the implementation explicitly evaluates the 22
  required moment components, the 27 scalar basis functions, and, when present,
  the fixed $27 \rightarrow 16 \rightarrow 16 \rightarrow 1$ readout network.

  During molecular dynamics, LAMMPS provides the Kokkos device neighbor list
  directly to the pair style.  For each central atom, a Kokkos team traverses its
  neighbors, evaluates the spline radial functions, and accumulates the moment
  descriptors in team scratch memory.  The spline coefficients are packed in the
  order
  \[
      [\mathrm{species\ pair}]
      [\mathrm{grid\ index}]
      [\mu]
      [c_0,c_1,c_2,c_3],
  \]
  so that the coefficients required by each neighbor edge are contiguous in
  memory.  This avoids the advanced tensor indexing and large intermediate edge
  tensors used in the original PyTorch implementation.

  Forces in the LAMMPS/Kokkos implementation are computed by a hand-written
  reverse pass rather than by PyTorch automatic differentiation.  The pair style
  first evaluates
  \[
      \frac{\partial E_i}{\partial B_{\alpha}^{(i)}},
  \]
  propagates this derivative analytically through the basis contractions to obtain
  \[
      \frac{\partial E_i}{\partial M_{\mu,\nu}^{(i)}},
  \]
  and then revisits each neighbor edge to compute the derivative of the
  spline-MTP moment with respect to $\mathbf{r}_{ij}$.  The central-atom force is
  accumulated by team reduction, while the neighbor-side force is accumulated
  using device atomics.  This eliminates the runtime cost and memory overhead of
  constructing a general autograd graph.

  The main kernels executed in each timestep are: moment accumulation, atomic
  energy and adjoint evaluation, force evaluation, and, when requested, global
  virial reduction.  Global stress is computed directly inside the pair style from
  the accumulated virial tensor,
  \begin{equation}
      \boldsymbol{\sigma}
      =
      -\frac{\mathbf{W}}{V},
  \end{equation}
  where $\mathbf{W}$ is the global virial and $V$ is the simulation-cell volume.
  This direct stress path avoids the generic LAMMPS
  $\mathbf{f}\cdot\mathbf{r}$ virial reconstruction and avoids routing stress
  output through \texttt{compute pressure} and \texttt{fix ave/time}.

  The LAMMPS/Kokkos implementation changes only the execution path, not the model
  definition.  The same spline coefficients, readout weights, species mapping,
  and energy shifts are used as in the original PyTorch checkpoint.  Small
  numerical differences between the two implementations are expected because the
  Kokkos implementation uses single-precision GPU arithmetic and a different
  parallel reduction order.

\clearpage

\section{Details for high-entropy alloy example}

\begin{figure}[h]
\includegraphics{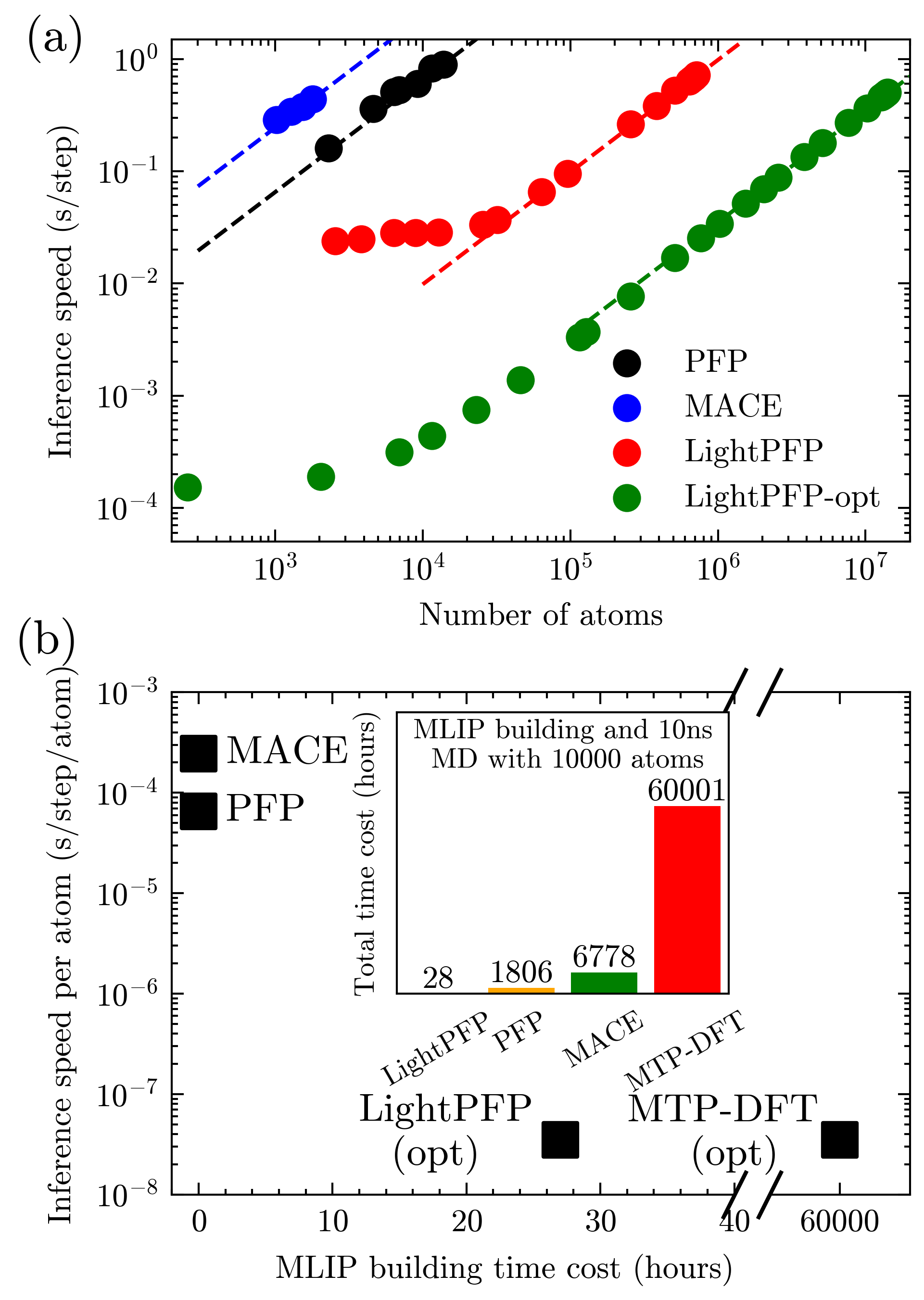}
\caption{(a) Molecular dynamics (MD) computational speed with \ce{AlCoCrFeNi} high-entropy alloy as a function of number of atoms for MLIPs: PFP, MACE, LightPFP and LightPFP-opt. (b) Trade-off between the overall time spent on MLIP building for \ce{AlCoCrFeNi} high-entropy alloy, including data collection and model training, and MD computational speed for PFP, LightPFP, MACE, and MTP. Inset: the total time cost to complete both MLIP building and a 10 ns MD simulation of a 10,000-atom system With PFP, LightPFP, MACE, and MTP.}
\label{fig:HEA_benchmark}
\end{figure}

\begin{figure}[h]
\includegraphics{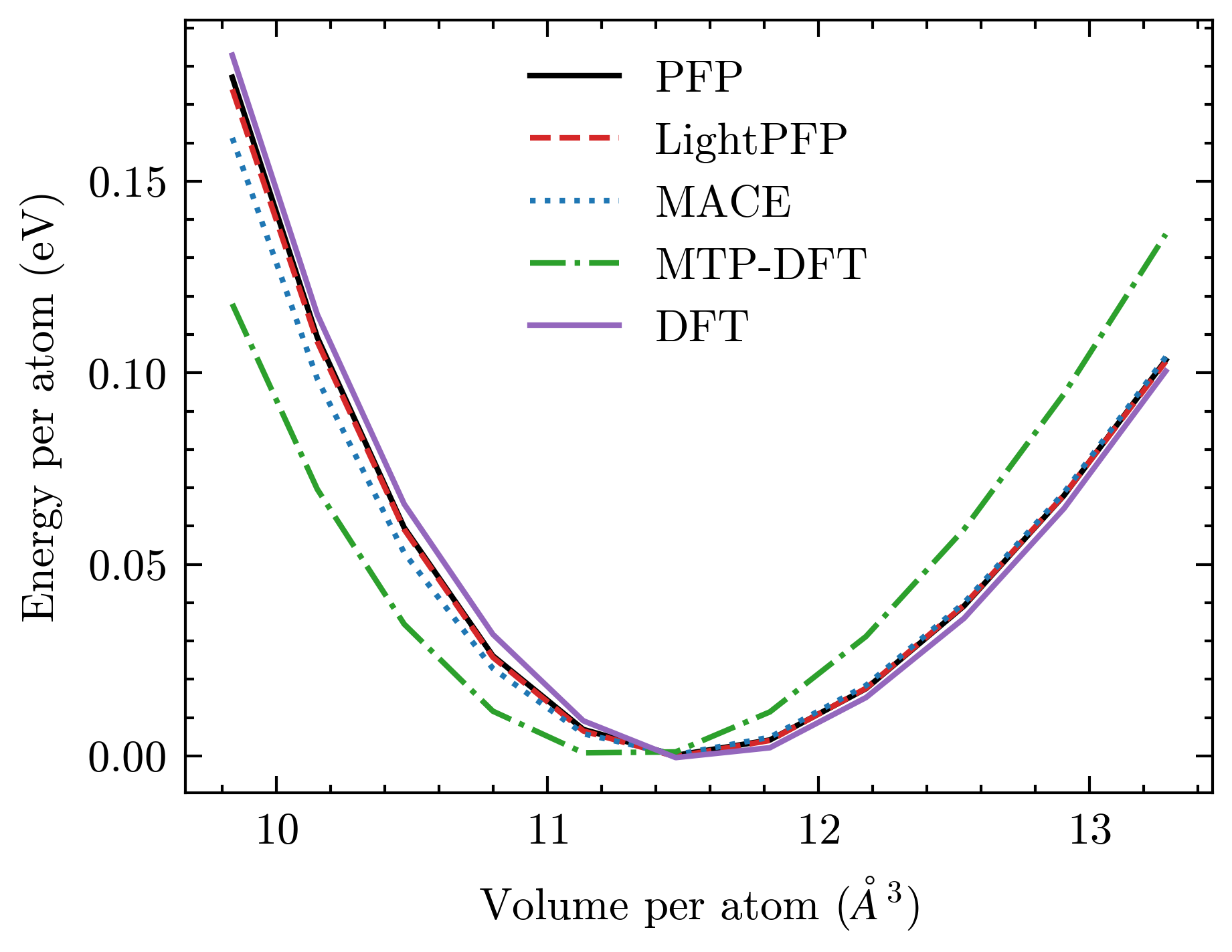}
\caption{Equation of states of \ce{AlCoCrFeNi} high-entropy alloy calculated by DFT, PFP, LightPFP, MACE and MTP}
\label{fig:HEA_eos}
\end{figure}

Detailed results for the high-entropy alloy (HEA) case study are presented here. Figure~\ref{fig:HEA_benchmark} compares the speed of LightPFP with that of other models. Figure~\ref{fig:HEA_eos} presents the equation of state (EOS) of the HEA as computed by different models. Table \ref{table:hea_dataset} summarizes the datasets used for the training of LightPFP and MTP-DFT models.

\begin{table}[h]
\centering
\caption{Composition of the \ce{AlCoCrFeNi} high-entropy alloy dataset.}
\begin{tabular}{llcc}
\hline
\textbf{Type of} & \textbf{Sampling} & \textbf{Number of} & \textbf{Number of} \\ 
\textbf{structure} & \textbf{method} & \textbf{structures} & \textbf{atoms} \\ 
\multicolumn{4}{c}{\textbf{LightPFP Dataset (labeled by PFP)}} \\ \hline
crystal & substitution+MD & 2040 & 206040 \\ 
boundary & substitution+MD & 6200 & 1083760 \\ 
slab & substitution+MD & 1398 & 66816 \\
\textbf{Total} & & \textbf{9638} & \textbf{1356616} \\ \hline
\multicolumn{4}{c}{\textbf{MTP Dataset (labeled by DFT)}} \\ \hline
crystal & substitution+MD & 531 & 42484 \\ 
boundary & substitution+MD & 286 & 9152 \\ 
slab & substitution+MD & 195 & 8724 \\ 
\textbf{Total} & & \textbf{1012} & \textbf{60360} \\ \hline
\end{tabular}
\label{table:hea_dataset}
\end{table}

\begin{figure}[h!]
\includegraphics[width=\columnwidth]{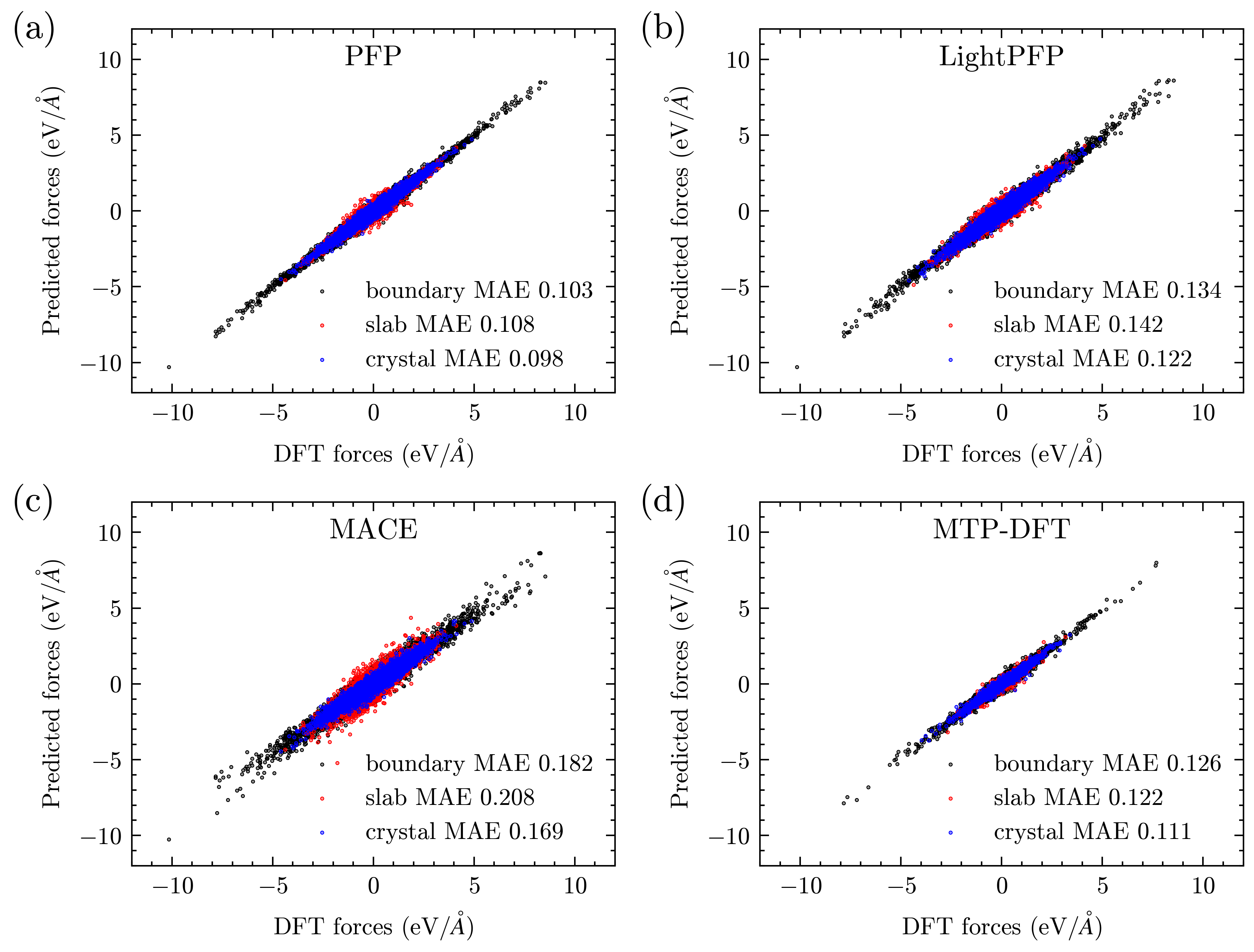}
\caption{Parity plot of DFT forces againes predicted forces by different MLIPs, (a) PFP; (b) LightPFP; (c) MACE and (d) MTP}
\label{fig:HEA_force_error}
\end{figure}

\clearpage

\section{Details for Transfer-Learned LightPFP for MgO}

To verify that the few-shot transfer learning improves the physical 
representation of MgO beyond simply shifting energy baselines, we 
computed radial distribution functions (RDFs) for both solid and 
liquid phases.

\begin{figure}[h!]
\includegraphics[width=\columnwidth]{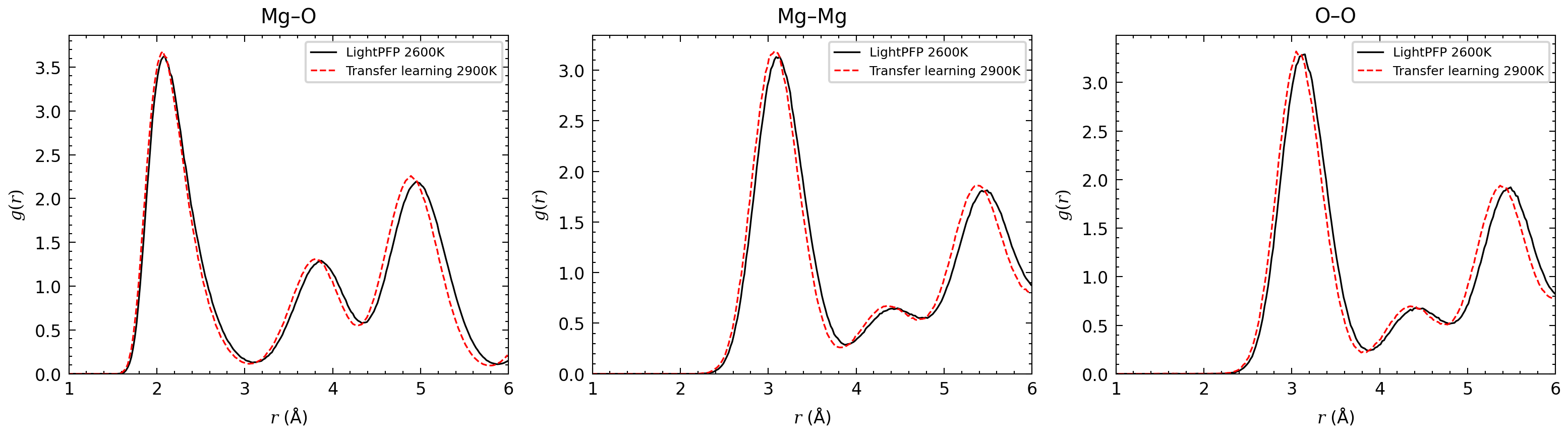}
\caption{Radial distribution functions of solid MgO comparing LightPFP before and after few-shot transfer learning. (a) Mg-O RDF, (b) Mg-Mg RDF, (c) O-O RDF. Black dashed lines: LightPFP before transfer learning (PBE-based); Red bash lines: LightPFP after transfer learning with 10 r²SCAN structures. Both models reproduce the characteristic features of crystalline rocksalt structure: sharp first Mg-O peak at ~2.10 Å corresponding to octahedral coordination. The near-identical RDFs indicate that transfer learning preserves the structural description while correcting energetics. Configurations sampled from MD trajectory at 2600 K for LightPFP and 2900 K for transfered LightPFP since they both below the melting point.}
\label{fig:solid_MgO_rdf}
\end{figure}

\begin{figure}[h!]
\includegraphics[width=\columnwidth]{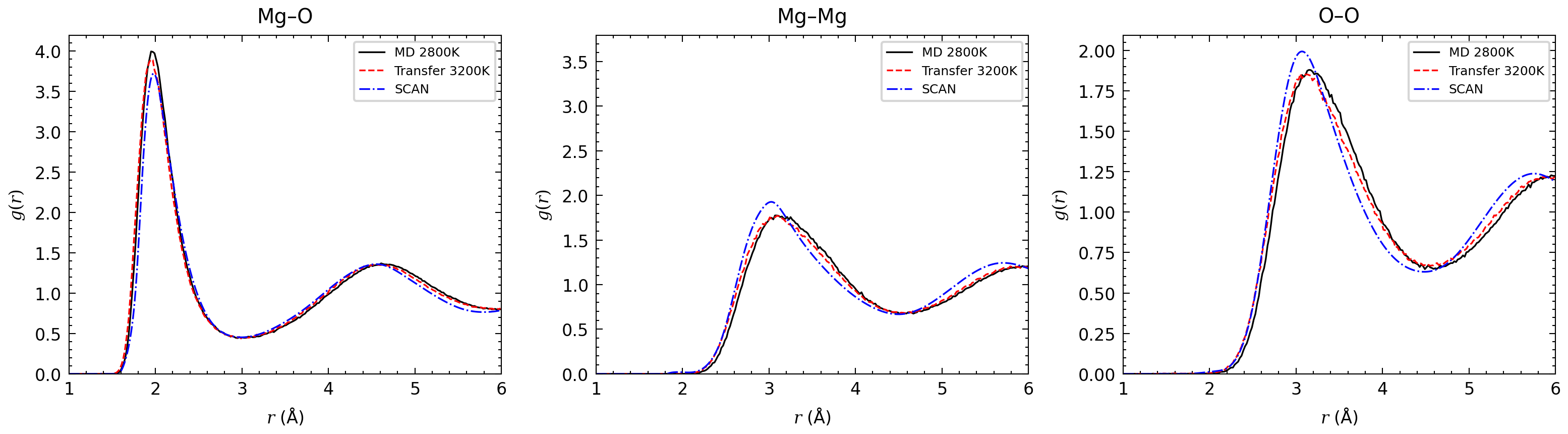}
\caption{Radial distribution functions of liquid MgO 
validating transfer-learned LightPFP against high-level theory. 
(a) Mg-O RDF, (b) Mg-Mg RDF, (c) O-O RDF. Black dashed lines: LightPFP before transfer learning (PBE-based); Red bash lines: LightPFP after transfer learning with 10 r²SCAN structures; blue dot dash line: SCAN MD data from Ref \cite{lee2022ab}. 
The liquid-phase RDFs show characteristic features of molten MgO: 
broadened first Mg-O peak at ~2.1 Å with reduced height compared to 
solid phase, indicating loss of strict octahedral order. 
The transfer-learned model shows good agreement with SCAN 
reference calculations in peak positions, heights, 
and widths. Configurations sampled from MD trajectory at 2800 K for LightPFP and 3200 K for transfered LightPFP since they both above the melting point.}
\label{fig:liquid_MgO_rdf}
\end{figure}

\clearpage

\section{3-stages training method}

We propose a three-stage training strategy for the LightPFP model. In Stage I, the optimization focuses on fitting forces; in Stage II, the emphasis shifts to energy and stress; and in Stage III, the loss terms are balanced so that the energy, force, and stress losses are of comparable magnitudes. This is achieved by progressively adjusting the coefficients of the energy ($\alpha$), force ($\beta$) and stress ($\gamma$) terms in the loss function.

\begin{figure}[h]
\includegraphics[width=\columnwidth]{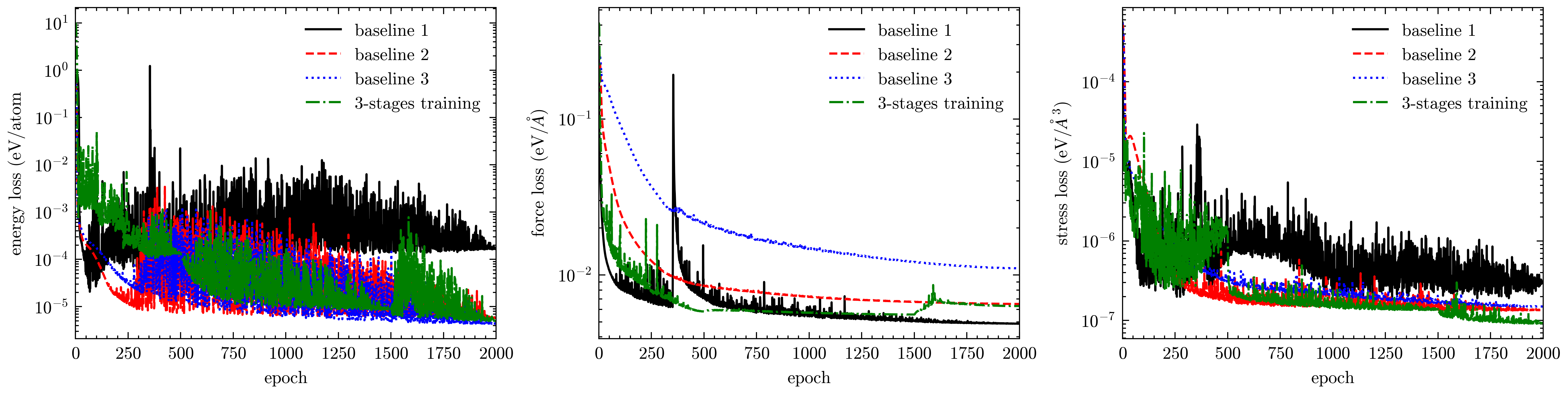}
\caption{Learning curves for different training strategies.}
\label{fig:3_stg_loss}
\end{figure}

To assess the effectiveness of this strategy, we trained on the \ce{Li6PS5Cl} dataset and conducted three fixed-weight baselines. Each baseline uses constant loss weights equal to those employed in one stage of the three-stage schedule: Baseline 1 ($\alpha=10^{-5}, \beta=10.0, \gamma=10^{-5}$), Baseline 2 ($\alpha=1.0, \beta=0.1, \gamma=10.0$), and Baseline 3 ($\alpha=26.2, \beta=0.034, \gamma=1383.1$). The evolution of the losses over epochs is shown in Figure \ref{fig:3_stg_loss}, and the final losses are summarized in Table \ref{tab:3_stg_loss}.

\begin{table}[ht]
\centering
\begin{tabular}{lcccc}
\hline
 & baseline 1 & baseline2 & baseline3 & 3-stages-training \\
\hline
energy & $1.67\times10^{-4}$ & $5.76\times10^{-6}$ & $4.39\times10^{-6}$ & $4.96\times10^{-6}$ \\
forces & 0.00486 & 0.00651 & 0.0110 & 0.00631 \\
stress & $3.03\times10^{-7}$ & $1.35\times10^{-7}$ & $1.50\times10^{-7}$ & $9.28\times10^{-8}$ \\
\hline
\end{tabular}
\caption{Comparison of the final energy, force, and stress losses across training strategies.}
\label{tab:3_stg_loss}
\end{table}

Baseline 1 fails to fit the energy accurately, yielding the largest energy loss, while Baseline 3 fails to fit forces accurately. Baseline 2 offers a more balanced trade-off; however, its energy, force, and stress losses are all larger than those achieved by the three-stage training. Overall, the proposed three-stage schedule provides the best balance across the three targets, with near-optimal energy and stress losses and competitive force accuracy.

\clearpage

\section{Active learning method}

Active learning is a powerful approach for developing accurate and efficient interatomic potentials in molecular dynamics simulations.  Here is a brief introduction of the active learning workflow we used for the "Dry etching of SiO$_2$" example and several other examples in the Supplementary Materials \Romannum{2} (see Fig.~\ref{fig:active_learning_workflow}):

1. Initial Dataset: A simple initial dataset is necessary for active learning. The initial dataset does not need to be large and robust.

2. Model Training: The initial LightPFP model is trained with this initial dataset.

3. Exploration: The LightPFP model is then used to drive molecular dynamics (MD) simulations, exploring new configurations and areas of the potential energy surface.

4. Quality Check: At certain MD steps, check the accuracy of LightPFP. Calculate the energy, forces, and stress of the MD snapshot using PFP, and compare these with the LightPFP predictions.

5. Data Selection: Include the MD snapshot in the training dataset if the discrepancy between PFP and LightPFP is greater than the minimum threshold and less than the maximum threshold.

6. Model Update: After several MD simulations are finished or cease based on other criteria, update the LightPFP model with the dataset collected in current and previous iterations and the initial dataset.

7. Iteration: Steps 3 to 6 are repeated iteratively. Each iteration improves the potential’s accuracy and extends its applicability.

This active learning approach allows for the efficient development of accurate potentials by focusing computational resources on the most informative data points, ultimately resulting in a potential that can reliably reproduce the behavior of the target system across a wide range of conditions.

\begin{figure}[]
\includegraphics[width=0.5\columnwidth]{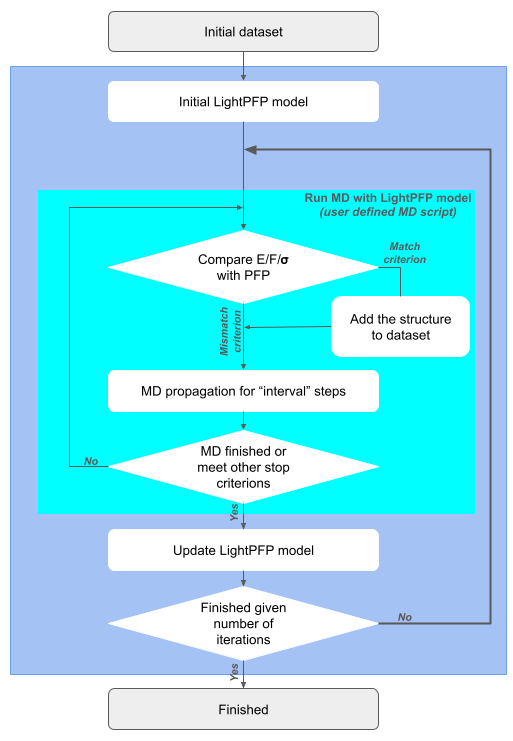}
\caption{Illustration of active learning workflow. Reproduced from the Matlantis manual (accessed 12 Nov 2025).}
\label{fig:active_learning_workflow}
\end{figure}

\paragraph{Sampling threshold}
We uses minimum/maximum thresholds to collect high-quality training data. For more efficient training data collection, it also performs early stopping of MD simulations or PFP-based sampling upon detecting outliers. Both PFP and LightPFP are used to calculate the energy, forces, and stress of given MD snapshots. The errors of LightPFP w.r.t PFP are used to determine if certain MD snapshots should be collected. The valid error range is defined by both a minimum (lower bound) and a maximum threshold (upper bound). When the error of an MD snapshot is very large, exceeding the upper bound, it indicates an unreasonable structure that is not beneficial for training. Continuing MD from this point can lead to more structures that are not valuable. In such cases, the MD simulation will stop early. In the "Dry etching of SiO$_2$" example, error checking is performed every 100 steps, and the selection criterion is: energy error in between 5.0 and 40.0 times of energy MAE of current using LightPFP model; force error (largest atomic error in the structure) is in between 1.5 and 50 eV/A. 
Note that the thresholds used here can be viewed as practical heuristics, and the optimal values may vary across applications.

\paragraph{MD early stop}
As mentioned, the MD simulation will stop early when the discrepancy between PFP and LightPFP is very large. This indicates that the MD has reached a configuration where the current LightPFP model is unreliable. While structures with huge errors compared to PFP are not useful for training, the structures leading up to such structures, typically several MD steps before, are critical. Learning from these preceding structures helps prevent the MD from evolving into unphysical configurations. Our workflow provides two mechanisms for this: (1) Back-tracking Method: When MD stops due to large prediction errors, the algorithm checks previous MD steps using a binary search until a training structure with the error in the specified region is found. MD snapshots are cached to facilitate this process.
(2) PFP-based fallback: When MD stops early, the simulation rolls back to the previous checkpoint and continues using PFP instead of the LightPFP model for several more additional samples. In the "Dry etching of SiO$_2$" example, we collect 5 additional training structures based on PFP when MD failed.

\begin{figure}[]
\includegraphics[width=\columnwidth]{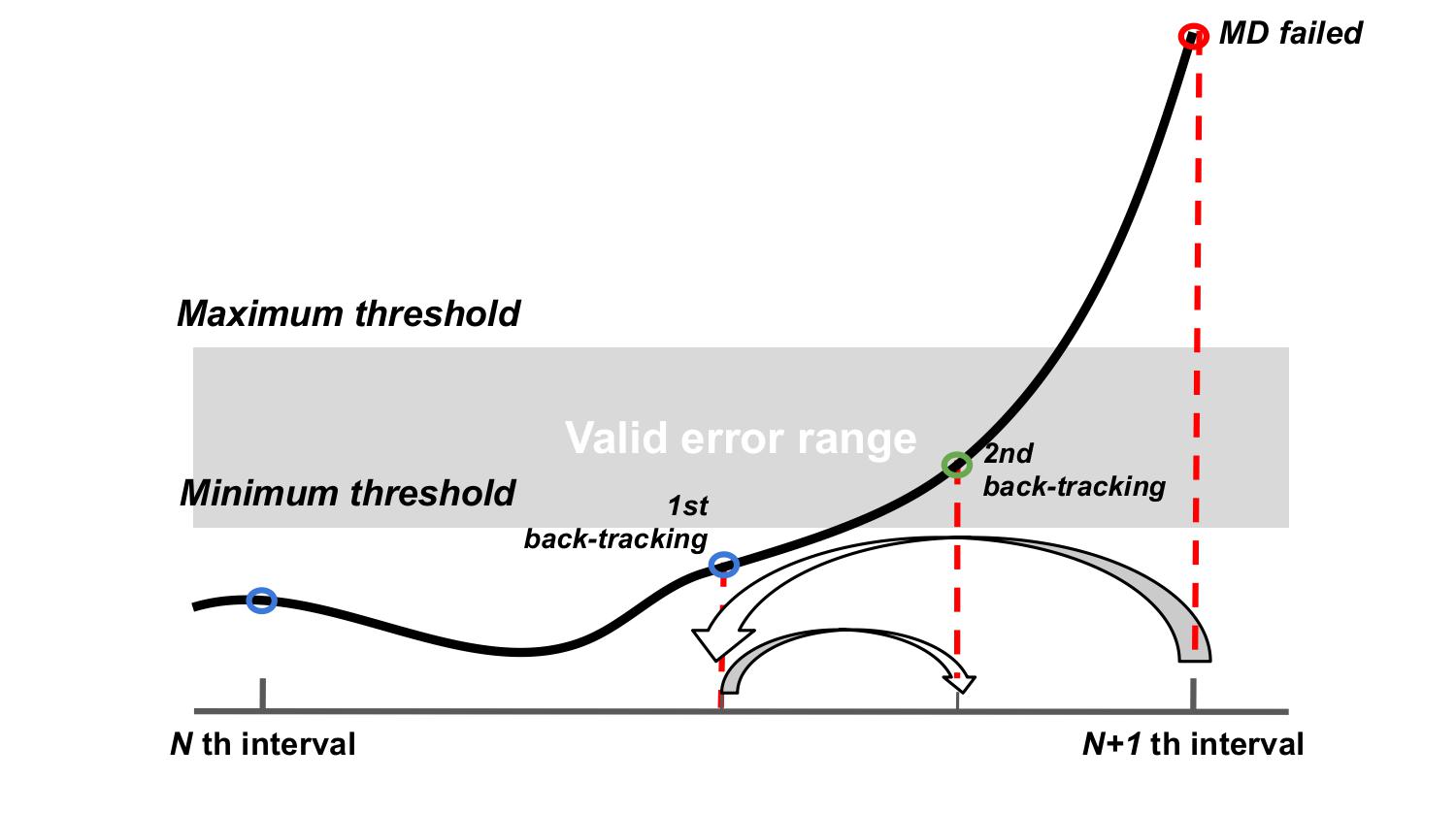}
\caption{Back-tracking mechanism of active learning when MD failed. Reproduced from the Matlantis manual (accessed 12 Nov 2025).}
\label{fig:active_learning_backtrack}
\end{figure}

\paragraph{Model update}

The model is updated in each iteration with the latest dataset and all previous datasets.
To accelerate active learning, the total number of epochs for model training is adjusted according to the size of whole datasets, and the training time is kept roughly constant. This mechanism is designed to handle the gradually increasing dataset during the active learning iterations. In the "Dry etching of SiO$_2$" example, we fixed the time cost for each model update to 0.5 hour. 

\clearpage

\section{Dataset of PFP}
The PFP model is a universal machine-learning interatomic potential pretrained on multiple large-scale DFT datasets that were deliberately constructed to cover broad chemical and structural spaces, rather than only relaxed ground-state structures. In the recent PFP v8 description, the training data include the PFP-R2SCAN dataset, which contains approximately $3 \times 10^6$ r2SCAN-calculated structures covering 70 elements and multiple domains including molecules, bulk crystals, slabs/surfaces, and disordered structures; the previously developed PFP-PBE/+U dataset, which contains approximately $6 \times 10^7$ PBE/PBE+U structures covering 96 elements and spans molecules, bulk crystals, slabs, disordered systems, clusters, and adsorption configurations; the PFP-WB97XD molecular dataset with approximately $6 \times 10^6$
 structures; and the OC20 adsorption dataset with approximately $10^8$ surface-adsorbate structures. The original PFP crystal dataset was designed to include diverse periodic configurations such as stable and hypothetical bulk crystals, non-optimized and metastable structures, cell-volume and cell-shape perturbations, randomly displaced atomic configurations, clean slabs, adsorbate-covered surfaces, finite clusters, molecular structures in periodic cells, high-temperature molecular dynamics snapshots, and disordered multi-element structures. The training labels include DFT total energies and atomic forces, with Bader charges used as auxiliary information where available. Since these datasets were generated using different DFT reference levels, PFP employs a calculation-mode label to learn multiple potential-energy surfaces within a single model, avoiding direct mixing of inconsistent energy references while retaining cross-domain transferability. This combination of broad elemental coverage, diverse material classes, equilibrium and far-from-equilibrium structures, and high-level r2SCAN reference data provides the basis for the strong zero-shot reliability of PFP in downstream tasks involving crystals, surfaces, adsorption systems, molecular environments, diffusion pathways, and finite-temperature dynamics.

\clearpage

% \section{DFT calculation conditions}

% Spin-polarized density functional theory (DFT) calculations for the \ce{Li6PS5Cl} and high-entropy alloy were performed using the Perdew–Burke–Ernzerhof (PBE) exchange–correlation functional, as implemented in the Vienna Ab initio Simulation Package (VASP, version 6.4.0) with GPU acceleration. The projector augmented-wave (PAW) method and a plane-wave basis set were employed with a kinetic energy cutoff of 520 eV. The choice of pseudopotentials follows the PFP dataset\cite{takamoto2022towards}. The k-point meshes were generated according to the lattice parameters, corresponding to a k-point density of 1000 k-points per reciprocal atom.

% MgO was calculated using the r$^2$SCAN meta-GGA functional, implemented in the same Vienna Ab initio Simulation Package (VASP, version 6.4.0). The r$^2$SCAN calculations were enabled in VASP by activating the Meta-GGA option within the input configuration. The kinetic energy cutoff was again set to 520 eV.  The Brillouin zone was sampled using a k-point grid generated with a KSPACING parameter of $0.5\,\text{\AA}^{-1}$., ensuring adequate convergence of total energies.

\bibliography{aipsamp}% Produces the bibliography via BibTeX.

% \section*{Supplementary Movies and Data}
% Movie S1: Description and playback speed.
% Dataset S1: Description, variables, and file format.

\title{Supplementary Information \Romannum{2} for: LightPFP: A Lightweight Route to Ab Initio Accuracy at Scale}

\author{Wenwen Li}
 \email{wenwenli@preferred.jp}
\affiliation{Preferred Networks Inc., Tokyo, Japan.}

\author{Nontawat Charoenphakdee}
 \email{nontawat@preferred.jp}
\affiliation{Preferred Networks Inc., Tokyo, Japan.}

\author{Yong-Bin Zhuang}
\affiliation{Preferred Networks Inc., Tokyo, Japan.}

\author{Ryuhei Okuno}
\affiliation{Preferred Networks Inc., Tokyo, Japan.}

\author{Yuta Tsuboi}
\affiliation{Preferred Networks Inc., Tokyo, Japan.}

\author{So Takamoto}
\affiliation{Preferred Networks Inc., Tokyo, Japan.}

\author{Junichi Ishida}
\affiliation{Matlantis Corporation, Tokyo, Japan.}

\author{Iori Kurata}
\affiliation{Preferred Networks Inc., Tokyo, Japan.}

\author{Ju Li}
\affiliation{Department of Materials Science and Engineering, Massachusetts Institute of Technology, Cambridge, MA 02139, USA}
\affiliation{Department of Nuclear Science and Engineering, Massachusetts Institute of Technology, Cambridge, MA, USA
}

\date{\today}

\maketitle

\tableofcontents % Generates the table of contents

\newpage

\section{Application 1: Simulation of interfacial structures of Pt (111)/benzene}

The main purpose of this example is to demonstrate how to build and validate the LightPFP model to simulate the solid-liquid interface using the Pt/benzene interface as a model system. By leveraging LightPFP's enhanced speed and accuracy, researchers can gain insights into solid-liquid interface behavior and expand its applications in materials science, chemistry, and nanotechnology. 

\subsection{Student fine-tuning}
\label{app1:fine-tuning}
To train the LightPFP model, we compiled a diverse set of structures with their energies, forces, and stresses labeled using the PFP. The training dataset encompassed various components: bulk Pt, Pt (111) slab, bulk benzene, and Pt (111)/benzene interface structures.

\begin{table}[ht]
\centering
\begin{tabular}{l p{0.6\textwidth} c c}
\hline
System & Methods & Num of & Num of \\
& & structures & Atoms \\
\hline
Pt & Cell compression/Stretching/Deforming \newline
Atom Displacement \newline
Vacancy \newline
Surface and MD (NPT 500--1500 K) & \multirow{4}{*}{2249} & \multirow{4}{*}{111,290} \\
Pt (111) & MD (NPT 500--1000 K) & 640 & 43,680 \\
Benzene & Cell compression/Stretching \newline
MD (NPT 300--400 K; NVT 500--1500 K) & \multirow{2}{*}{984} & \multirow{2}{*}{177,120} \\
Interface & MD (NPT 300--800 K; NVT 500--1750 K) & 5400 & 2,632,800 \\
\hline
\end{tabular}
\end{table}

The dataset was randomly split into training and testing datasets, with 90\% comprising the training dataset and the remaining 10\% constituting the testing dataset.

\begin{figure}
\includegraphics[width=\columnwidth]{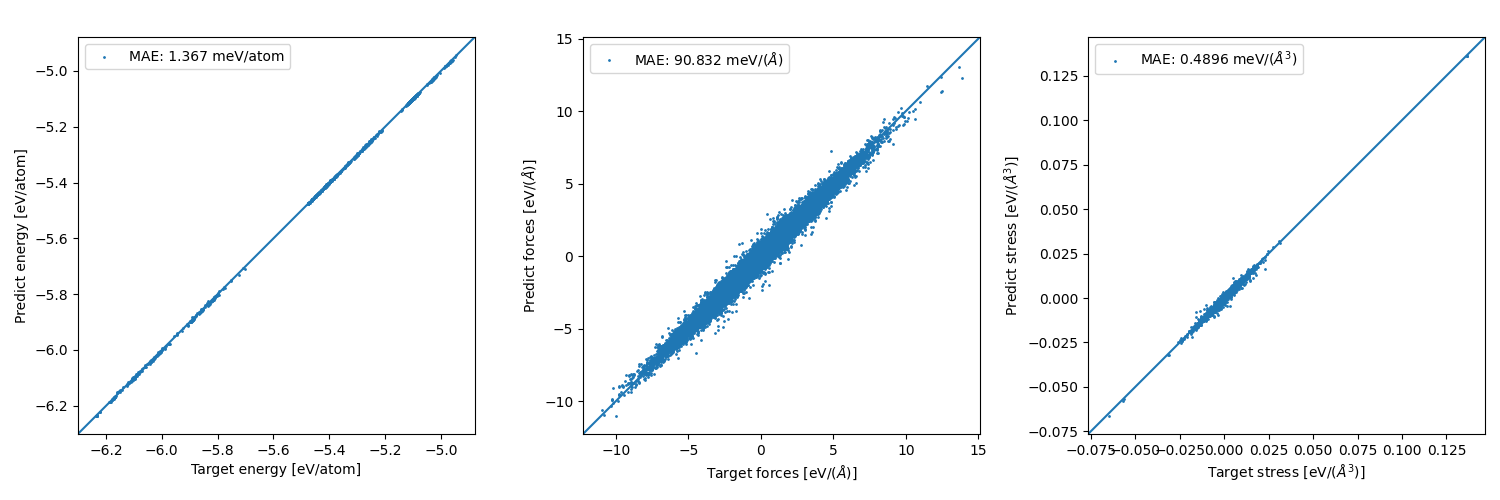}
\caption{Parity plots for energy, forces, and stress on the test set: PFP ground‑truth values (x‑axis) vs LightPFP model predictions (y‑axis).}
\label{fig:appendix_1_model_accuracy}
\end{figure}

\subsection{Evaluation using PFP}
\label{app1:eval-by-pfp}
To validate the performance of the LightPFP models, we performed MD simulations on a small Pt (111)/benzene interface system and compared the trajectories with PFP.

The initial simulation box has dimensions of $38.58 \,\text{\AA} \times 33.10 \,\text{\AA} \times 101.32 \,\text{\AA}$, with cell angles of $90^\circ$, $90^\circ$, and $120^\circ$. It consists of a total of 9,072 atoms, including 1,512 Pt atoms and 630 benzene molecules. We have chosen a relatively smaller structure in order to compare the results with PFP.

The initial structure undergoes a 20 ps equilibrium at a temperature of 300.0 K at first, using the NVT ensemble. Subsequently, NPT MD simulations are conducted for 100 ps at 300.0 K and 1 bar, utilizing the NPT ensemble. MD snapshots are saved for future analysis.

\begin{figure}
\includegraphics[width=\columnwidth]{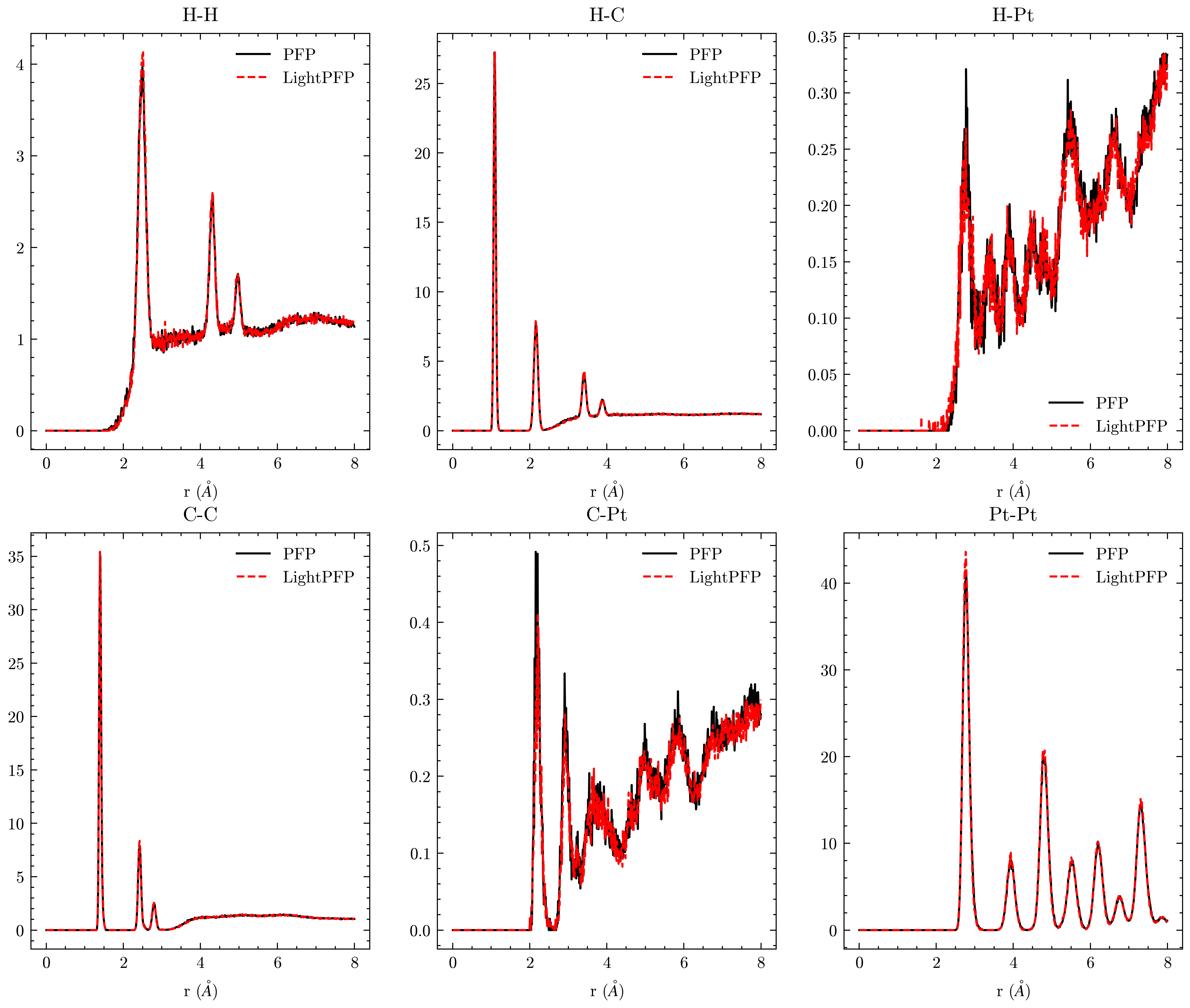}
\caption{Comparison of radial distribution functions from LightPFP and PFP trajectories}
\label{fig:appendix_1_rdf}
\end{figure}

The density estimated from LightPFP NPT MD (5.08 g/cm³) agrees well with that from PFP (5.06 g/cm³).
To calculate the radial distribution function, we used the snapshots of the MD trajectory taken after 50 ps. 
The resulting radial distribution function is shown in Fig ~\ref{fig:appendix_1_rdf}. 
We discovered that the results obtained from both the LightPFP and PFP trajectories exhibit a high degree of agreement.

\begin{figure}[htbp]
\centering
\includegraphics[width=0.5\textwidth]{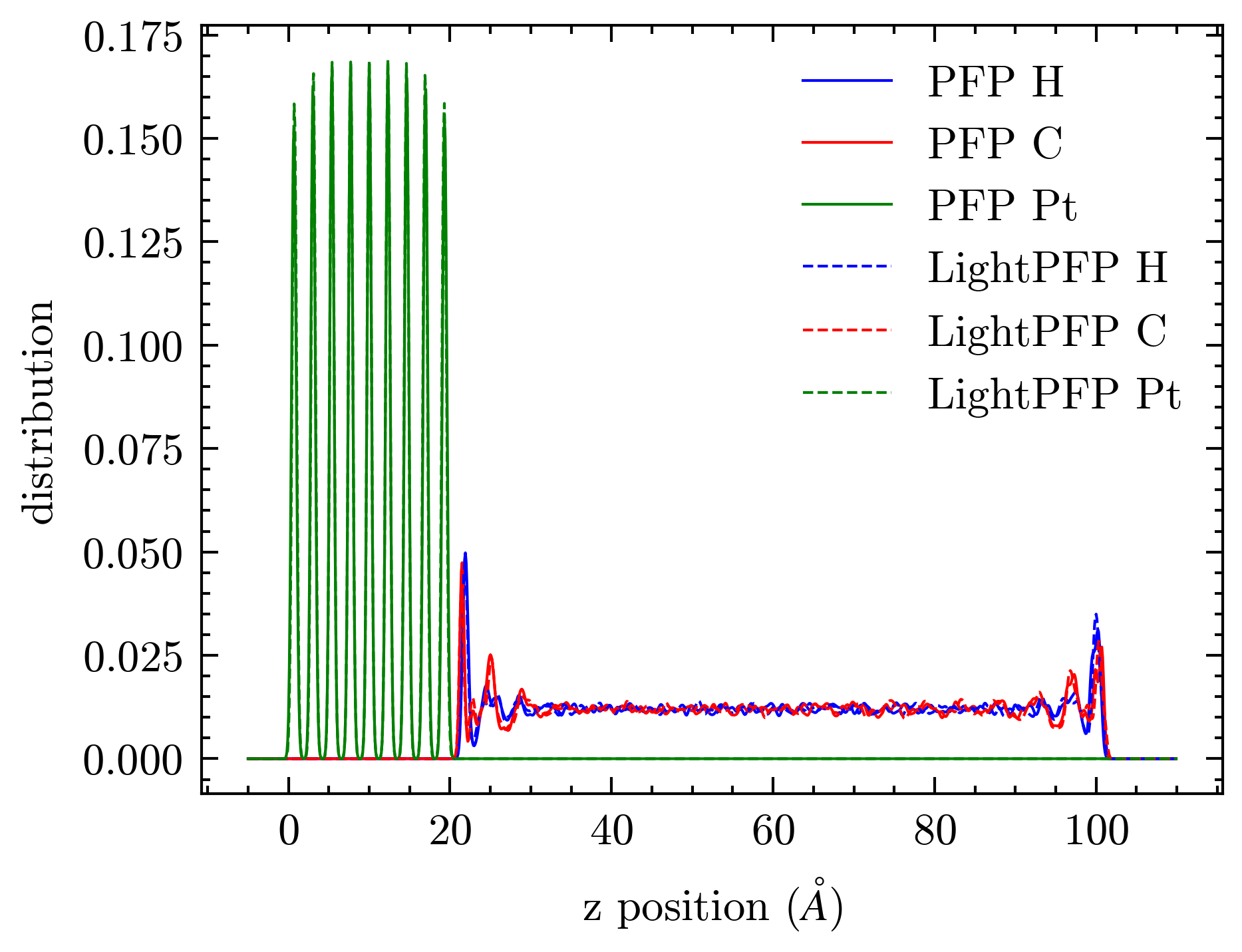}%
\hfill
\includegraphics[width=0.5\textwidth]{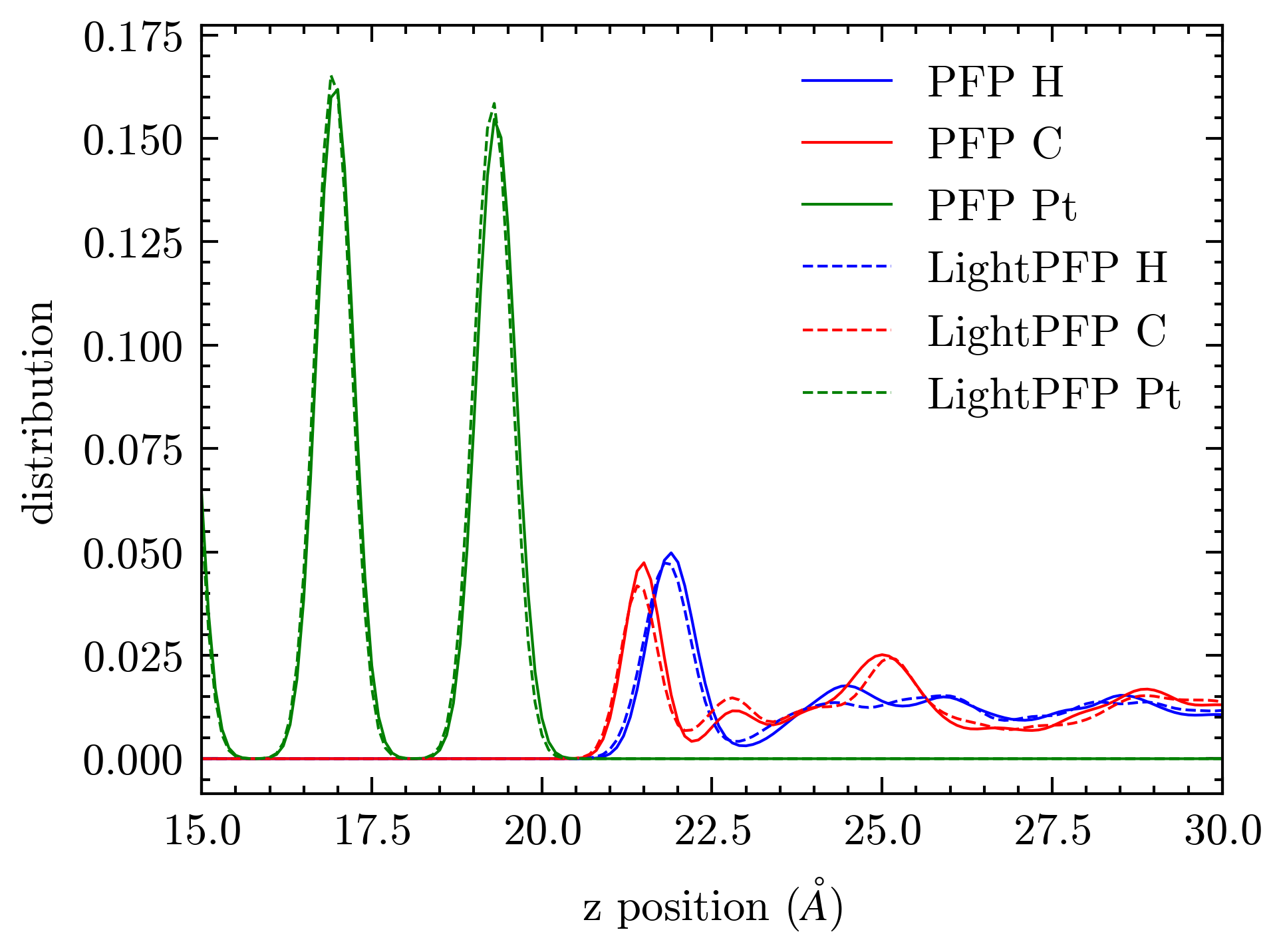}
\caption{Spatial distribution of Pt, C and H atoms along the z direction. (left) Whole simulation box. (right) Zoomed-in view at the interface.}
\label{fig:appendix_1_distribution}
\end{figure}

To characterize the atomic distribution along the z axis (normal to the interface), we sampled MD trajectory snapshots after 50 ps and generated a z-position density via Gaussian broadening (width 0.25 \text{\AA}). The resulting distribution is plotted in Fig.~\ref{fig:appendix_1_distribution}.

\begin{figure}[htbp]
\centering
\includegraphics[width=0.32\textwidth]{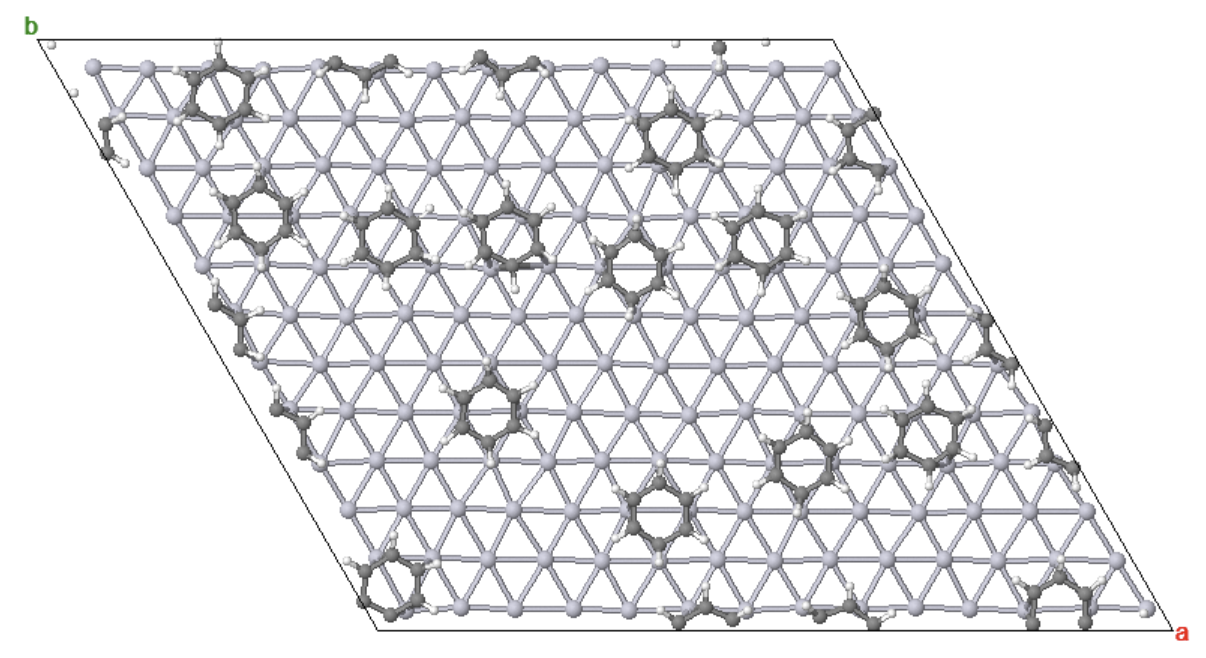}\hfill
\includegraphics[width=0.32\textwidth]{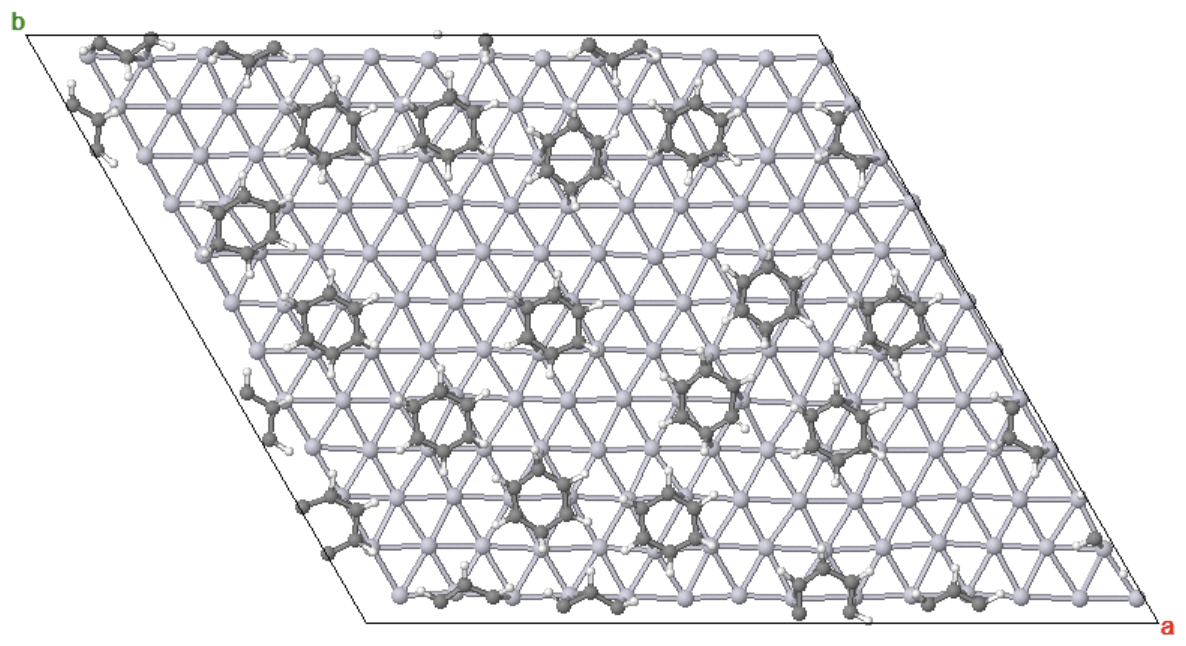}\hfill
\includegraphics[width=0.32\textwidth]{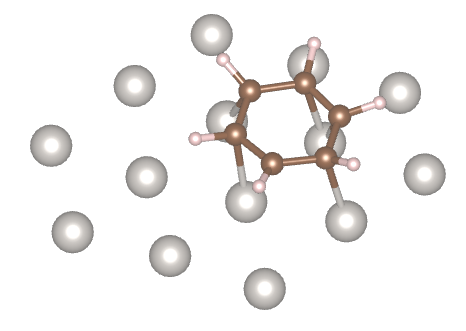}
\caption{The benzene adsorption on Pt surface (left) LightPFP (middle) PFP (right) The bri30 adsorption site of the benzene molecule.}
\label{fig:appendix_1_adsorption}
\end{figure}

The density profile in Fig~\ref{fig:appendix_1_distribution} shows several peaks in the H and C elements near the Pt surface (around 20 \text{\AA}), indicating that the benzene structure is significantly different from the uniform liquid phase due to adsorption with the Pt surface. As we move further away from the Pt surface, the interfacial liquid structure gradually transitions towards a uniform liquid phase. According to the figure, the thickness of the interfacial layer is approximately 15 \text{\AA}.

The position and intensity of the peaks in the LightPFP model were compared with the PFP result, and they matched each other across most regions. This alignment demonstrates that the LightPFP model effectively captures and represents the structural characteristics of the solid/liquid interface.

Symmetric peaks can be observed around $z=100\,\text{\AA}$, indicating the presence of the same liquid-solid interface due to the periodic boundary condition. However, these peaks exhibit slight fluctuations in shape due to noise originating from cell shape changes in the NPT MD simulation.

In Fig ~\ref{fig:appendix_1_adsorption}, we can observe the benzene molecules that have adsorbed onto the Pt surface. Specifically, in the LightPFP MD simulation, 18 benzene molecules were found to be adsorbed onto the Pt surface within the specified area. In the PFP MD simulation, on the other hand, there were a total of 21 benzene molecules observed to be adsorbed onto the Pt surface within the same area. In conclusion, the LightPFP MD simulation reproduced the surface coverage rate of benzene on the Pt surface well.

Fig.~\ref{fig:appendix_1_adsorption}(right) illustrates the adsorption structure of a single benzene molecule. The bri30 conformation, in which the center of the benzene molecule is located on the bridge site of the Pt surface, was found to be the most energetically stable in first-principles calculations\cite{morin2004chemisorption}. 
Interestingly, we observed that almost all the molecules were adsorbed in the bri30 site in both the LightPFP and PFP simulations. This result is consistent with the findings of the first-principle calculations.

\clearpage

\section{Application 2: Miscibility of water, benzene and heptane}

In this example, we use LightPFP to investigate the miscibility of binary liquid mixtures among water, benzene and heptane via large-scale molecular dynamics simulations using LightPFP. 

\subsection{Student fine-tuning}
\label{app2:fine-tuning}
We trained LightPFP on a collection of datasets designed to cover both homogeneous and demixed liquid environments. The initial dataset comprised nine classes of configurations: (1) bulk water, (2) bulk benzene, (3) bulk heptane, (4) homogeneous water/benzene mixtures, (5) homogeneous water/heptane mixtures, (6) homogeneous heptane/benzene mixtures, and (7) explicit liquid–liquid interfaces for water/benzene, (8) water/heptane, and (9) heptane/benzene. To efficiently obtain interfacial training data, we constructed liquid–liquid interface geometries and sampled them by short MD, rather than relying on spontaneous demixing from homogeneous starting states. The latter would require prohibitively long trajectories to capture phase-separated configurations because phase separation proceeds via slow nucleation, coarsening, and domain growth. By seeding and sampling interfacial structures, the training set explicitly exposed the model to the distinct local environments present at liquid–liquid boundaries.

Starting from this initial model, we performed active learning to further improve accuracy and robustness. For each of the nine system types, we ran 20 ps MD simulations over 280-350 K, monitored model performance, and selectively augmented the training set with configurations with high disagreement w.r.t. PFP. The resulting LightPFP model was then used for the large-scale MD simulation.

\subsection{Large-scale MD simulation}
\label{app1:large-scale-md}
Three systems are considered: (1) 17,280 water molecules with 4,320 benzene molecules; (2) 17,280 water molecules with 2,160 heptane molecules; and (3) 4,320 benzene molecules with 2,160 heptane molecules. A 1 ns molecular dynamics simulation was performed for each system. According to experimental data, water and heptane, water and benzene are immiscible at room temperature, while heptane and benzene are miscible \cite{IUPACNISTSolubilityDB} For the water/benzene and water/heptane mixtures, spontaneous liquid–liquid phase separation is clearly observed in the MD trajectories, with the two immiscible components forming distinct phases. In contrast, no phase separation is observed for the benzene/heptane mixture, as these liquids are mutually miscible. The MD simulation results are all consistent with experimental observations.

\begin{figure}
\includegraphics[width=0.48\textwidth]{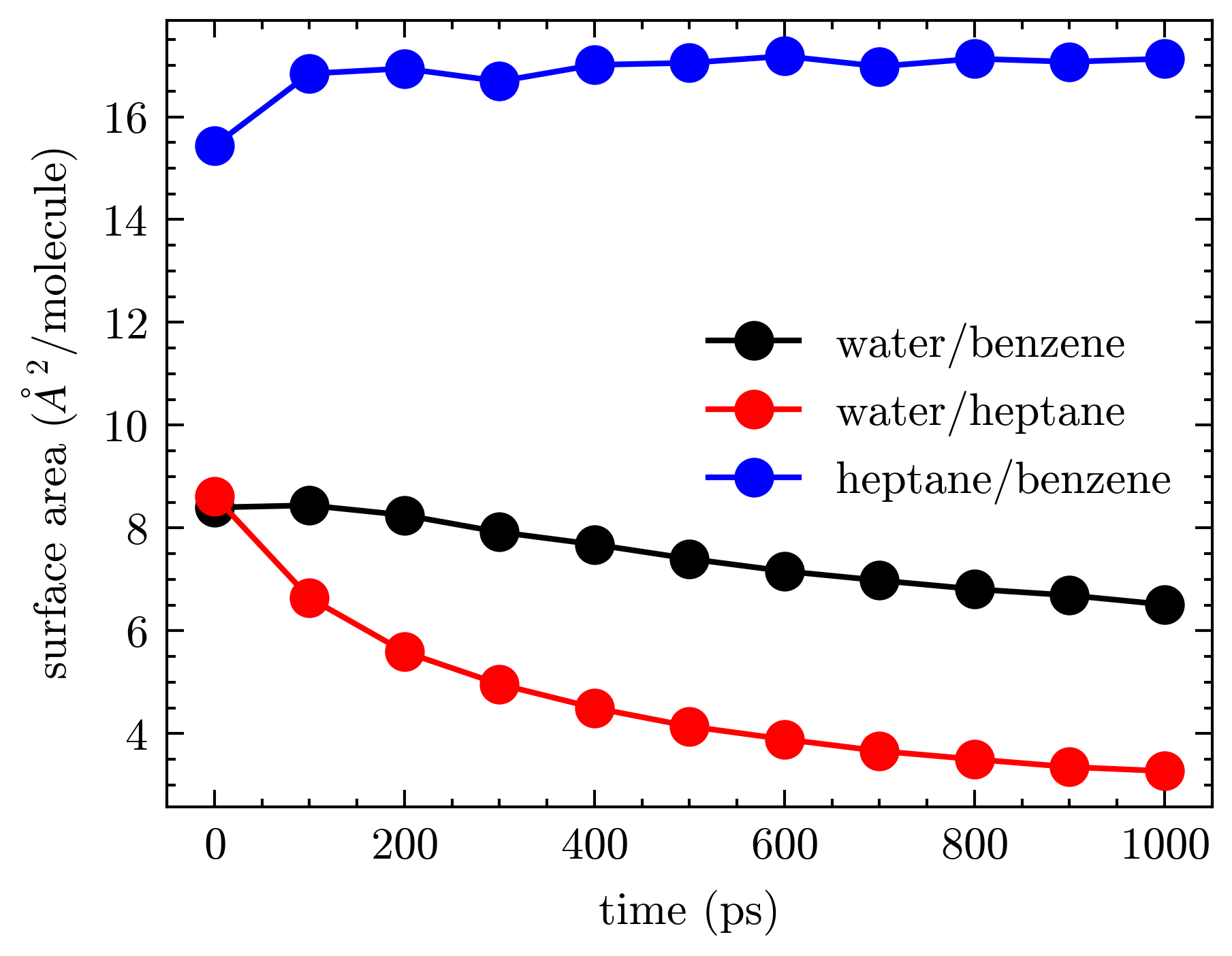}
\caption{Surface area of liquid-liquid interface.}
\label{fig:appendix_2_surface_area}
\end{figure}

\begin{figure}
\includegraphics[width=0.8\textwidth]{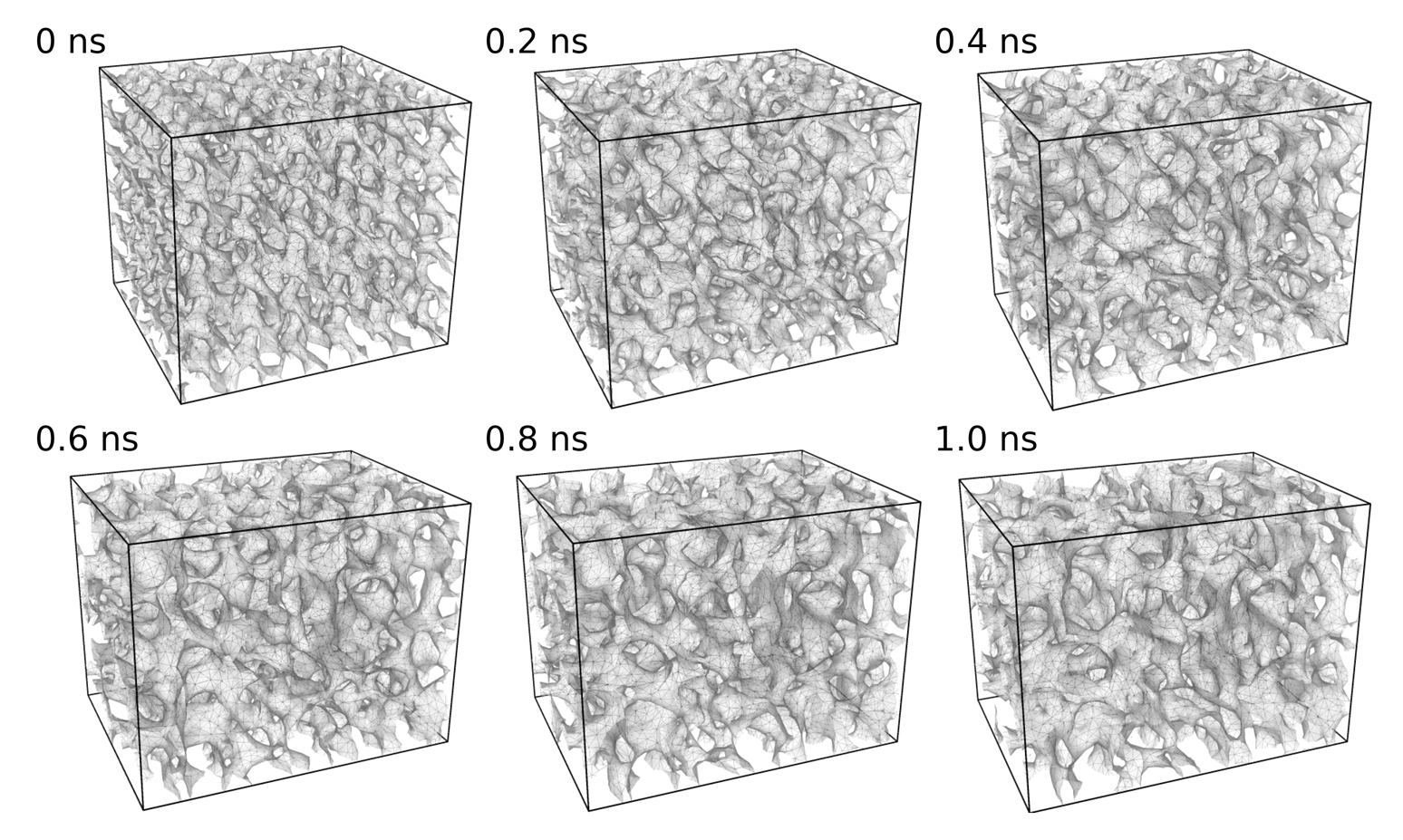}
\caption{Liquid-liquid interface between water and benzene}
\label{fig:appendix_2_water_benzene}
\end{figure}

\begin{figure}
\includegraphics[width=0.8\textwidth]{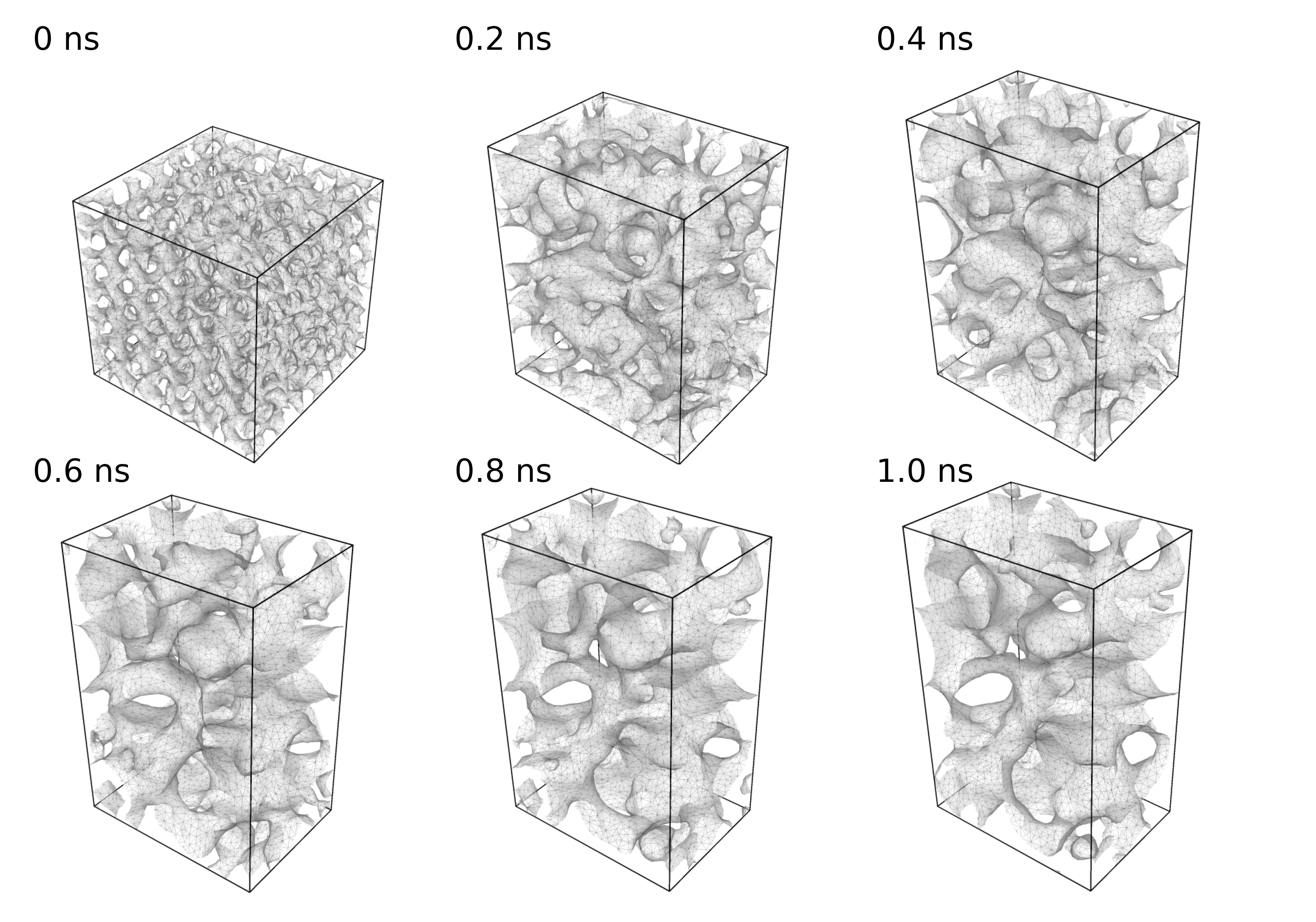}
\caption{Liquid-liquid interface between water and heptane}
\label{fig:appendix_2_water_heptane}
\end{figure}

\begin{figure}
\includegraphics[width=0.8\textwidth]{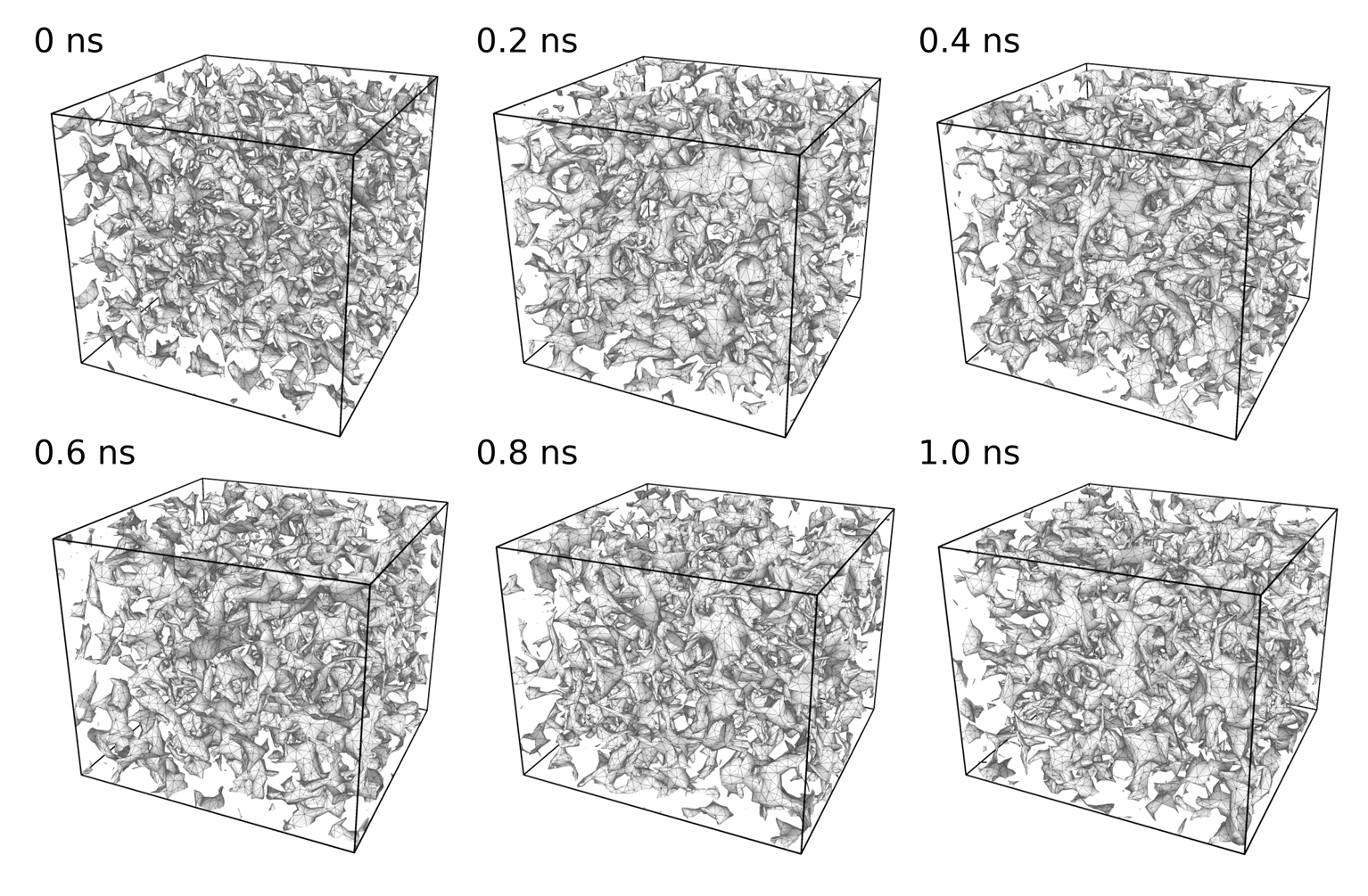}
\caption{Liquid-liquid interface between heptane and benzene}
\label{fig:appendix_2_heptane_benzene}
\end{figure}

To quantify demixing, we analyzed the MD snapshots and computed the liquid–liquid interfacial area using OVITO’s construct surface mesh modifier with the alpha-shape method~\cite{stukowski2014computational}. For each saved frame, the two species were identified, a triangulated interface was generated, and the total interfacial area was recorded. The time evolution of this area is plotted in Fig.~\ref{fig:appendix_2_surface_area}. For the water/benzene and water/heptane mixtures, the interfacial area decreases rapidly and then approaches a low, nearly steady value, indicating fast coarsening and macroscopic phase separation. In contrast, for the benzene/heptane mixture, the interfacial area remains essentially unchanged from the outset over the entire 1 ns window, consistent with the absence of demixing. In addition, the experimental data show that the solubility of heptane in water is lower than that of benzene. In MD simulations, we also found that the interface area decreases more rapidly in the heptane-water system, reflecting faster nucleation and phase transition dynamics\cite{pubchem_heptane_solubility,IUPACNISTSolubilityDB}.
Figures~\ref{fig:appendix_2_water_benzene},~\ref{fig:appendix_2_water_heptane}, and~\ref{fig:appendix_2_heptane_benzene} show representative interface morphologies at different times, providing a visual corroboration of these trends: the liquid–liquid interface in water/benzene and water/heptane smooths and recedes over time, whereas no well-defined interface emerges in benzene/heptane.

\clearpage

\section{Application 3: Interface thermal resistance between {Ni} and DPO/BP}

In this example, we use LightPFP to quantify interfacial thermal transport between a Ni (111) surface and the diphenyl oxide (DPO)/biphenyl (BP) eutectic heat‑transfer fluid, a widely used medium in parabolic trough concentrated solar power systems with a maximum operating temperature near $400^\circ\text{C}$. Our target property is the thermal conductivity across the metal–fluid interface, which governs heat exchange efficiency in receiver tubes and heat exchangers\cite{carrillo2024probing}.

\subsection{Student fine-tuning}
\label{app3:fine-tuning}
We trained LightPFP on a dataset tailored to capture both bulk and interfacial environments relevant to the Ni–DPO/BP system. The initial configurations encompassed four classes: bulk Ni, a Ni (111) slab exposing the surface, bulk liquid DPO/BP, and explicit Ni (111) and DPO/BP interfaces. Each class was sampled via short molecular dynamics and light “rattle” perturbations to diversify local environments. MD sampling covered NVT simulations at 500 K, 1000 K, and 1500 K and NPT simulations at 300 K, 400 K, and 500 K. The high‑temperature NVT trajectories were chosen to generate randomized, higher‑energy configurations that improve the robustness and stability of the trained model; in contrast, the NPT sampling was restricted to lower temperatures to avoid unphysical density fluctuations and cell distortions that can arise at very high temperatures under barostat control. For bulk liquid DPO/BP and for Ni (111)-DPO/BP interfaces, we also manually created multiple distinct atomic configurations prior to short MD “structure sampling.” This manual seeding was used to enrich the distribution of molecular orientations in the training set, because orientational rearrangements in the liquid and at the metal–liquid interface relax slowly in MD, especially in the 300–500 K range.

Starting from this base model, we performed active learning to improve accuracy and robustness over the temperature range of interest. For each configuration class, we ran short MD trajectories spanning 300–500 K under complementary conditions: 5 ps NVT, 10 ps NPT, and 30 ps reverse non-equilibrium MD (rNEMD) to explicitly sample heat-flow states. We also varied setup details (for example, MD temperature, slab and liquid layer thickness and the rNEMD swapping frequency, etc.) to enrich coverage of interfacial motifs. Configurations exhibiting high model disagreement were iteratively labeled and added to the training set, and the query–retrain loop was repeated until validation metrics stabilized.

\subsection{Evaluation using PFP}
\label{app3:eval-by-pfp}

\begin{figure}
\includegraphics[width=0.48\textwidth]{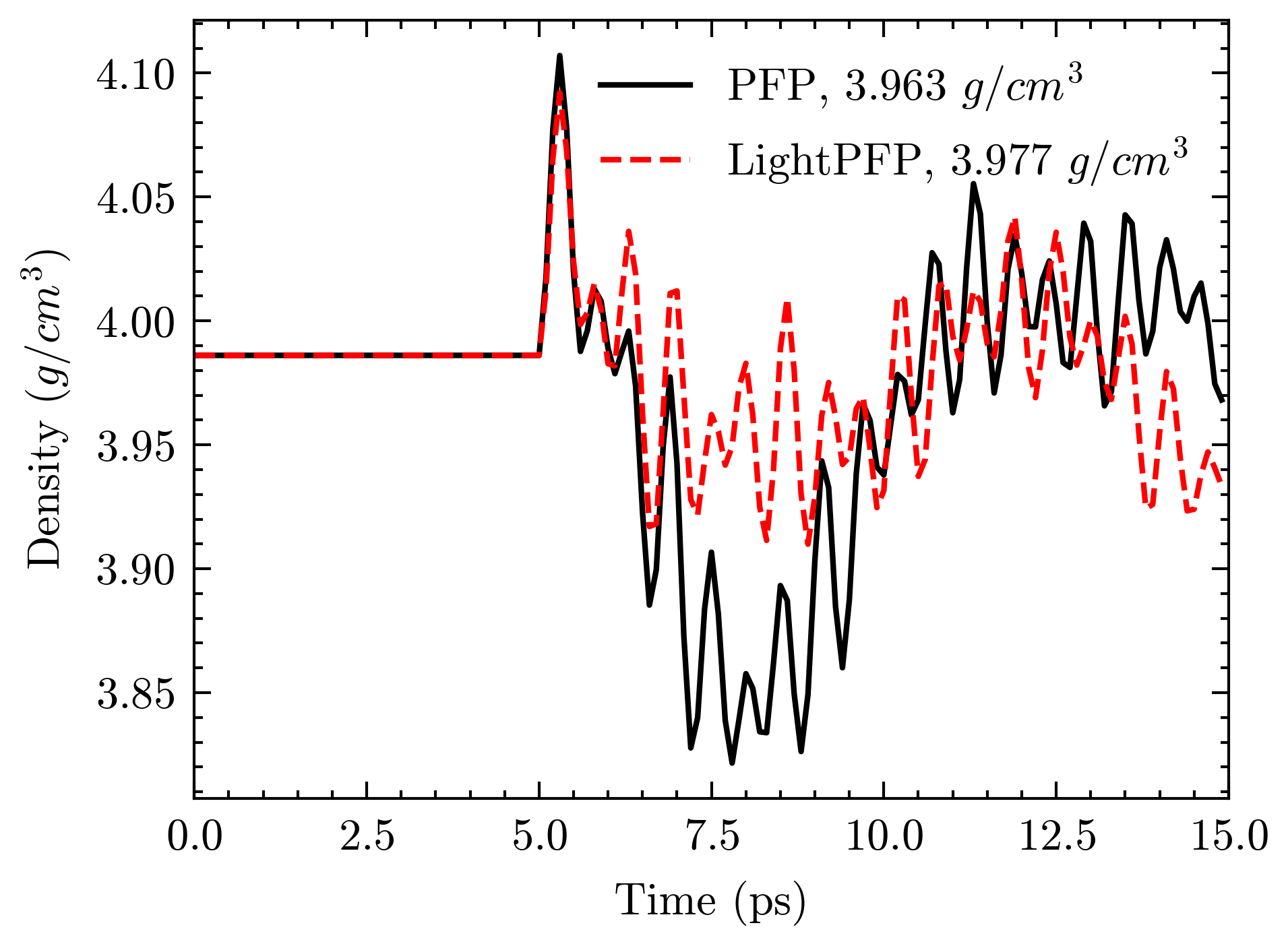}
\caption{Density of Ni and DPO/BP interface structure at 300 K}
\label{fig:appendix_3_density}
\end{figure}

\begin{figure}
\includegraphics[width=0.8\textwidth]{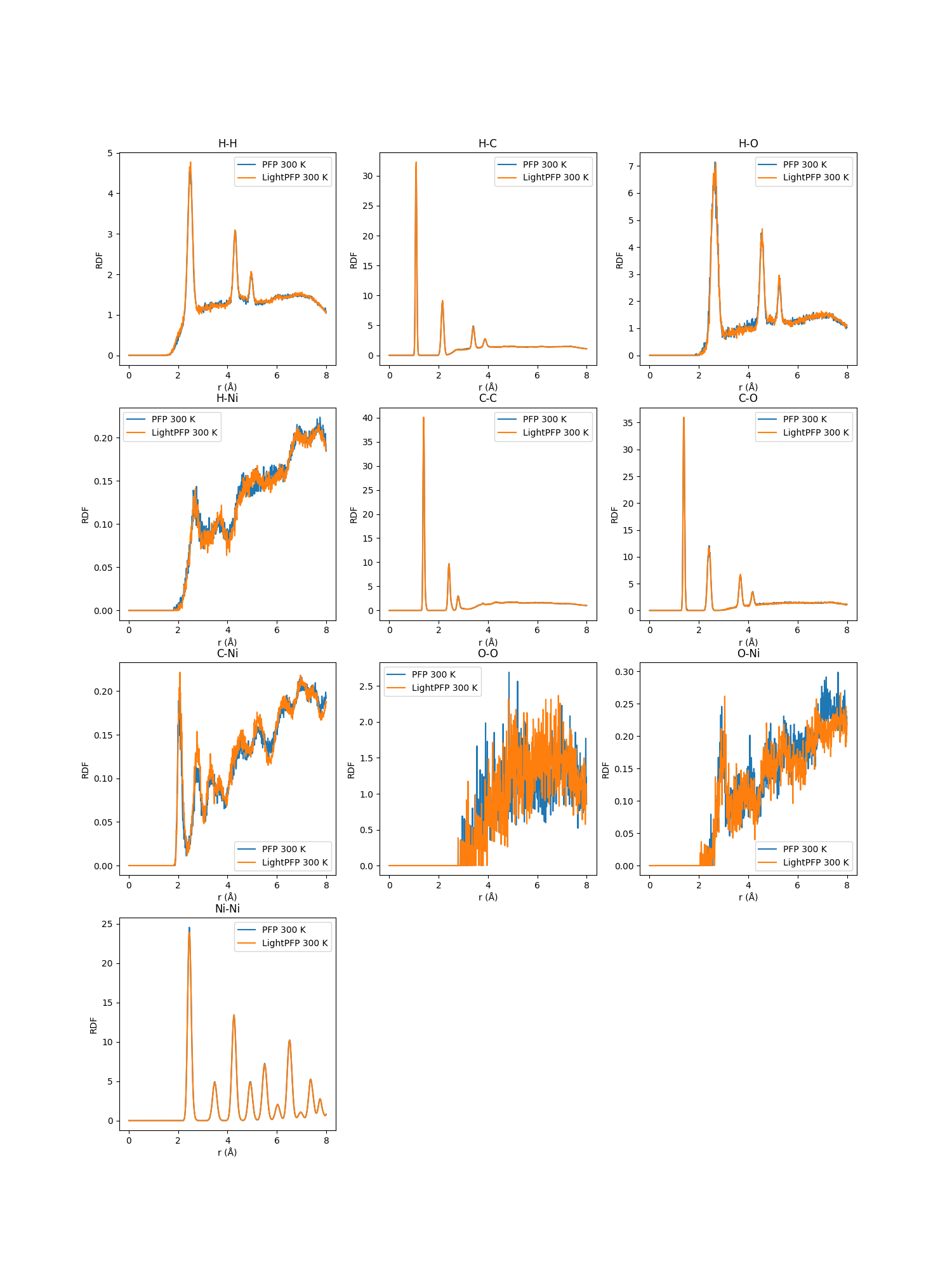}
\caption{Radial distribution function of Ni and DPO/BP interface structure at 300 K}
\label{fig:appendix_3_rdf}
\end{figure}

\begin{figure}
\includegraphics[width=\textwidth]{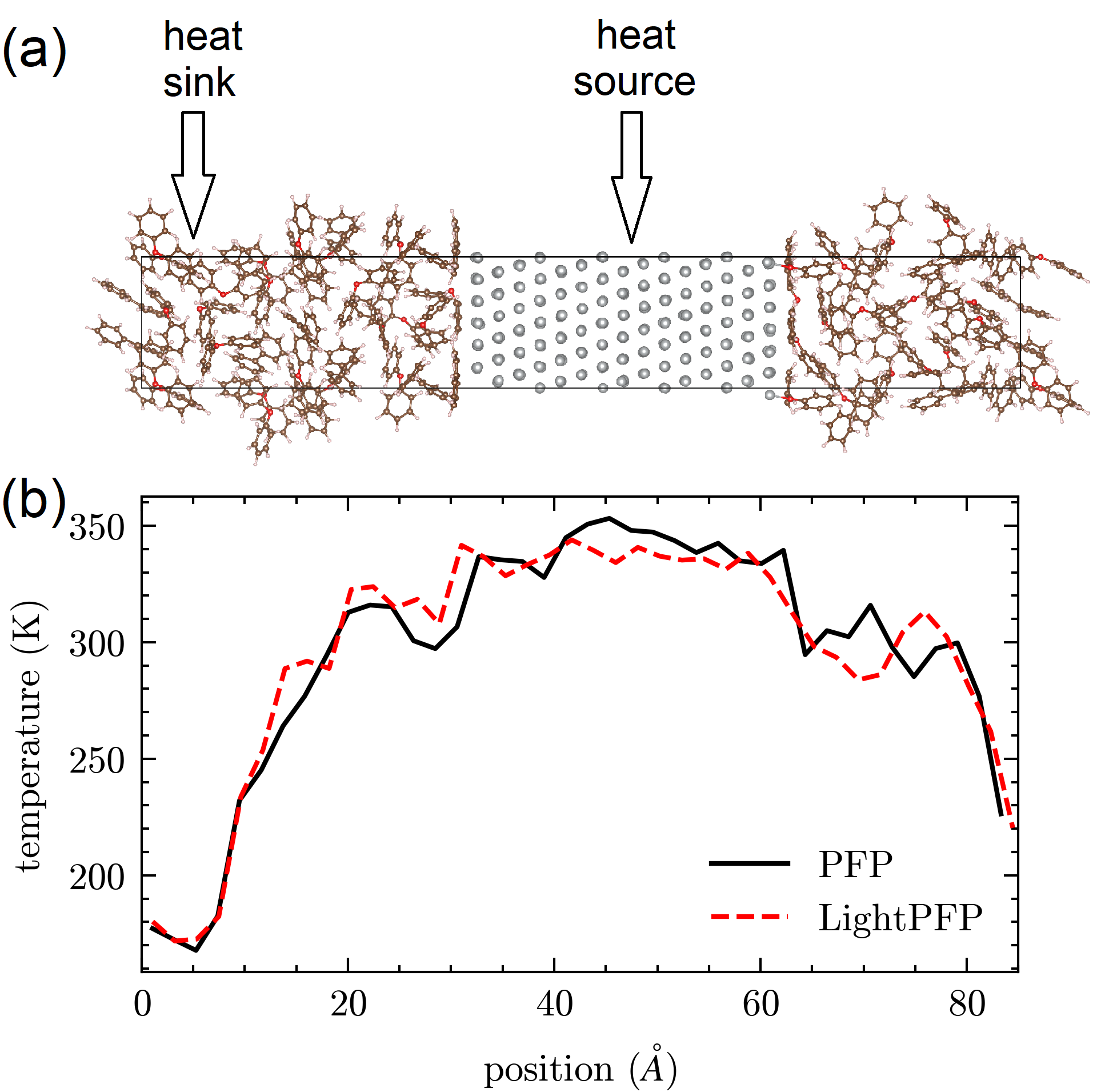}
\caption{Temperature profile across Ni and DPO/BP interface}
\label{fig:appendix_3_temperature_profile}
\end{figure}

We assessed the resulting LightPFP model against a reference PFP potential under matched conditions. The model reproduces key equilibrium and non-equilibrium observables: system densities (Fig~\ref{fig:appendix_3_density}) and radial distribution functions (Fig.~\ref{fig:appendix_3_rdf}) for interfacial systems, as well as steady-state temperature profiles obtained from reverse non-equilibrium MD across Ni (111)-DPO/BP interfaces (Fig.~\ref{fig:appendix_3_temperature_profile}). The close agreement with PFP across these metrics supports the use of LightPFP for the interfacial thermal conductivity calculations reported below.

\clearpage

\section{Application 4: Viscosity of n-decane}

In this study, we target the shear viscosity of liquid \textit{n}-decane via reverse non-equilibrium molecular dynamics (rNEMD) using the Müller–Plathe momentum-exchange scheme \cite{hess2002determining}. To be more specific, a liquid is equilibrated at the target thermodynamic state and the simulation box is partitioned into slabs along the gradient direction, with two slabs designated as momentum source and sink. At fixed intervals, the particle with the largest positive flow-direction velocity in the source slab and the particle with the most negative component in the sink slab exchange their flow-direction velocities, imposing a constant momentum flux. The spatially resolved velocity profile is accumulated and time-averaged; its linear region away from the exchange slabs provides the velocity gradient. The imposed flux is computed from the cumulative exchanged momentum divided by cross-sectional area and simulation time (accounting for periodic shear planes). 
The shear viscosity is then $\eta = \frac{J}{\mathrm{d}v/\mathrm{d}z}$, where $J$ is cumulative exchanged momentum, $v$ is the velocity of atoms and $z$ is atomic position along z axis.

Important practical considerations accompany rNEMD. Because accessible simulation times are limited, rNEMD often relies on velocity gradients far larger than those used in experiments, which can alter molecular orientation (e.g., induce partial alignment along the flow) and thereby modify the intrinsic shear response. To reduce the artificially imposed gradient, one can lower the exchange frequency or elongate the simulation box along the gradient direction. Lowering the exchange frequency weakens the driving but requires longer MD sampling to resolve the slope of the velocity profile accurately, particularly for high-viscosity liquids. Increasing the box length similarly reduces the gradient at a given momentum flux but raises computational cost. This makes uMLIP require longer time to compute viscosities for many molecules, whereas the LightPFP method becomes more useful due to its better computational efficiency. In this example, we selected \textit{n}-decane as the case study and trained its LightPFP model. We computed the viscosity of \textit{n}-decane at high temperatures under different computational conditions using both PFP and LightPFP as a validation of LightPFP. Because viscosity is lower at high temperature, PFP also yields good results. We then used LightPFP to compute the viscosity of n-decane at room temperature, ultimately obtaining values that closely match experiment.

\subsection{Student fine-tuning}
\label{app4:fine-tuning}
We trained LightPFP for \textit{n}-decane with a dataset focused on bulk liquid environments. Several initial liquid structures were generated with realistic densities to cover different molecule orientations. The initial dataset combined short MD segments and “rattle” perturbations to diversify local configurations: NVT trajectories at 500 K, 1000 K, and 1500 K; NPT trajectories at 300 K, 400 K, 500 K, and 600 K; and rattle displacements are sampled from normal distribution with standard deviation of 0.10 \text{\AA} and 0.15 \text{\AA}. As in our other applications, high-temperature NVT sampling was used to generate randomized, higher-energy configurations that improve robustness, whereas NPT sampling was focused at moderate temperatures to avoid unphysical density excursions and cell distortions that can occur under aggressive barostatting at very high temperatures.

Starting from this base model, we executed an active learning loop tailored to the viscosity task. For bulk liquid \textit{n}-decane, we ran short trajectories under complementary ensembles and non-equilibrium driving: 5 ps NVT, 10–20 ps NPT, and 20–30 ps rNEMD. rNEMD simulations followed the slab-based momentum-exchange approach, with systematic variation of setup details (swap interval, swap slab thickness, and cell aspect ratio) to probe sensitivity and enrich configurational coverage. Configurations exhibiting large model disagreement were labeled and added to the training set; the query–retrain cycle was repeated until validation metrics stabilized.

\subsection{Evaluation using PFP}
\label{app4:eval-by-pfp}

After obtaining the LightPFP model, we ran the rNEMD calculation using PFP and LightPFP. The calculation method is as follows: First, the size of initial simulation box is 17.3x17.3x51.9 angstrom contains 48 \textit{n}-decane molecules. Then, the MD is performed in thee NVT ensemble for 5 ps at 400, 450 and 500 K, and then, NPT ensemble for 20 ps at the same temperature, 1 bar to achieve equilibrium status. After that, the rNEMD is performed, the simulation box is divided into 20 slabs. For each temperature, the momenta exchange is performed for each 100 or 500 fs. The rNEMD is performed for 50 ps to achieve a stable momenta profile across the simulation box.

\begin{figure}
\includegraphics[width=0.8\textwidth]{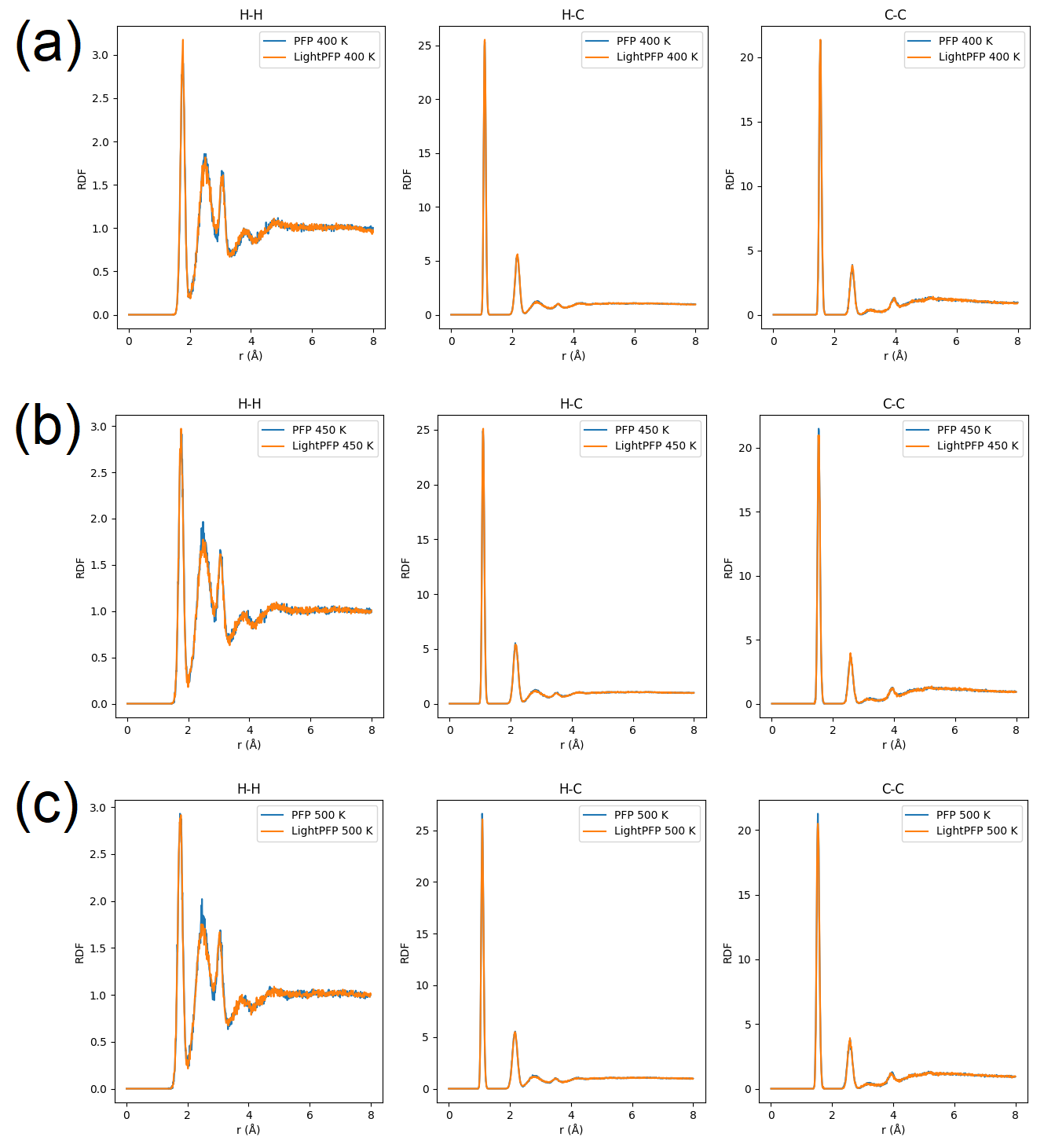}
\caption{Radial distribution function of n-decane. (a) 400 K (b) 450 K and (c) 500 K}
\label{fig:appendix_4_rdf}
\end{figure}

The accuracy of LightPFP is evaluated by comparing several properties with PFP results. As shown in table \ref{table:decane_density} at 400, 450 and 500 K is obtained from the NPT-MD part of trajectory. The predicted liquid densities agree well with PFP with an average absolute deviation of 0.0076 g/cm$^3$. 
The radial distribution functions for \textit{n}-decane also obtained from the MD frames took from the NPT part. The results are plotted in Fig \ref{fig:appendix_4_rdf} showing a close agreement with PFP over the same temperature range.

\begin{table}
\centering
\caption{Density of n-decane at different temperatures}
\begin{ruledtabular}
\begin{tabular}{lccc}
%\hline
Temperature (K) & PFP (g/cm$^3$) & LightPFP (g/cm$^3$) & Error (g/cm$^3$)\\ 
\hline
400 & 0.689 & 0.675 & 0.014\\ 
450 & 0.647 & 0.650 & 0.003\\ 
500 & 0.611 & 0.617 & 0.006\\ 
\end{tabular}
\end{ruledtabular}
\label{table:decane_density}
\end{table}

\begin{figure}
\includegraphics[width=\textwidth]{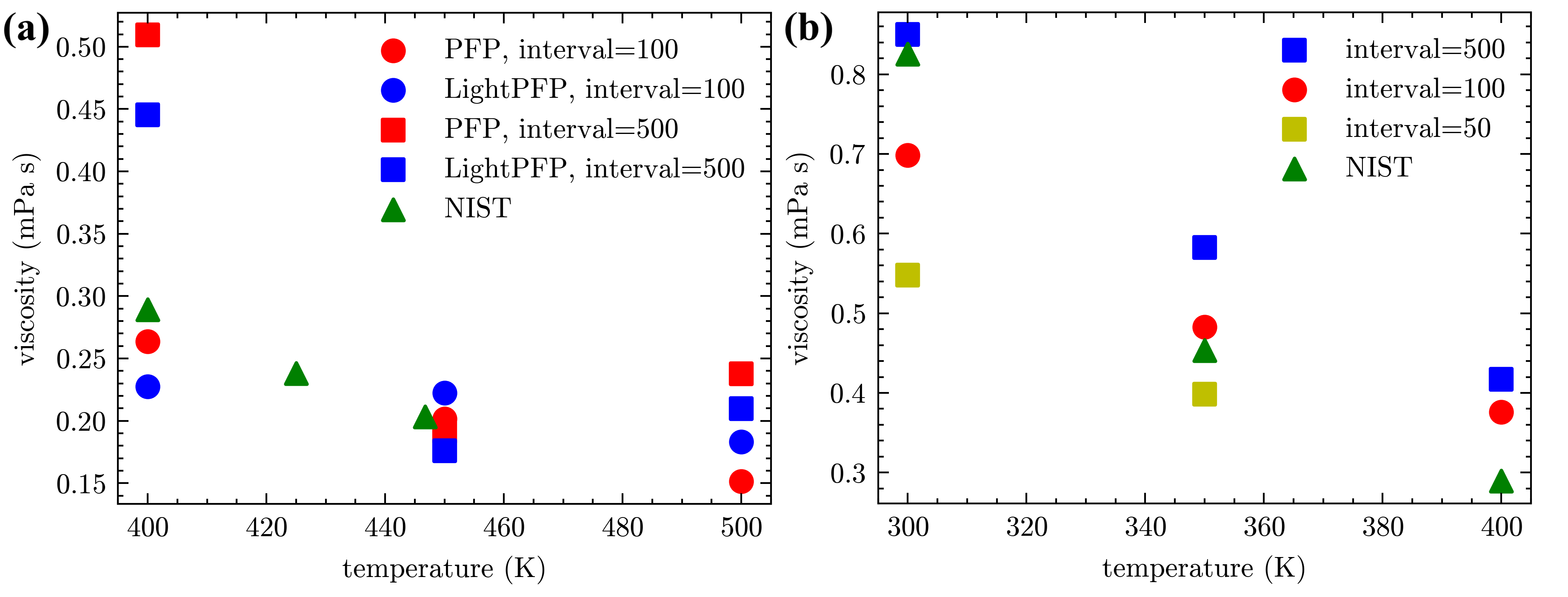}
\caption{Viscosity of n-decane}
\label{fig:appendix_4_viscosity}
\end{figure}

At last, we compared the viscosity results calculated by LightPFP and PFP (Fig.~\ref{fig:appendix_4_viscosity}(a)). First, at 450 K, both PFP and LightPFP yield the same viscosity of 0.2 mPa·s for exchange intervals of 100 fs and 500 fs. All results are consistent with experimental data \cite{NIST_SRD69_nDecane_TFREEZE} at 446.75 K. At 400 K, viscosities computed with a 500 fs momentum-exchange interval were higher than those with a 100 fs interval for both PFP and LightPFP. Under both intervals, PFP and LightPFP were in close agreement. The MD-derived viscosities spanned 0.23–0.51 mPa·s across the tested conditions, and the experimental value of 0.29 mPa·s falls within this range.

\subsection{Large-scale MD simulation}
\label{app4:large-scale-md}
After confirming of the LightPFP model by comparing the high-temperatures viscosity with PFP and experimental results, we tried to run large-scale MD simulation to get more accuract viscosity at low temperature. The simulation box size is 34.6 x 34.6 x 103.8 Angstrom, which contains 384 \textit{n}-decane molecules. We run the MD simulation at 300, 350 and 400 K. The MD protocol is same as the above one, except we prolong the rNEMD simulation to 1 ns, to get accurate statistic results. The results is shown in Fig.~\ref{fig:appendix_4_viscosity}(b). The viscosity is estimated to be 0.85 mPa s at 300 K which is very close to the experiment value, 0.83 mPa s.

\clearpage

\section{Application 5: Crack propagation in graphene nanoribbon}

In this example, we assess the capability of LightPFP to describe fracture propagation in graphene nanoribbon via molecular dynamics\cite{jung2023artificial}. 

\subsection{Student fine-tuning}
\label{app5:fine-tuning}
The training data were constructed to span two-dimensional carbon environments under mechanical loading, from pristine structures to defect-containing systems. The initial dataset included defect-free graphene sheets and graphene nanoribbons (GNRs) with armchair (AC) and zigzag (ZZ) edges, as well as structures with a triangular hole that served as crack initiators. To sample elastic responses broadly, we (a) performed NVT MD at 500 K, 1000 K, and 1500 K; (b) applied small homogeneous deformations: ±5\% strain to both diagonal (uniaxial/biaxial) and off-diagonal (simple shear) components of the simulation cell. To further diversify local environments, atomic positions were rattled with Gaussian noise of 0.1 \text{\AA} standard deviation. No cracks were included in this initial dataset; thus, the base model learned from pristine bonding environments across temperatures, strain states, and edge types without explicit exposure to fracture.

Starting from this model, we carried out active learning to improve robustness in high-strain regimes. Using the initial structures (graphene sheets and AC/ZZ GNRs), we ran strain-controlled MD with a “deform extension” protocol that incrementally altered the cell shape to increase strain every fixed number of MD steps. This procedure drives the systems into strongly non-linear regimes where bond stretching, bond angle distortions, and incipient bond breaking occur. During these runs, we monitored model performance and selectively augmented the training set with configurations exhibiting large errors or anomalous forces/energies under increasing deformation. In addition, we included strained configurations of ribbons with pre-introduced triangular holes to expose the model to local stress concentration fields characteristic of crack tips and to the chemistry of bond scission in sp2 carbon. The resulting LightPFP model was then used for fracture simulations.

\subsection{Evaluation using PFP}
\label{app5:eval-by-pfp}
\begin{figure}
\includegraphics[width=\textwidth]{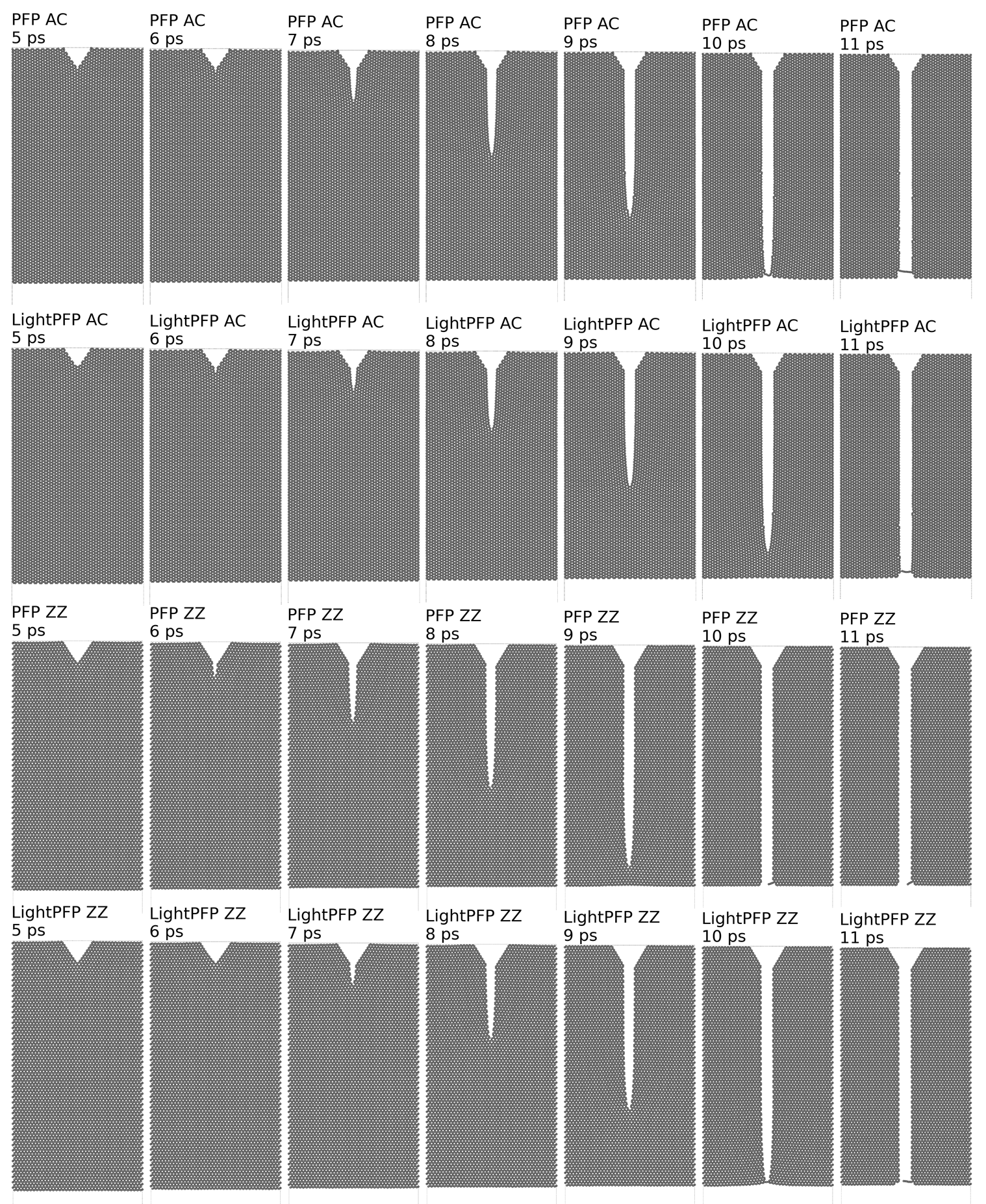}
\caption{Illustration of crack propagation in graphene nanoribbons. Each column displays molecular dynamics snapshots at 5, 6, 7, 8, 9, 10, and 11 ps, corresponding to strains of 0.05, 0.06, 0.07, 0.08, 0.09, 0.10, and 0.11, respectively. From top to bottom, the four rows show PFP MD of AC GNR, LightPFP MD of AC GNR, PFP MD of ZZ GNR, and LightPFP MD of ZZ GNR.}
\label{fig:appendix_5_GNR_crack}
\end{figure}

For evaluation, we simulate crack initiation and growth in AC and ZZ GNRs at 300 K by introducing a triangular hole on one ribbon edge to serve as a notch. Uniaxial tension was applied along the ribbon axis (x direction), with the Green-Lagrange strain increased at a rate of $10^{-5}$ per femtosecond. Each trajectory was propagated for 50 ps, reaching a total strain of 0.5. Identical protocols were run with both the original PFP potential and LightPFP for direct comparison. The MD snapshot during the crack propagation process is shown in Fig.~\ref{fig:appendix_5_GNR_crack}.

Both PFP and Light predict crack initiation at about 6 ps (strain is 0.06) in AC GNR. For ZZ GNR, the PFP shows crack initiation at about 6 ps while LightPFP is a little bit slower. LightPFP reproduces the spatial pattern of bond breaking and the subsequent crack path observed with PFP: bonds fail first near the notch where stress concentrates, and the crack advances into the ribbon width under continued loading. The straight crack is reproduced by LightPFP in both AC and ZZ GNR. The predicted morphology and sequence of fracture events are consistent across edge types, with no spurious branching or unphysical healing observed in LightPFP. 

\begin{figure}
\includegraphics[width=0.5\textwidth]{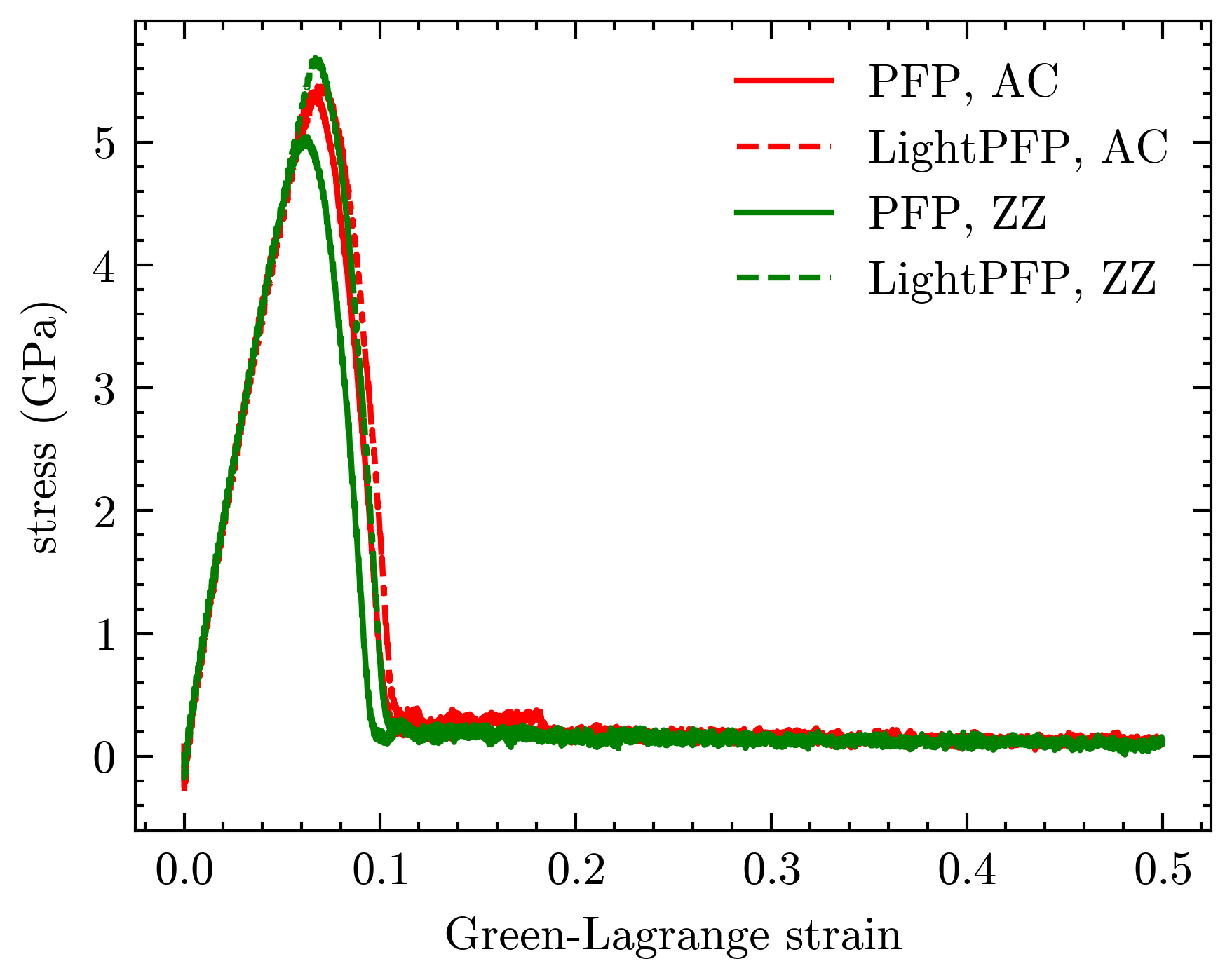}
\caption{Strain stress curve of AC and ZZ GNR in crack propagation process}
\label{fig:appendix_5_strain_stress}
\end{figure}

The stress–strain responses computed from the virial stress is plotted in Fig.~\ref{fig:appendix_5_strain_stress}. The curve shows close agreement between PFP and LightPFP over the entire loading history, including the elastic state where stress and strain change linearly, the crack propagation state where stress drops rapidly, and the state where stress is 0 after fracture. 

Taken together, these results demonstrate that LightPFP, trained without explicit cracks and refined via strain-driven active learning, accurately captures the initiation and propagation of fractures in graphene nanoribbons. The agreement in crack onset time, crack morphology, and stress–strain behavior indicates that LightPFP attains PFP-level fidelity for fracture simulations while retaining its computational efficiency.

\clearpage

\section{Application 6: Friction of \ce{Fe2O3} surface with lubricant and fatty acid surfactant}

\begin{figure}
\includegraphics[width=\textwidth]{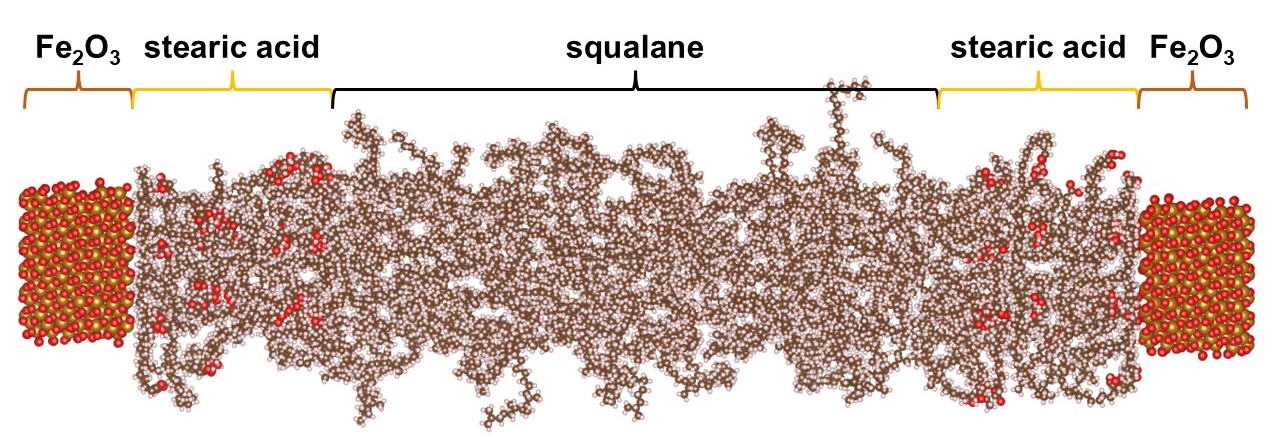}
\caption{An atomistic structure of \ce{Fe2O3} - lubricant - surfactant system}
\label{fig:appendix_6_atomic_structure}
\end{figure}

In this example, we use LightPFP to investigate lubrication and shear responses at iron oxide–organic interfaces by simulating friction between two \ce{Fe2O3} slabs separated by a multilayer film of fatty-acid surfactants and a squalane lubricant \cite{doig2014structure}. Specifically, the simulation box comprises a five-layer stack: an \ce{Fe2O3} slab, a monolayer of stearic acid or oleic acid, a squalane layer, a second monolayer of stearic or oleic acid, and a second \ce{Fe2O3} slab. Figure \ref{fig:appendix_6_atomic_structure} shows the atomistic structure of this system. This configuration results in a complex multilayer structure, making the collection of training data more challenging.

\subsection{Student fine-tuning}
\label{app6:fine-tuning}
We trained LightPFP on a curated dataset comprising (1) bulk \ce{Fe2O3} crystals, (2) \ce{Fe2O3} slabs, (3) bulk liquids of stearic acid, oleic acid, and squalane, (4) solid-liquid interfaces between \ce{Fe2O3} and each of stearic acid, oleic acid, and squalane, and (5) liquid-liquid interfaces between squalane and stearic acid, and between squalane and oleic acid. We generated diverse initial structures within each category and used them to sample training configurations. Multiple distinct initial configurations were employed to enhance the diversity of the training set.

To construct the initial dataset, we combined molecular-dynamics sampling with rattle perturbations. We performed MD in both NVT and NPT ensembles. NVT sampling at elevated temperatures (500, 1000, and 1500 K) broadened the range of conformations and interfacial arrangements, whereas NPT sampling at 300, 400, and 500 K targeted thermodynamic states relevant to friction simulations and yielded realistic organic-layer densities. To improve robustness to rare distortions, we applied Gaussian-distributed displacements (standard deviations 0.10 and 0.15 \text{\AA}) to the initial structures, generating physically plausible yet diverse local environments.

Because friction can induce large molecular deformations and uncommon contact geometries that are rare under equilibrium sampling, we augmented the model via active learning focused on shear. In each iteration, we (i) fixed the bottom part of the system, (ii) imposed a controlled lateral displacement on the top part to generate steady sliding, and (iii) evolved the remaining atoms with NVT dynamics. Configurations exhibiting large energy/force discrepancies were added to the training set, and the model was retrained. We conducted 10 active-learning rounds: iterations 1–5 used smaller interface cells containing both iron oxide–molecule solid–liquid interfaces and surfactant–lubricant liquid–liquid interfaces to rapidly accumulate diverse contact motifs, whereas iterations 6–10 employed larger cells to capture cooperative rearrangements under shear.

\subsection{Evaluation using PFP}
\label{app6:eval-by-pfp}
The final LightPFP model was used to simulate the full five-layer stack: \ce{Fe2O3} slab – stearic acid – squalane film – stearic acid – \ce{Fe2O3} slab. We validated the model by performing friction MD with both PFP and LightPFP. The initial system contained 4,604 atoms. Sliding simulations were run for 300 ps with a top-slab velocity of 30 m/s, while the bottom slab was fixed. The system temperature was maintained at 300 K using a thermostat. The same sliding protocol was applied with PFP, and the resulting trajectories and final structures were compared.

\begin{figure}
\includegraphics[]{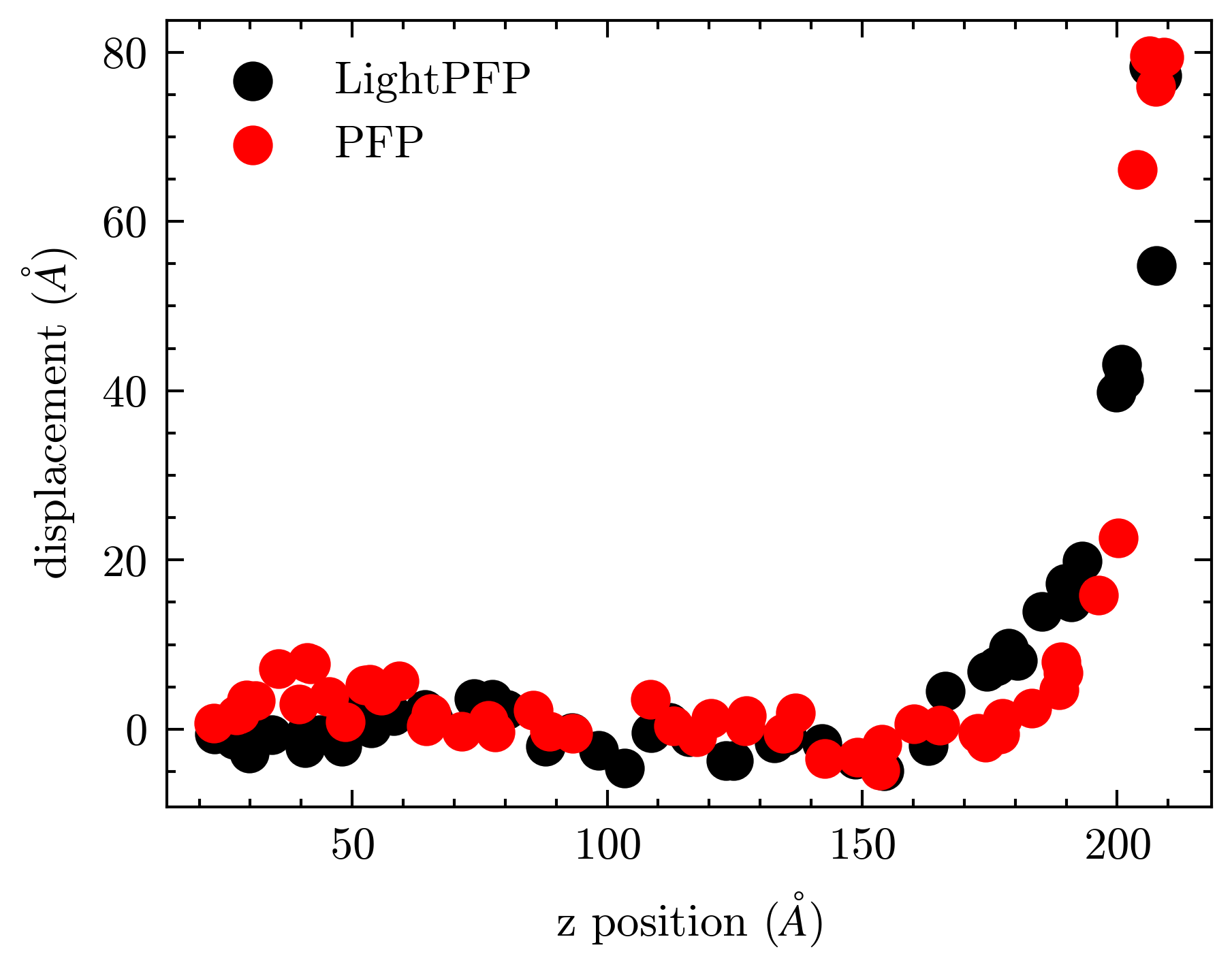}
\caption{Molecular displacements during 300 ps of sliding for the Fe2O3 / fatty acid / squalane / fatty acid / Fe2O3 stack.}
\label{fig:appendix_6_displacement}
\end{figure}

\begin{figure}
\includegraphics[width=0.48\textwidth]{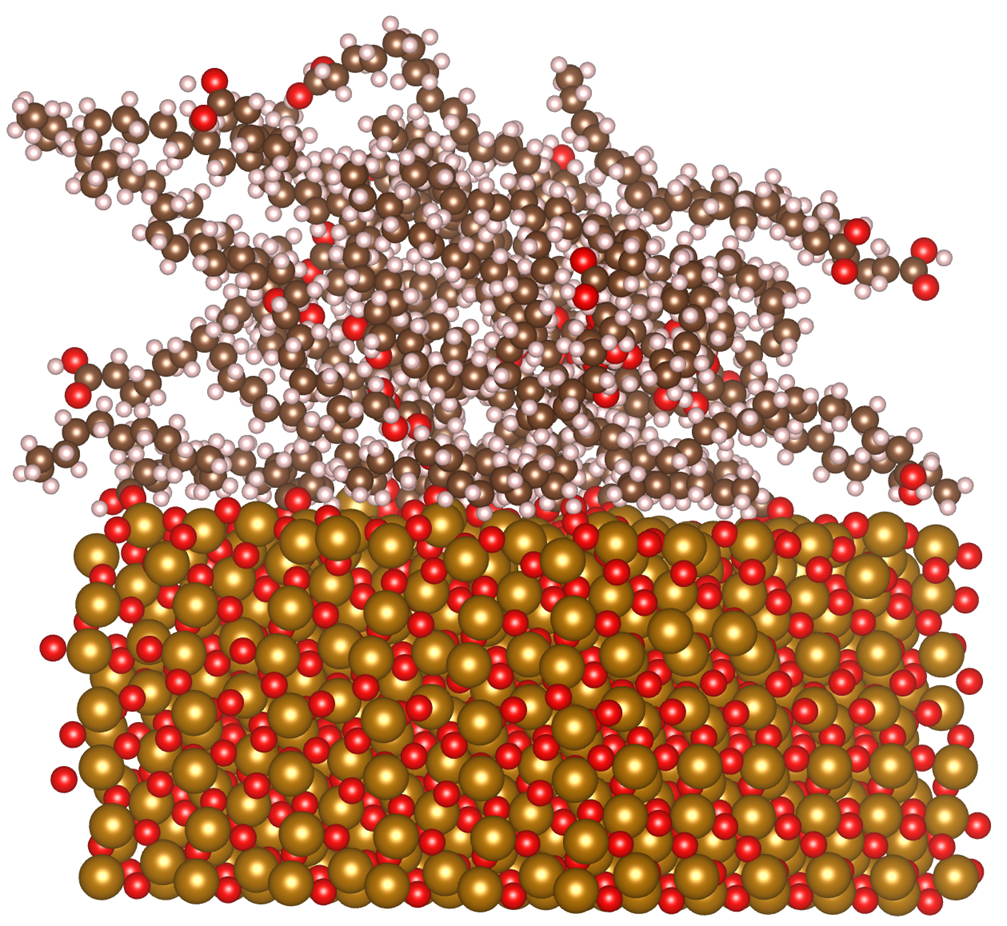}
\caption{Final-frame morphology after 300 ps of sliding.}
\label{fig:appendix_6_morphology}
\end{figure}

Figure \ref{fig:appendix_6_displacement} shows the displacement of molecules over 300 ps of sliding. As the upper \ce{Fe2O3} slab moves, nearby molecules are entrained and translate in the sliding direction. Both PFP and LightPFP predict that molecules adjacent to the moving \ce{Fe2O3} layer travel by approximately 80 \text{\AA} due to interfacial friction, with displacements that decay with depth into the surfactant and lubricant layers. Figure \ref{fig:appendix_6_morphology} depicts the morphology of the moving \ce{Fe2O3} slab and adjacent molecules in the final frame. Interfacial shear substantially stretches the molecular chains and aligns them along the sliding direction. These results support the ability of LightPFP to capture coupled solid–organic interfacial mechanics and shear-induced ordering relevant to boundary lubrication.

\clearpage

\section{Application 7: Diffusion behavior in Polymer Ionic Liquid}

In this example, we use LightPFP to investigate anion diffusion in a neat polymer ionic liquid (PIL), taking poly (ethyl vinyl imidazolium) paired with \ce{PF6-} as a representative system\cite{luo2020molecular}. 
Poly (ethyl vinyl imidazolium) forms a positively charged polymer network, while discrete \ce{PF6-} anions occupy the interstitial regions of the polymer matrix, collectively giving rise to the characteristic ion-conducting behavior of PILs.

\subsection{Student fine-tuning}
\label{app7:fine-tuning}
We curated a training set to expose LightPFP to the relevant local environments of the target PIL across a broad range of densities, temperatures, and chain conformations. To efficiently cover conformational diversity, we generated multiple initial structures rather than relying on a single long trajectory: starting from distinct packings accelerates exploration of chain orientations, local packing motifs, and ion coordination states. Structural assembly was performed with a dedicated routine that constructs mixed-oligomer boxes comprising monomers, dimers, trimers, 5-mers, and 7-mers of poly (ethyl vinyl imidazolium), together with the stoichiometric number of \ce{PF6-} anions to satisfy charge neutrality. Molecules were placed with randomized positions and orientations to maximize initial configurational diversity.

Initial data generation combined short molecular dynamics (MD) sampling and stochastic perturbations. MD trajectories were run in both NVT and NPT ensembles. For NVT, we used elevated temperatures (500 K, 1000 K, 1500 K) to accelerate configurational de-correlation and broaden the coverage of local structures. For NPT, we sampled temperatures (400 K, 500 K, 600 K) representative of the intended application window to capture realistic densities and coordination statistics. In addition, a rattle procedure applied Gaussian displacements (standard deviations of 0.10 and 0.15 \text{\AA}) to further diversify local atomic environments and improve robustness. The initial LightPFP model is trained from this dataset.

Starting from this model, we carried out active learning to enhance accuracy and stability under production conditions. Each active-learning iteration consisted of multiple short MD jobs, each seeded from a newly generated structure via the above mentioned procedure. For each job, we randomly selected a temperature between 300 and 700 K, ran a 5 ps NVT trajectory followed by a 50 ps NPT trajectory, and attempted data collection every 100 MD steps. Configurations identified as poorly described by the current model were labeled and added to the training set, after which the model was retrained.

\subsection{Evaluation using PFP}
\label{app7:eval-by-pfp}
To validate LightPFP for the target PIL, we performed MD comparisons against PFP. First, we generated reference PFP trajectories at 400, 500, and 600 K with NPT ensemble. Using identical initial configurations, ensemble settings, and integrator parameters, we then repeated the simulations with LightPFP and saved all trajectories for post hoc analysis. 

\begin{table}
\centering
\caption{Density of polymer ionic liquid}
\label{tab:appendix_7_density}
\begin{tabular}{lrr}
\hline
Temperature (K)& PFP (g/cm$^3$) & LightPFP (g/cm$^3$) \\
\hline
300 K & 1.394 & 1.389 \\
400 K & 1.374 & 1.372 \\
500 K & 1.351 & 1.356 \\
\hline
\end{tabular}
\end{table}

\begin{figure}
\includegraphics[width=0.8\textwidth]{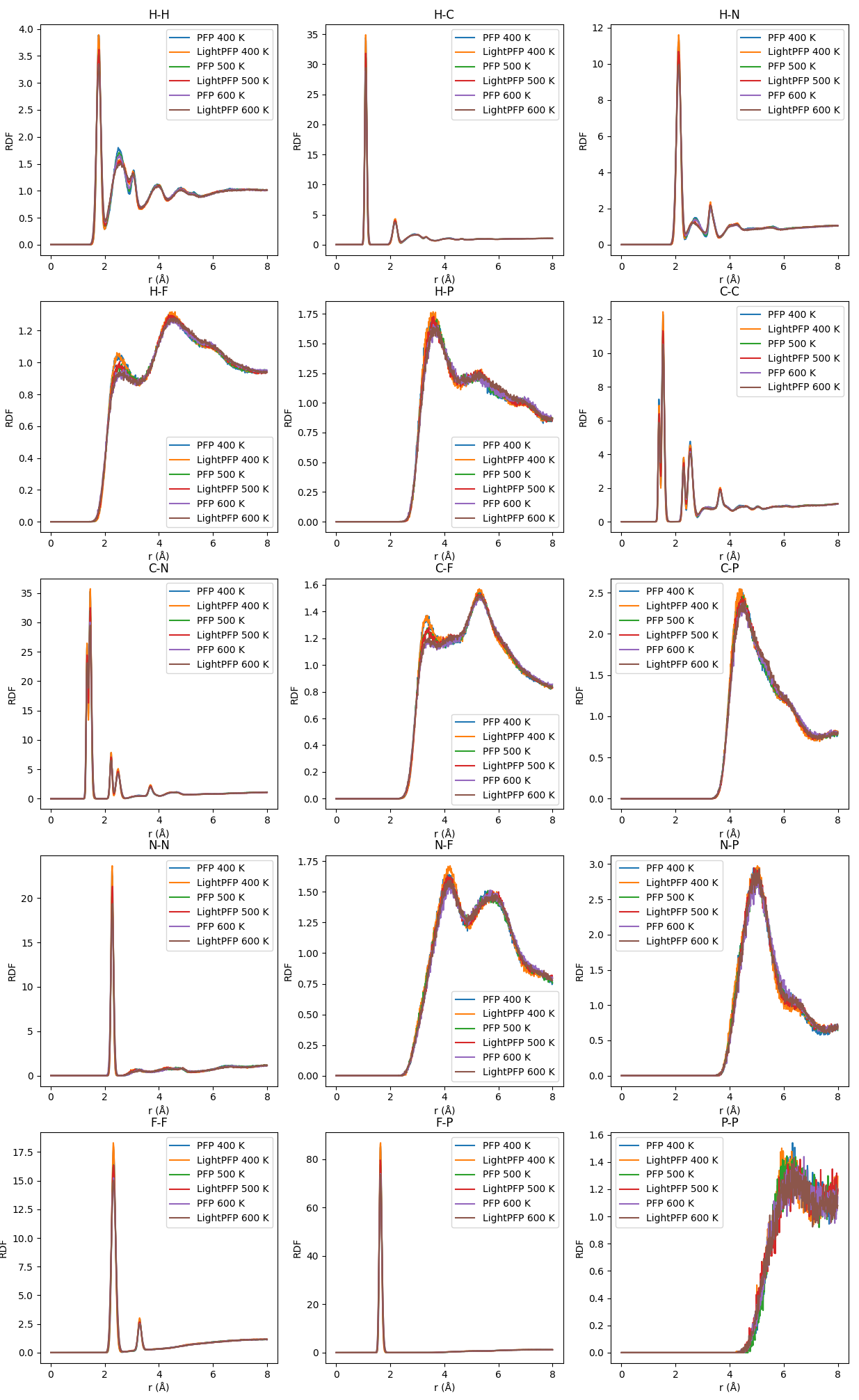}
\caption{Radial distribution function of polymer ionic liquid}
\label{fig:appendix_7_rdf}
\end{figure}

\begin{figure}[h!]
    \centering
    \begin{subfigure}[b]{0.48\textwidth}
        \centering
        \includegraphics[width=\linewidth]{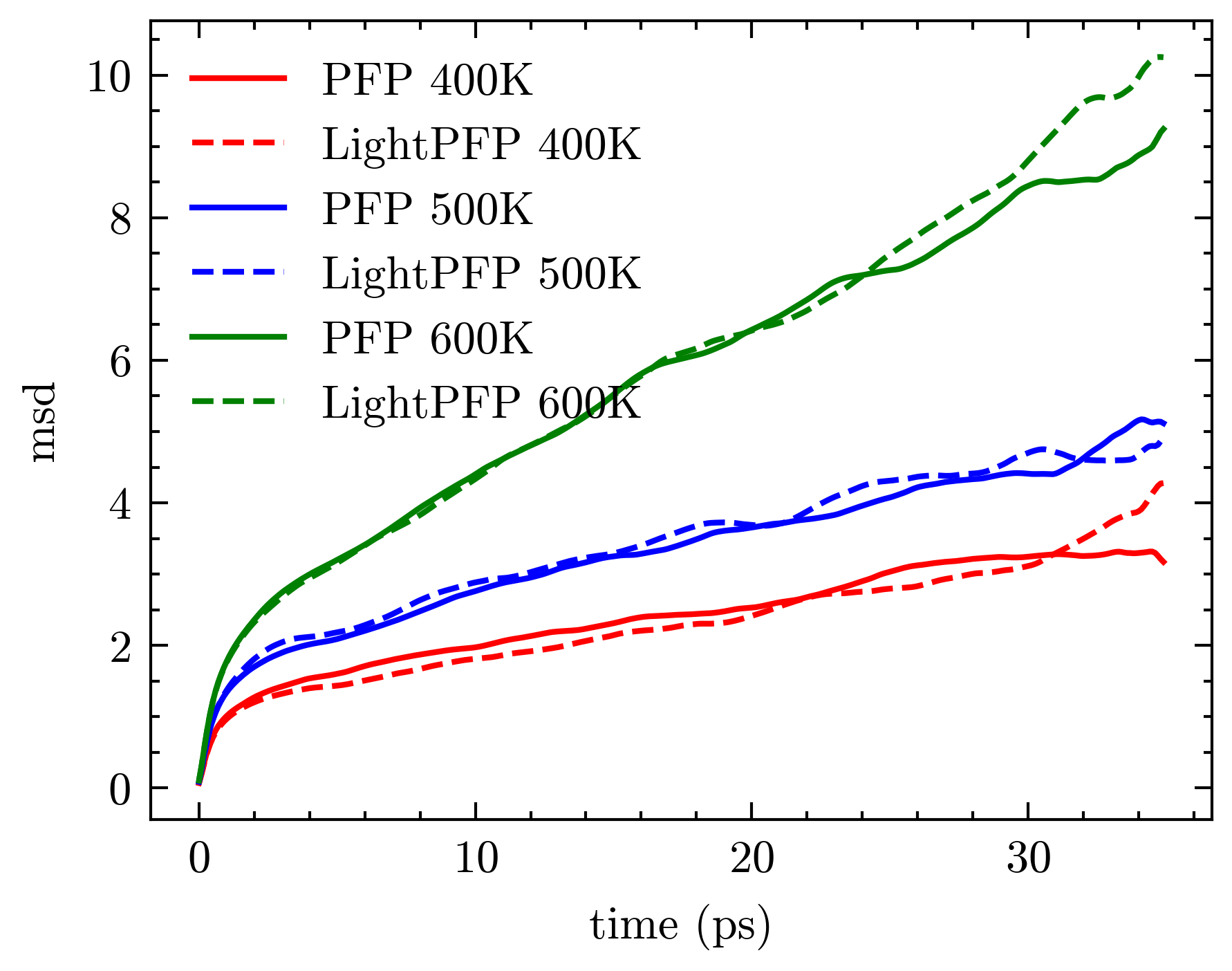}
    \end{subfigure}
    \hfill % Adds horizontal space between subfigures
    \begin{subfigure}[b]{0.48\textwidth}
        \centering
        \includegraphics[width=\linewidth]{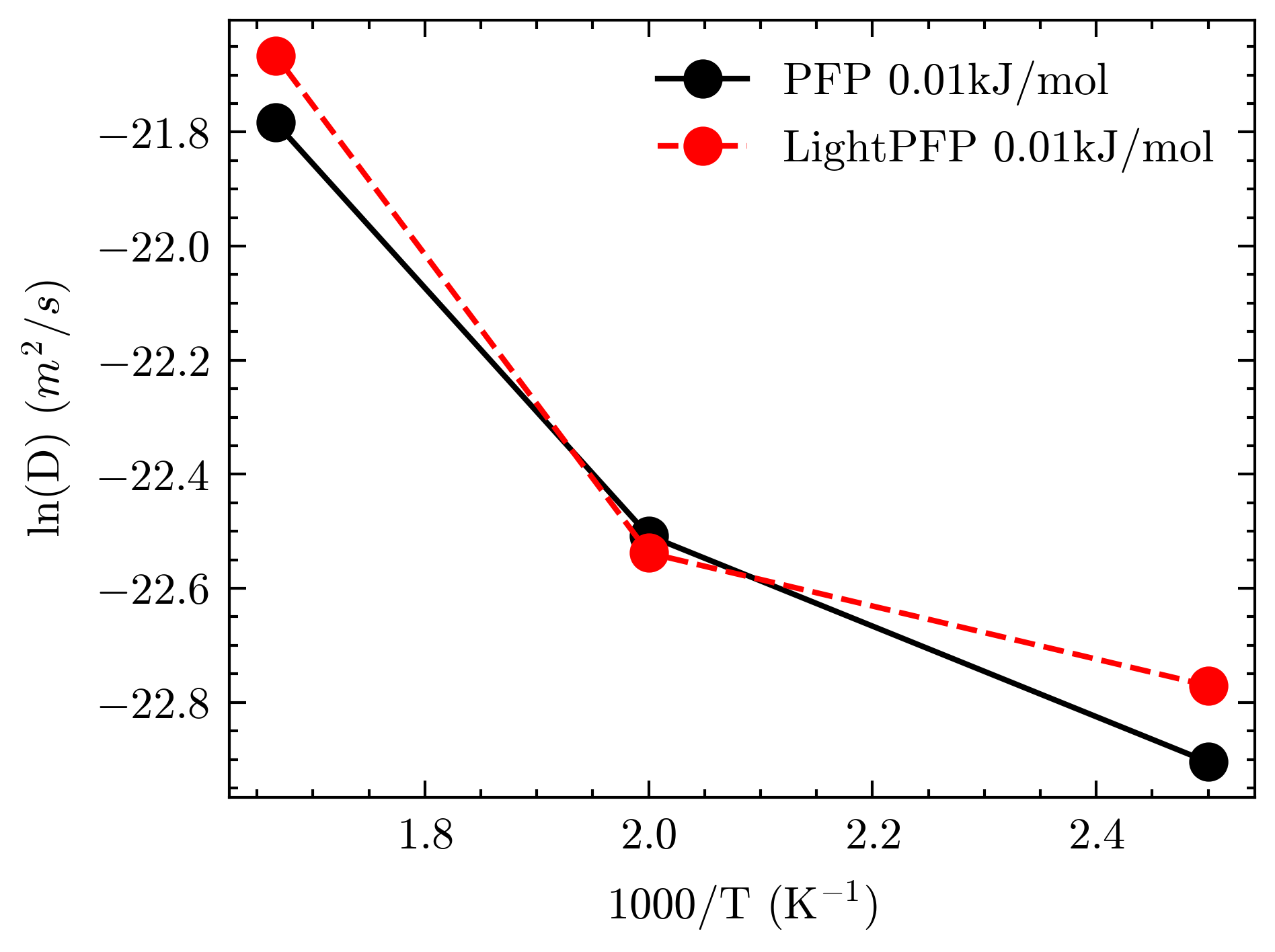}
    \end{subfigure}
    \caption{Diffusion behavior of \ce{PF6-} anion (left) mean squared displacement of \ce{PF6-} anion at 300 K, 400 K and 500 K (right) Arrhenius plot of \ce{PF6-} anion diffusion coefficient}
    \label{fig:appendix_7_diffusion}
\end{figure}

We computed equilibrium densities from the last 10 ps NPT trajectories. The densities at 300 K, 400 K, and 500 K listed in Table \ref{tab:appendix_7_density}. LightPFP densities closely track PFP across this range. The radial distribution functions (RDFs), which were accumulated over the last 10 ps of each trajectory at 300 K, 400 K, and 500 K, are shown in Fig.~\ref{fig:appendix_7_rdf}. LightPFP reproduces the positions and heights of the PFP peak, indicating consistent local coordination and packing.

To probe anion mobility, we monitored phosphorus atoms (proxies for \ce{PF6-}) and computed mean squared displacement (MSD), as shown in Fig. \ref{fig:appendix_7_diffusion}. MSD curves from LightPFP agree well with those from PFP, indicating consistent diffusive behavior in the polymer matrix. In addition, the diffusion coefficients D(T) were extracted from the MSD and fitted to an Arrhenius form to estimate the activation energy for \ce{PF6-} transport. The activation energy from LightPFP is 10.46 kJ/mol, in close agreement with the PFP value of 10.82 kJ/mol. We note that the limited number of temperatures constrains the precision of these estimates; the results are presented primarily to demonstrate consistency between the two models.

\clearpage

\section{Application 8: Mechanical property of \ce{SiO2}–\ce{P2O5}–\ce{Al2O3}–\ce{Na2O} glass}

In this example, we use LightPFP to investigate composition–property relationships and the mechanical response of multicomponent oxide glasses in the \ce{SiO2}–\ce{P2O5}–\ce{Al2O3}–\ce{Na2O} system, as a stringent test on amorphous materials\cite{qian2022ab}. We consider glasses with compositions \ce{SiO2}:(79.69-x) mol\%, \ce{P2O5}:x mol\%, \ce{Al2O3}:13.79 mol\%, and \ce{Na2O}:15.52 mol\%, with x ranging from 0 to 50 mol\%.

\subsection{Student fine-tuning}
\label{app8:fine-tuning}
To build the initial training set, we combined crystalline and disordered configurations. First, crystal structures of \ce{SiO2}, \ce{P2O5}, \ce{Al2O3}, and \ce{Na2O} were obtained from the Materials Project to capture characteristic, energetically stable local environments (e.g., tetrahedral Si). Second, random packings were generated by placing atoms uniformly at random in periodic simulation boxes subject to a minimum interatomic separation, avoiding atomic overlap while spanning a broad space of disordered motifs. From these crystalline and random initial structures, we performed NPT MD at 500 K, 1000 K, and 1500 K, and sampled uncorrelated snapshots across densities and coordination states to train the initial LightPFP model.

Active learning was then used to refine accuracy for glassy states. For each composition, we initiated melt–quench protocols from random packings: (i) NVT MD at 2000 K for 10 ps to fully melt and randomize the network; (ii) linear cooling from 2000 K to 500 K at 0.1 K/fs to form the glass; and (iii) additional MD at 500 K to relax the structure. During these runs, we monitored model reliability and selectively augmented the training set with configurations exhibiting large force/energy discrepancies relative to the reference PFP, iterating retraining until convergence. This pipeline exposes the model to topological rearrangements (bond breaking/formation, modifier-induced non-bridging oxygens) and the broad spectrum of local environments that emerge during melt–quench.

\subsection{Evaluation using PFP}
\label{app8:eval-by-pfp}
\begin{figure}
\includegraphics[width=0.5\textwidth]{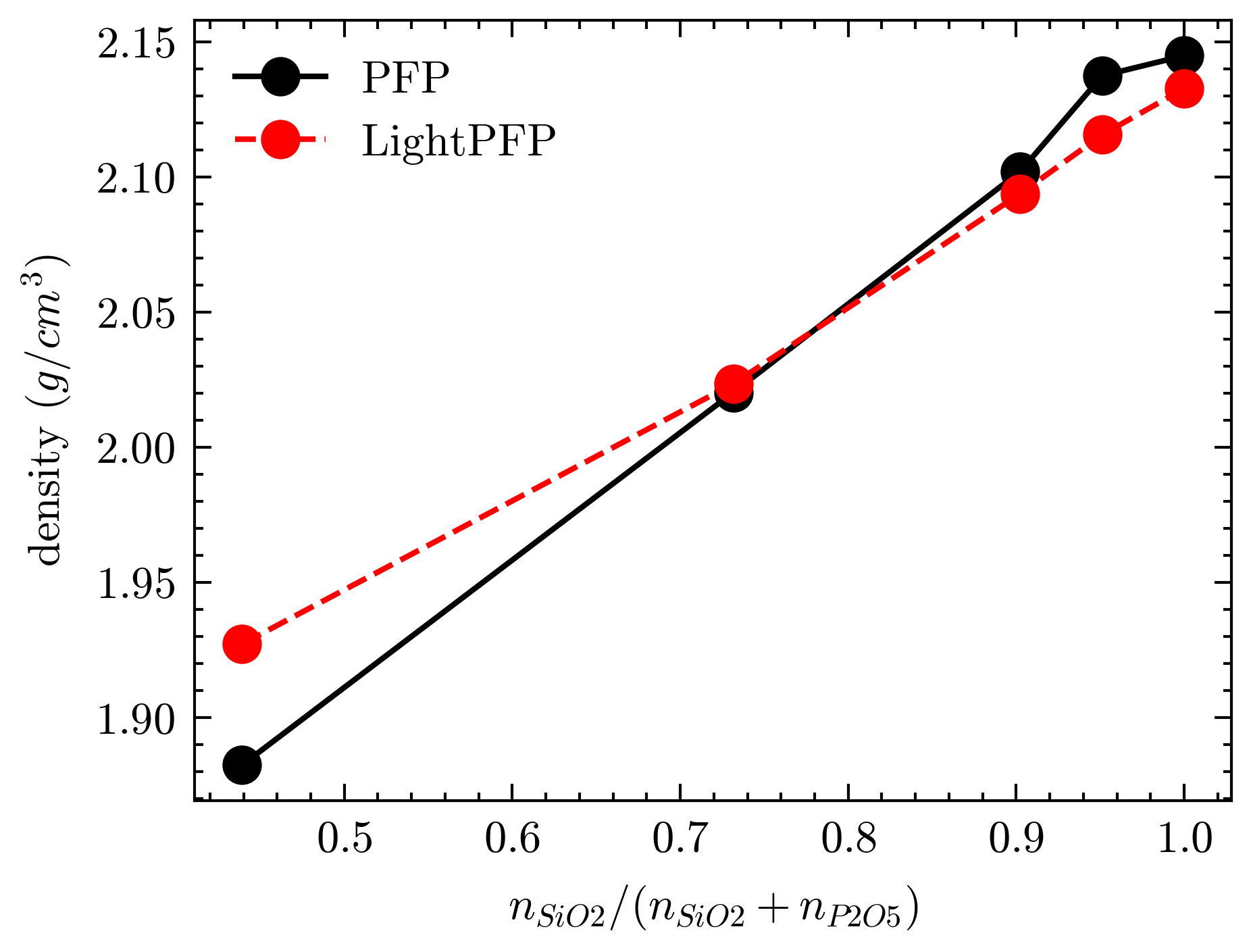}
\caption{Density of (82-x)\ce{SiO2}-x\ce{P2O5}-16\ce{Al2O3}-18\ce{Na2O} glass}
\label{fig:appendix_8_density}
\end{figure}

For evaluation, we generated glass structures at five representative compositions, i.e. (82-x)\ce{SiO2}–x\ce{P2O5}–16\ce{Al2O3}–18\ce{Na2O}, where x = 0, 4, 8, 22, 46. Identical melt–quench protocols were performed with both the PFP and LightPFP. Specifically, each system was equilibrated in the melt and then quenched at 0.02 K/fs to the 500K, followed by NPT relaxation to determine the density. The resulting densities are plotted in Fig. \ref{fig:appendix_8_density}. In both PFP and LightPFP simulations, the density increases with \ce{SiO2} content. The mean absolute error (MAE) of the LightPFP-predicted density across the five compositions is 0.014 $g/cm^3$.

\begin{figure}
\includegraphics[width=0.8\textwidth]{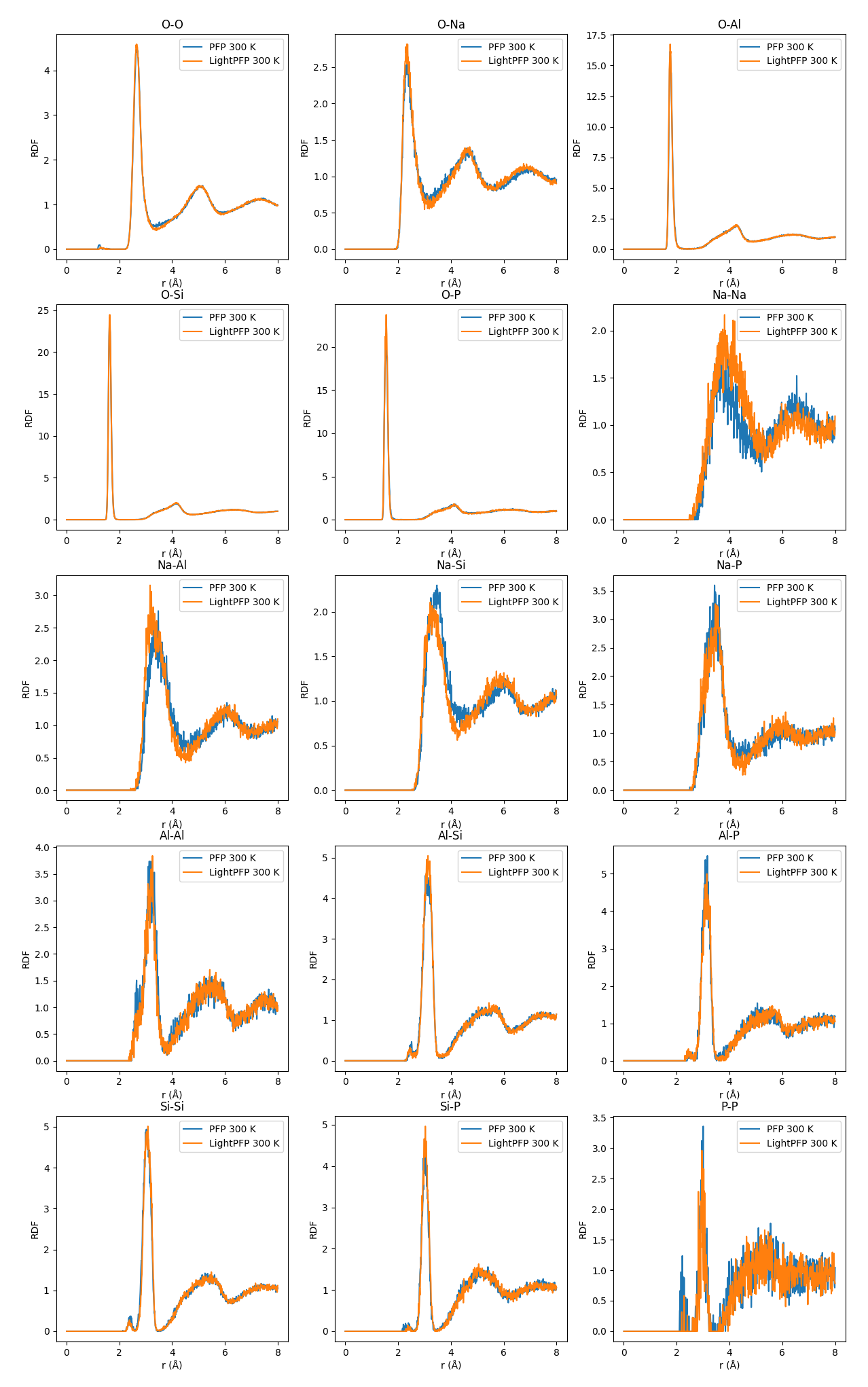}
\caption{Radial distribution function of 74\ce{SiO2}-16\ce{Al2O3}-18\ce{Na2O}-8\ce{P2O5}}
\label{fig:appendix_8_rdf}
\end{figure}

In addition, We evaluated the glass structure using the RDF. Figure \ref{fig:appendix_8_rdf} shows the RDF for the composition 74\ce{SiO2}–16\ce{Al2O3}–18\ce{Na2O}–8\ce{P2O5}. LightPFP is in good agreement with PFP across the principal pair correlations.

\begin{table}[htbp]
\centering
\caption{Elastic properties of 74\ce{SiO2}-16\ce{Al2O3}-18\ce{Na2O}-8\ce{P2O5}}
\begin{tabular}{lrrrr}
\hline
& LightPFP & PFP & abs error & relative error \\
\hline
C11 & 53.312 & 55.138 & 1.825 & 0.033 \\
C12 & 18.077 & 15.643 & 2.434 & 0.156 \\
C13 & 18.169 & 15.699 & 2.470 & 0.157 \\
C22 & 58.004 & 56.435 & 1.570 & 0.0278 \\
C23 & 19.364 & 16.745 & 2.619 & 0.156 \\
C33 & 58.283 & 58.618 & 0.335 & 0.00571 \\
C44 & 19.709 & 20.521 & 0.812 & 0.0396 \\
C55 & 19.283 & 20.065 & 0.782 & 0.0390 \\
C66 & 18.493 & 19.606 & 1.113 & 0.0568 \\
bulk modulus & 31.163 & 29.580 & 1.583 & 0.0535 \\
shear modulus & 19.081 & 20.168 & 1.087 & 0.0539 \\
Young's modulus & 47.540 & 49.299 & 1.759 & 0.0357 \\
Poisson ratio & 0.246 & 0.222 & 0.0235 & 0.106 \\
\hline
\end{tabular}
\label{tab:appendix_8_elastic}
\end{table}

To assess mechanical response, we computed the elastic stiffness tensor for the representative glass 74\ce{SiO2}–16\ce{Al2O3}–18\ce{Na2O}–8\ce{P2O5} with both PFP and LightPFP. Results are listed in Table \ref{tab:appendix_8_elastic}, including major components of elastic tensor and derived bulk, shear, and Young’s moduli, as well as Poisson’s ratio.

These results indicate that LightPFP can robustly learn and transfer the structural and mechanical behavior of complex multicomponent oxide glasses from PFP, while accommodating substantial composition variation from silica-rich to phosphate-rich regimes.

\clearpage

\section{Application 9: Heterogeneous grain boundary between FCC Cu and BCC Mo}
\label{sec:app9}

In this example, we use LightPFP to investigate the energetics of grain boundaries between FCC Cu and BCC Mo via large-scale atomistic simulations.

\subsection{Student fine-tuning}
\label{app9:fine-tuning}
\begin{figure}
\includegraphics[width=0.5\textwidth]{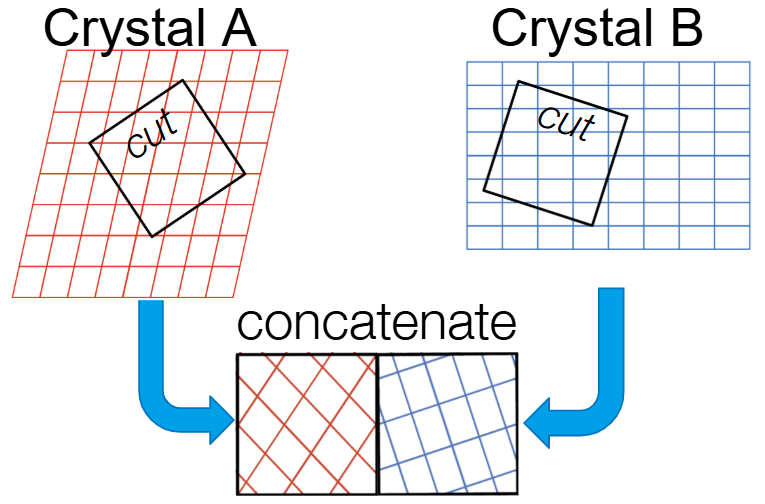}
\caption{Illustration of cut-and-concatenate method}
\label{fig:appendix_9_illustration}
\end{figure}

We constructed initial Cu/Mo bicrystals using the cut-and-concatenate method, starting from ideal FCC Cu and BCC Mo crystals. The method is shown in Fig. \ref{fig:appendix_9_illustration}. Simply speaking, the method takes two input crystal structures (i.e., Cu and Mo), cuts out cubic fragments at arbitrary positions and orientations, and then stitches them together to form grain-boundary–like structures. Unlike predefined low-energy grain boundaries structures, the generated structures are highly diverse. They often exhibit large lattice mismatch, and because the original periodic order is disrupted, numerous defects are introduced. This is not a drawback; rather, it is advantageous for training MLIPs, as it provides more off-equilibrium data samples and thereby improves the stability and robustness of the MLIP.

The initial database was assembled via short MD sampling on these crystal and grain boundary structures. Specifically, we performed NVT trajectories at 500 K, 1000 K, and 1500 K to introduce thermal disorder and local reconstructions, and NPT trajectories at 300 K, 400 K, 500 K, and 600 K to allow relaxation and sampling of strain-accommodated interfacial structures. The initial LightPFP model is trained on this dataset.

Starting from this initial model, we performed active learning to improve accuracy and robustness. For one initial Cu/Mo bicrystal structure generated by cut-and-concatenate method, we ran MD from 300 K to 1000 K using a two-step cycle: (1) 2 ps in the NVT ensemble to enable short-time reconstructions, followed by (2) 5 ps in the NPT ensemble to capture stress relaxation and incipient structural transformations. After that, configurations were relaxed by geometry optimization. Frames identified as low accuracy were added to the training set and the model was retrained. Iterating this procedure produced a LightPFP model that faithfully describes bulk phases and interface structure.

\subsection{Evaluation using PFP}
\label{app9:eval-by-pfp}
\begin{figure}
\includegraphics[width=0.5\textwidth]{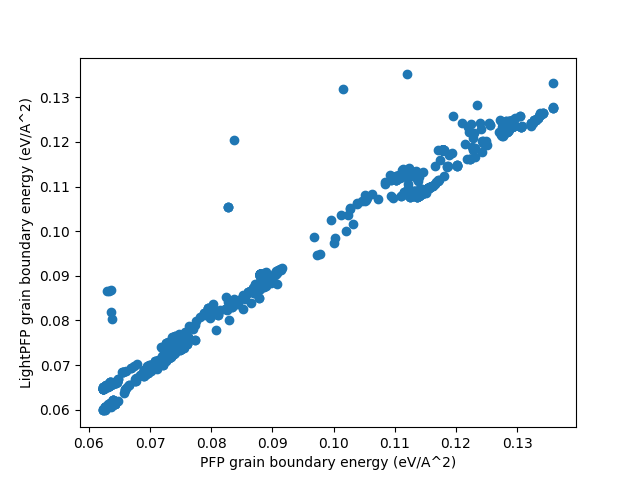}
\caption{Comparison of grain boundary energy of 1000 Cu/Mo interface structures calculated by PFP and LightPFP}
\label{fig:appendix_9_gb_energy}
\end{figure}

For evaluation, we generated 1000 low-strain CSL Cu/Mo grain-boundary structures using pymatgen, covering a broad range of misorientations and in-plane shifts. For each structure, we performed structure optimization first, and then computed the grain-boundary energy with both the reference PFP model and LightPFP. The grain-boundary energy was obtained by subtracting appropriate bulk reference energies for FCC Cu and BCC Mo from the total energy of the bicrystal and normalizing by the interfacial area (accounting for the two interfaces in periodic slabs). The resulting parity plot is shown in Fig. \ref{fig:appendix_9_gb_energy}. Across the 1000-member test set, LightPFP achieves good agreement with PFP for the vast majority of boundaries.

Overall, these results demonstrate that LightPFP can accurately predict Cu/Mo grain-boundary energies over a broad structural space while matching the reference PFP model for most configurations, validating its use for high-throughput screening and large-scale interfacial simulations.

\clearpage

\section{Application 10: Micelle simulation}

In this example, we investigated micelle formation using LightPFP. Micellization involves collective structural rearrangements of surfactant molecules, where hydrophilic groups orient toward the aqueous phase and hydrophobic tails aggregate to minimize their exposure to water.

\begin{figure}[h!]
\includegraphics[width=0.5\textwidth]{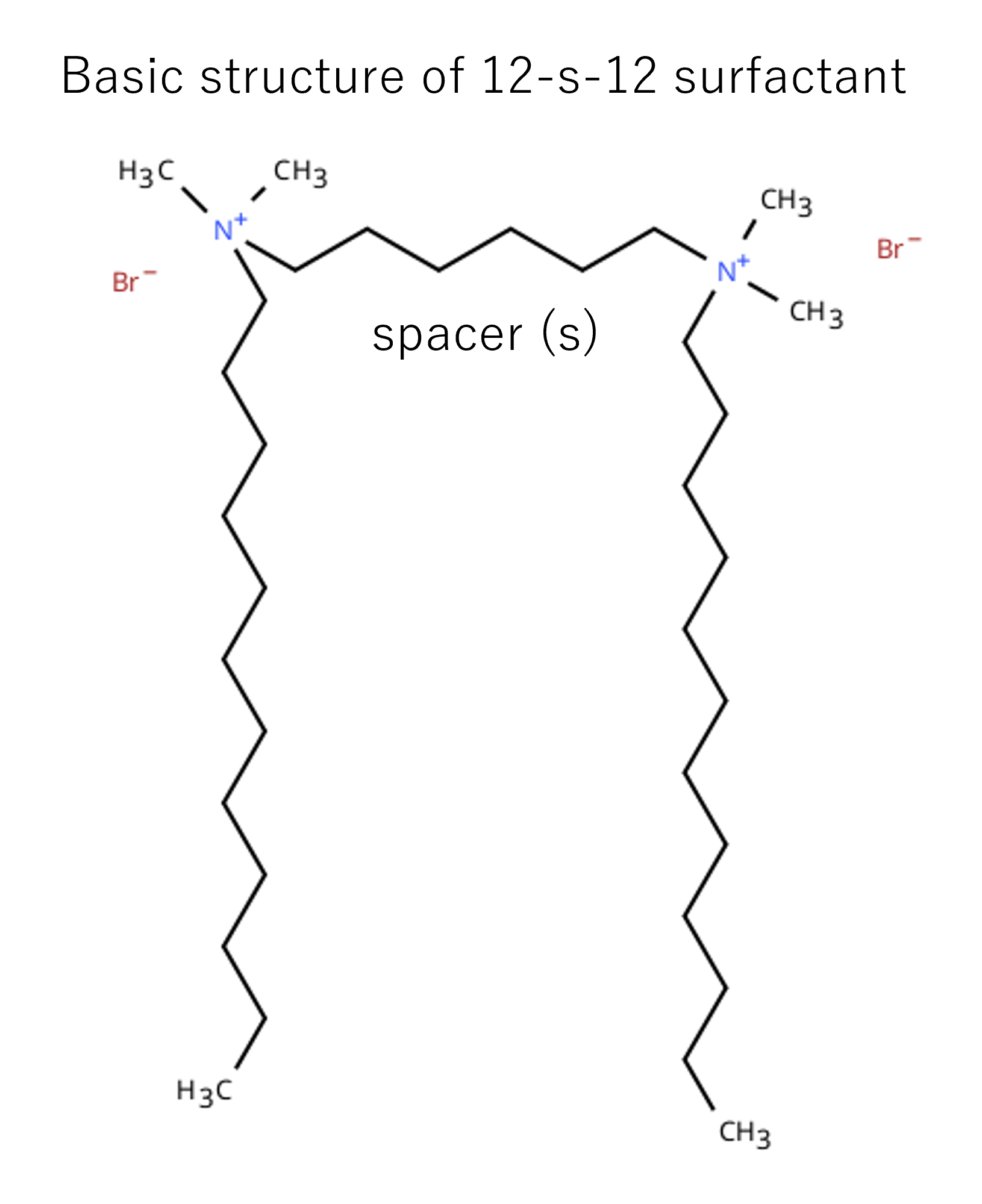}
\caption{Structure of 12-$s$-12 gemini surfactant with dual hydrophobic tails, dual cationic heads, and bromide counter-ions.}
\label{fig:appendix_10_gemini_structure}
\end{figure}

As a representative system, we selected a gemini surfactant with dual hydrophobic tails and dual cationic headgroups, accompanied by bromide counter-ions (Br$^-$). The 12–s–12 family (where $s$ denotes the spacer length) is known to form micelles efficiently due to the structural characteristics of its twin hydrophobic tails (see Fig.~\ref{fig:appendix_10_gemini_structure}).

\subsection{Student fine-tuning}
\label{app10:fine-tuning}
Initial structures for the 12–s–12 surfactant were prepared from SMILES expressions. For each spacer length $s$, we independently constructed initial datasets by generating multiple solvated structures with different numbers of surfactant and water molecules. Separate LightPFP models were therefore prepared for each value of $s$.

During active learning, small systems were used in the early iterations (0–4), medium-sized systems in iterations 5–8, and the largest systems in iterations 9–12. Within each iteration, internal loops varied the number of surfactant molecules, the water content, and the temperature (300–500 K) to enhance configurational diversity. After completing active learning, the collected dataset was used for final model training to obtain the LightPFP model for production MD. The entire workflow required approximately half a day and was repeated for $s = 2, 6, 10$.

\subsection{Large-scale MD simulation}
\label{app10:large-scale-md}
\begin{figure}[h!]
\includegraphics[width=0.5\textwidth]{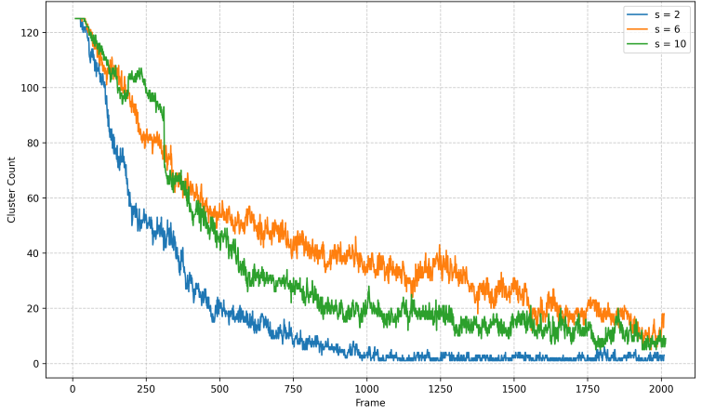}
\caption{Evolution of the number of clusters over time during 2 ns NPT simulations at 350 K for surfactants with spacer lengths $s$ = 2, 6, and 10.}
\label{fig:appendix_10_cluster}
\end{figure}

\begin{figure}[h!]
\includegraphics[width=\textwidth]{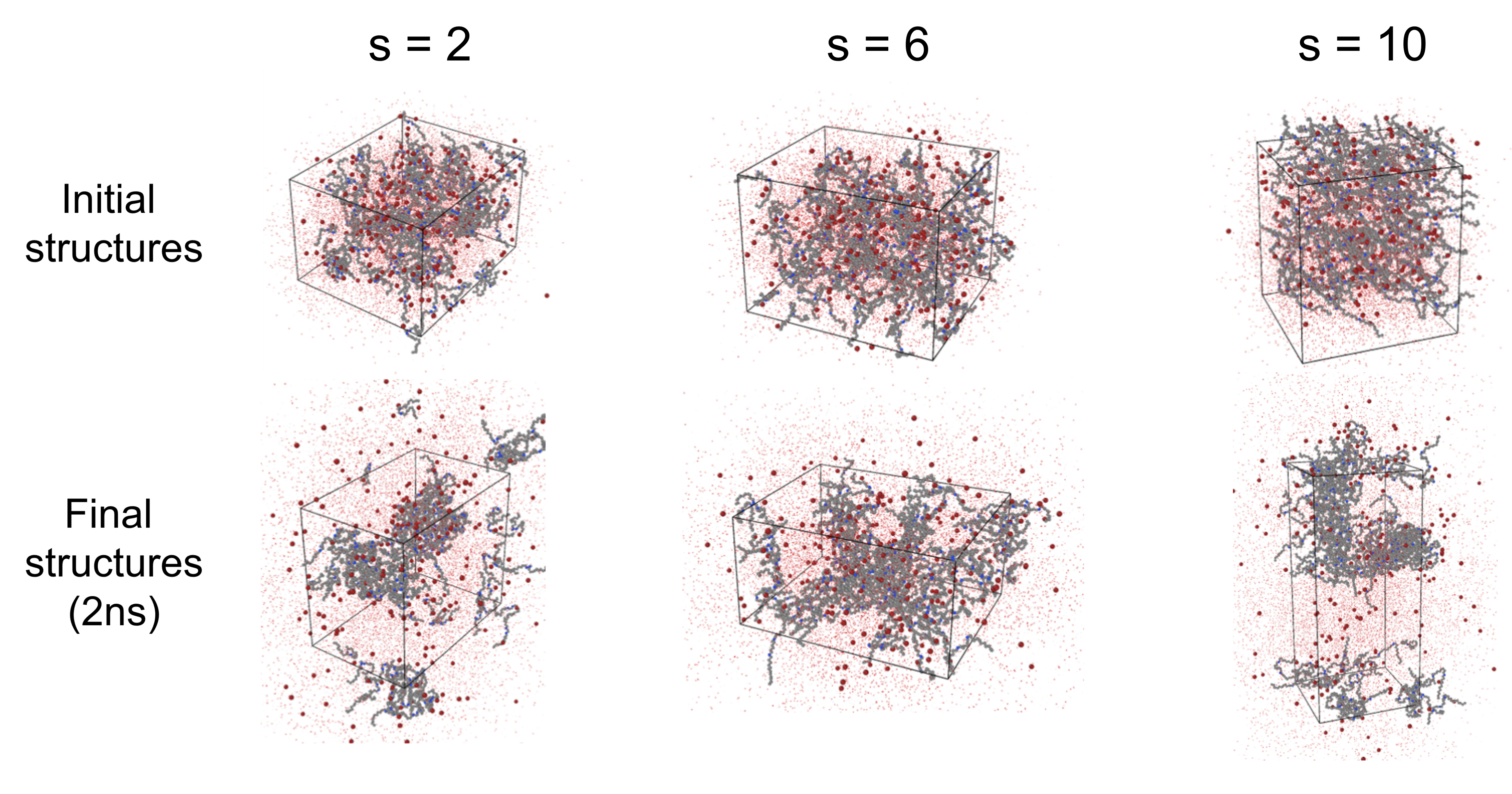}
\caption{Snapshots from MD simulations showing initial and final structures for systems with spacer lengths $s$ = 2, 6, and 10, demonstrating micelle formation.}
\label{fig:appendix_10_snapshots}
\end{figure}

Production simulations consisted of 2 ns NPT runs at 350 K for $s = 2, 6, 10$. Fig.~\ref{fig:appendix_10_cluster} shows the time evolution of the number of surfactant clusters. Initially, the cluster count equals the number of surfactant molecules; as aggregation proceeds, this count decreases. In all systems, the number of clusters gradually reduced, ultimately forming a large aggregate consistent with early-stage micelle formation. Representative initial and final snapshots are provided in Fig.~\ref{fig:appendix_10_snapshots}. The final structures exhibit the expected morphology, where hydrophilic headgroups orient outward and hydrophobic tails are sequestered inside the aggregate.

While 2 ns is insufficient to observe complete micelle formation, these results demonstrate that LightPFP can capture the essential aggregation behavior. Notably, the use of the R2SCAN mode was critical for stability: when PFP’s PBE+D3 mode was used instead, the simulations became unstable and eventually collapsed, likely due to the poorer description of water properties (e.g., RDFs, viscosity, density) at the PBE level. The improved accuracy of R2SCAN for aqueous systems appears to contribute significantly to the robustness of the MD simulations.

\clearpage

\section{Application 11: Chemical mechanical polishing of Si surface}
To illustrate the practical application of LightPFP, we present an example reproducing abrasive rolling during the chemical mechanical polishing (CMP) of crystalline silicon by a silica particle \cite{si2011abrasive}. The simulation box consists of a Si(100) slab and a spherical \ce{SiO2} particle placed above the slab. The simulation protocol follows the two-stage loading scheme commonly employed in molecular dynamics (MD) studies of CMP: first, an external normal (“down”) load is applied to the silica particle to bring it into contact with the silicon surface; then, a tangential (“driving”) load is applied to the same particle to induce rolling and sliding motion across the slab.

\subsection{Student fine-tuning}
\label{app11:fine-tuning}
The system construction and initial training dataset are intentionally minimal. Crystalline Si and \ce{SiO2} structures are generated, from which spherical \ce{SiO2} clusters are cut. Si and \ce{SiO2} slabs are then prepared with vacuum layers, and a representative solid/solid interface is assembled through the simple cut-and-concatenate procedure introduced in Section~\ref{sec:app9}. To populate the initial dataset, we sample crystalline structures using MD, rattle, compression, deformation, and vacancy methods, while non-crystalline and interfacial structures are sampled using MD and rattle only. This combination of bulk, surface, cluster, and interface configurations provides chemically diverse yet computationally inexpensive coverage for the first LightPFP fitting.

The active-learning protocol is used to generate a more robust LightPFP model. Active learning uses the same MD protocol as described above. The structure with large discrepancy with PFP will be detected and collected for further training. The active cycle is organized as a simple curriculum across particle sizes (small → medium → large), so that the model learns contact physics at increasing mechanical intensity. After the active acquisition phase the accumulated dataset is used for a single, consolidated retraining step to produce the final LightPFP model.

\subsection{Evaluation using PFP}
\label{app11:eval-by-pfp}
\begin{figure}
\includegraphics[width=0.5\textwidth]{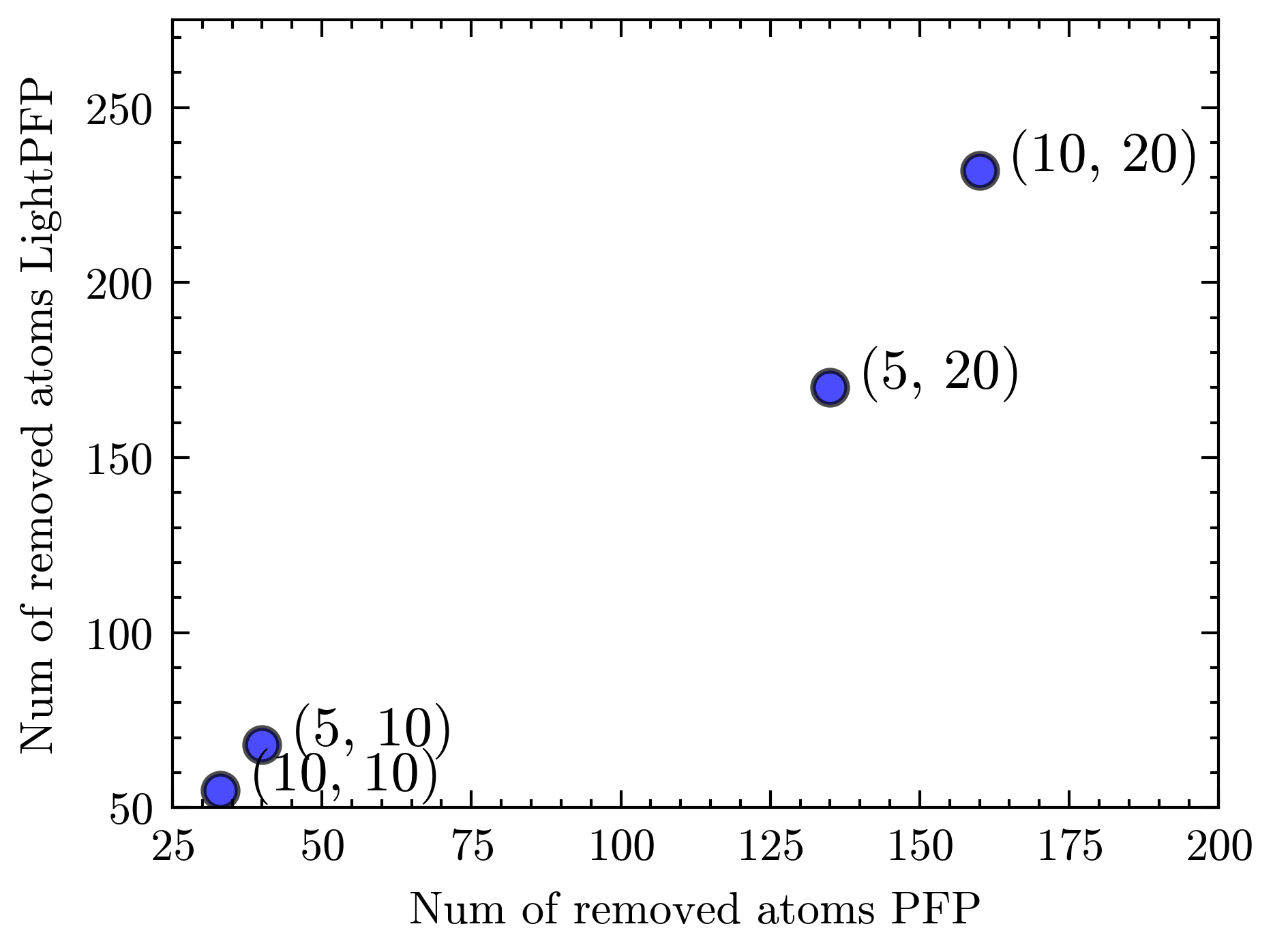}
\caption{Number of removed Si atoms from Si slab during MD simulation of chemical mechanical polishing. The down/driving force applied on particle is annotated.}
\label{fig:appendix_11_remove}
\end{figure}

For validation, we compare LightPFP with a PFP reference model using a small simulation cell containing 6133 atoms, following the same two-stage loading protocol. Four loading conditions were tested with down/driving forces of (5/10 eV/\text{\AA}), (10/10 eV/\text{\AA}), (5/20 eV/\text{\AA}), and (10/20 eV/\text{\AA}), each simulated for 200 ps. The number of removed atoms was determined by counting those displaced more than 2~\text{\AA} during the MD trajectory, as shown in Figure~\ref{fig:appendix_11_remove}. Both LightPFP and PFP exhibit consistent trends, with the number of removed atoms increasing in the order: (10/10) < (5/10) < (5/20) < (10/20). However, LightPFP consistently predicts a higher number of removed atoms than PFP under all loading conditions. This difference likely arises from subtle variations in surface interaction strength. Despite this quantitative deviation, the qualitative agreement in trend demonstrates that LightPFP accurately reproduces the underlying contact and material removal physics while offering improved sensitivity to interfacial dynamics.

\subsection{Large-scale MD simulation}
\label{app11:large-scale-md}
\begin{figure}
\includegraphics[width=\textwidth]{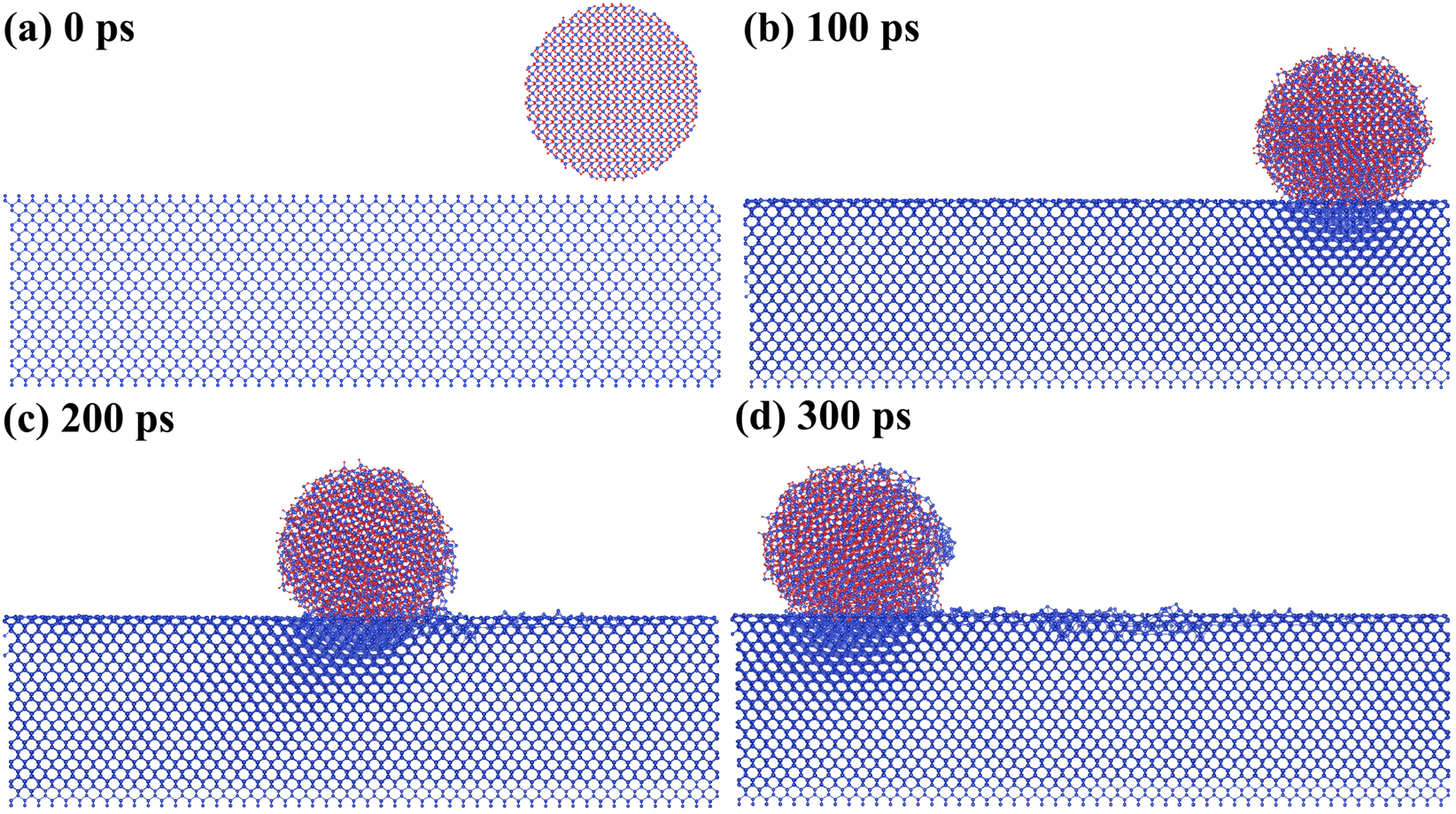}
\caption{Time evolution of the molecular dynamics simulation illustrating the chemical mechanical polishing (CMP) of crystalline Si by a 5 nm \ce{SiO2} particle. Panels (a)–(d) correspond to 0, 100, 200, and 300 ps, respectively, showing progressive rolling motion, and surface atom removal on the Si slab.}
\label{fig:appendix_11_cmp}
\end{figure}

Finally, we demonstrate scalability by simulating the polishing of Si by a 5~nm-diameter silica particle in a dry environment. This production system comprises 59,266 atoms and is evolved for 0.6~ns under a 50~eV/\text{\AA} normal load and a 100~eV/\text{\AA} driving load. The snapshots of the MD trajectory are presented in Figure~\ref{fig:appendix_11_cmp}, showing the motion of the \ce{SiO2} particle and the corresponding polishing process.

\clearpage

% \bibliography{refs}
\bibliography{aipsamp}